\documentclass[prd,amsfonts,onecolumn,superscriptaddress,aps,nofootinbib,11pt]{revtex4-1}

\pdfoutput=1
\usepackage[utf8]{inputenc}
\usepackage{amsmath,amssymb,mathrsfs,bbm}
\usepackage{graphicx}
\usepackage{enumerate}
\usepackage{multirow}
\usepackage{float}
\usepackage[usenames,dvipsnames]{xcolor} 
\usepackage{hyperref}
\usepackage{slashed}
\usepackage{soul}
\usepackage[normalem]{ulem}
\hypersetup{colorlinks=true,
		breaklinks=true,
    linkcolor=Blue,          
    citecolor=Plum,        
    urlcolor=YellowOrange           
}

\newcommand{\met}{\not\!\!\! E_T}
\newcommand{\ptm}{\not\!\! p_T}

\newcommand{\mbbllvv}{M_{bb\ell\ell\nu\nu}}
\newcommand{\ptllvv}{p_T(\ell\ell\nu\nu)}
\newcommand{\x}{\mathbf{x}}
\newcommand{\z}{\mathbf{z}}
\newcommand{\tgt}{\mathbf{r}}

\begin{document}
\title{Variational Autoencoders for Regression: \\
Recovering Fully Leptonic $b\bar{b}W^+W^-$ in Di-Higgs Searches}

\author{Alexandre Alves}
\email{aalves@unifesp.br}
\affiliation{Departamento de Física, Universidade Federal de São Paulo, Diadema, 09913-030, Brazil}
\author{Eduardo da Silva Almeida}
\email{almeidae@ufba.br}
\affiliation{Departamento de Física do Estado Sólido, Universidade Federal da Bahia, R. Barão de Jeremoabo, Ondina, 40170-115, Salvador - Bahia, Brazil}
\author{Igor Neiva Mesquita}
\email{igor_neiva@usp.br}
\affiliation{Instituto de Física, Universidade de São Paulo, \\
R. do Matão 1371, 05508-090 São Paulo, Brazil}


\begin{abstract}
 The search for double Higgs production in $b\bar{b}W^+W^-$, where both $W$ bosons decay to leptons, has been rehabilitated as a good option to look for that key process to the Standard Model scalar sector study in the LHC. The missing neutrinos, however, hinder the reconstruction of useful information like the Higgs pair mass, which is very sensitive to the trilinear Higgs self-coupling. We present a solution to that problem using a Variational Autoencoder for Regression (VAER) to reconstruct the Higgs and top pairs decays $hh,t\bar{t}\to b\bar{b}W^+W^-\to b\bar{b}\ell^+\ell^{\prime -}\nu_\ell\bar{\nu}_{\ell^\prime}$. The algorithm predicts the invariant mass of non-resonant $hh$ irrespective of the trilinear coupling, even for events whose Higgs self-couplings were never presented to it. VAER is also able to identify a new Higgs resonance in an unsupervised way, showing generalization power for events not presented in its training phase. Finally, we demonstrate that VAER prediction is as useful to statistical inference as ground truth simulated distributions by computing a $\chi^2$ between trilinear coupling hypotheses based on binned invariant mass distributions of $b\bar{b}\ell^+\ell^{\prime -}\nu_\ell\bar{\nu}_{\ell^\prime}$.
\end{abstract}

\maketitle
\section{Introduction}
\label{sec:intro}

 A challenging problem in high-energy physics phenomenology is recovering information lost in collisions that produce feebly interaction particles that escape detection like neutrinos. In particular, for kinematics reconstruction, missing neutrinos pose a problem whenever we want to detect resonances or measure theory parameters that are sensitive to that kinematics. For example, to measure the $W$ boson mass, we rely only on the kinematic distributions of the charged lepton that accompany the neutrino in the leptonic decay mode since it is not possible to reconstruct its four-momentum in this case, and because two jet decay is plagued by overwhelming QCD backgrounds. In the absence of a resonant peak, the {\it transverse mass}, a $W$ mass-sensitive variable, is used to compare data against prediction. As an outcome, the $W$ mass is measured with much less precision than the $Z$ mass whose resonance peak is available through the lepton pair invariant mass~\cite{10.1093/ptep/ptac097}.
 
 Transverse mass is a typical feature that is engineered to substitute for the missing information that prevents us from building an optimal variable to measure a theory parameter. Many examples exist in other contexts. For instance, in models with dark matter, measurements of the intermediate particles that produce them are hindered, like in SUSY models, where sleptons might decay promptly to a charged lepton and a stable neutralino that escapes detection and carries away the information on the slepton's mass. Instead of a clear peak from where the mass can be read, one needs to look up the information in the endpoints~\cite{Lester:2006cf} of suitable kinematic distributions at the cost of precision. Other ingenious solutions and variables are devised to solve that kind of problem, but, of course, it would be much better if we could somehow recover the information lost to build the most sensitive variables to measurements. For a good review of kinematic variables engineering, see Ref.~\cite{Franceschini:2022vck}.

 In the SM context, missing particles also get in the way of accessing vital information. Among the SM measurements, the scalar potential is of prime importance, even more so now that gravitational wave astronomy opened up the possibility of giving hints about the electroweak phase transition~\cite{Caldwell:2022qsj}. Apart from that, anyway, new physics might lurk in deviations of the SM scalar parameters. The most straightforward way to access that information is by measuring the Higgs self-couplings in double and triple Higgs production at colliders. 
In the SM, the Higgs self-interactions, after electroweak symmetry breaking, are given by
\begin{equation}
    V(h) = \frac{1}{2}m_h^2 + \kappa_3\lambda_{SM} h^3+\frac{1}{4}\kappa_4\lambda_{SM}h^4
\end{equation}
where $\lambda_{SM} = m_h^2/2v^2 \approx 0.13$, and $m_h=125$ GeV, and $v=246$ GeV represent the SM Higgs mass and the vacuum expectation value. Here, $\kappa_3$ and $\kappa_4$ parametrize deviations from the SM values. As we are interested in studying trilinear self-couplings, we define $\kappa_3\equiv \kappa_\lambda$ from now on.

In the LHC, the prospects of detecting Higgs self-interactions in single channels until the end of the experiment are not particularly bright, especially for the quartic coupling. Only by combining several search channels a 68\% confidence limit (CL) of $0.57\leq \kappa_\lambda \leq 1.5$ can be reached~\cite{Cepeda:2019klc}. Currently, $-1 \lesssim \kappa_{\lambda} \lesssim 6$ \cite{ATLAS:2022jtk,CMS:2022dwd,atlascollaboration2023studies} at 95\% CL.

Among the decay channels for $hh$ studies, $b\bar{b}\gamma\gamma$ is the most promising one and dominates the combination, while $b\bar{b}W^+W^-$ and $b\bar{b}ZZ$ are the less important ones~\cite{Cepeda:2019klc}. Recently, however, the authors of Ref.~\cite{Kim:2018cxf} rehabilitated $b\bar{b}W^+W^-$ by computing new features that can efficiently discern between $b\bar{b}W^+W^-$, with leptonic $W$ bosons, from double Higgs and its backgrounds, mainly the $t\bar{t}$ events, increasing the statistical significance by a factor of $\sim 4$ and reaching $\sim 2.1\sigma$ after 3~ab$^{-1}$. This makes the fully leptonic $b\bar{b}W^+W^-$ as competitive as the best channels to look for $hh$.

The $hh$ production rate is sensitive to $\lambda$, and an inference of this parameter can be made by counting the number of events in excess of expected backgrounds. However, the dependence of the total cross section on $\lambda$ is polynomial, causing a twofold ambiguity in the determination of the trilinear coupling for a given number of measured events. That ambiguity will probably not be lifted at the 95\% CL even after 3 ab$^{-1}$ for a single experiment, so a combination of the ATLAS and CMS results is important~\cite{Cepeda:2019klc}. Better prospects are expected at the next linear collider generation~\cite{Roloff:2019crr,Contino:2013gna} where both the total rates and the shape of suitable distributions can be used to constrain the $\lambda$ parameter. 

In fact, the same strategy can be employed at hadron colliders. In this respect, the $hh$ invariant mass distribution shows good sensitivity to the $\lambda$ parameter due to the contributions from a triangle and a box diagram to the total amplitude. The exact dependence on the trilinear coupling and the top quark Yukawa coupling determines the interference pattern of the two contributions shaping the $hh$ mass. That shape can be used to further test the coupling hypotheses. However, in the case of final states where neutrinos are present, like fully leptonic $b\bar{b}W^+W^-$, for example, the $hh$ mass cannot be reconstructed. Moreover, detector and hadronization effects smear the $hh$ mass distributions, blurring the distinction between two sets of couplings and diminishing the advantage of using the shape of the distribution.

In this work, we propose a neural network solution -- a Variational Autoencoder for Regression (VAER) algorithm -- that addresses the difficulties in recovering the $hh$ and $t\bar{t}$ masses from the observable kinematics from detector-level events. We will show that VAER has a very good generalization power predicting distributions of events never presented at the learning phase of the algorithm both for non-resonant and resonant $hh$ production. We will demonstrate that the predicted distributions can be used for practical statistical purposes, for example, in a $\chi^2$ test between coupling hypotheses based on partonic binned $b\bar{b}\ell^+\ell^-\nu_\ell\bar{\nu}_{\ell^\prime}$ mass.  The proposed algorithm can be used in many other contexts, like dark matter searches and long-lived particles that escape detectors. It can also be used as an unfolding algorithm to discount for detector effects and difficulties brought by hadronization of jets once it learns the partonic underlying information from simulated events. Finally, we envisage applications to recover other variables hidden by information leakage, such as $W$ and $Z$ polarization studies and spin and mass measurements that need a full reconstruction of kinematic variables.

Our paper is organized as follows. In section~\ref{sec:vaer}, we describe the VAER algorithm; in section~\ref{sec:sim}, details of our simulations are provided; in sections~\ref{sec:non-res} and~\ref{sec:res}, our results for the non-resonant and the resonant $hh$ production are presented, respectively; in section~\ref{sec:conclusions}, we present our conclusions and an outlook of possible applications and future work using VAER.

\section{Variational Autoencoder for Regression}
\label{sec:vaer}

 The VAER algorithm was originally designed to predict the age of a person from the 3D structural brain magnetic resonance image~\cite{DBLP:journals/corr/abs-1904-05948}. The authors of that work also demonstrate that the regression task works even for tabular data representing other types of measurements of the brain. To understand how VAER works, we need to recall the basics of autoencoders and variational autoencoders.

 An autoencoder works by learning a dimensionally reduced representation of the data, encoding the original data, $\mathbf{x}$, into a latent space, $\mathbf{z}$, through a neural network $\mathbf{z}=E_\theta(\x)$, where $\theta$ represents the parameters of the neural net encoder. The encoder is stimulated to produce good latent representations of the original data by decoding the latent representation of the data back to $\mathbf{x}^\prime$ through another neural net $\mathbf{x}^\prime = D_\phi(\mathbf{z})$, where $\phi$ represents the parameters of the neural net decoder, and minimizing the dissimilarity  between $\mathbf{x}$ and $\mathbf{x}^\prime$, for example, their mean squared error
 \begin{equation}
     \underset{\theta,\phi}{\mathrm{argmin}} \frac{1}{|{\cal D}|}\sum_{\x\in{\cal D}} [\x-D_\phi(E_\theta(\x)]^2\; .
     \label{eq:vae_mse}
 \end{equation}

 
 The Variational Autoencoder (VAE) \cite{Kingma_2019}, by its turn, is a generative neural network model that learns the probability distribution of a dataset, ${\cal D}$. As such, it can be used to draw new instances from that distribution and that resemble the data.  The variational aspect of the algorithm refers to the probabilistic nature of the latent space. Instead of a static encoding, the encoder is built as a Gaussian function that learns the mean and the standard deviation of the data, that is, a neural net, $\mathbf{\mu}_\theta$, is trained to encode the multidimensional mean of the data set, and another neural net, $\mathbf{\sigma}^2_\theta$, to capture the variance of the dataset. This way, given a data point, $\mathbf{x}$, its latent representation is $\mathbf{z}\sim {\cal N}(\mathbf{x};\mathbf{\mu}_\theta(\x),\mathbf{\sigma}^2_\theta(\x))$. Once the latent representation has been learned, creating new instances is easy. Draw a $\mathbf{z}$ and decode it with the neural net decoder such that $\mathbf{x}^\prime = D_\phi(\mathbf{z})$ is a brand new instance, not contained in the dataset, but hopefully emulating a true member of ${\cal D}$. Notice that, in VAE, $E_\theta$ is probabilistic, but $D_\phi$, is deterministic.
 
 Let us start with the distribution of the data conditioned on a latent representation vector, $\mathbf{z}$, 
 \begin{equation} 
     P(\x)=\int_{\cal Z} P(\x|\z)p(\z) d\z = \mathbbm{E}_{\z\sim p(\z)}[P(\x|\z)]\; .
     \label{eq:marginal}
 \end{equation}

 We know neither the prior $p(\z)$ nor the likelihood $P(\x|\z)$, so we use neural networks to learn them from data. The problem is that this process is very inefficient as the majority of latent points are not likely to produce $\x$ that resembles the data. Instead, we can learn a function, $q_\phi(\z|\x)$, that is conditioned on $\x$ and write $P(\x)$ as
 \begin{equation}
     P(\x) = \mathbbm{E}_{\z\sim q_\phi(\z|\x)}[P_\theta(\x|\z)]\; .
     \label{eq:px}
 \end{equation}
 
 Here, $q_\phi(\z|\x)$ and $P_\theta(\x|\z)$ now represent the encoder and the decoder models, respectively. To produce a generative model, we just need to have a pdf for the latent space from which we draw latent vectors that can be decoded into instances that emulate drawing from $P(\x)$ itself. This can accomplished with $q_\phi(\z|\x)={\cal N}(\z;\mu_\phi(\x),\sigma^2_\phi(\x))$, where $\mu_\phi$ and $\sigma^2_\phi$ are modeled by neural networks. There is an important computational detail here, though: $\z$ should be randomly generated in the training phase, as Eq.~\eqref{eq:px} suggests, but backpropagation does not work in sampling nodes. 
 The solution is the {\it reparametrization trick}, calculating points of the latent space as $\z = \mu_\phi(\x)+\sigma_\phi(\x)\odot\epsilon,\; \epsilon\sim{\cal N}(0,1)$, with deterministic mean and variance. But how to learn the mean, $\mu_\phi$, and the variance, $\sigma^2_\phi$, models? 

 We calculate the following Kullbach-Liebler (KL) divergence \cite{KL}
 \begin{equation}
     D_{KL}(q_\phi(\z|\x)||P(\z|\x))=\mathbbm{E}_{\z\sim q_\phi}\left[\log\frac{q_\phi(\z|\x)}{P(\z|\x)}\right]=\mathbbm{E}_{\z\sim q_\phi}\left[\log q_\phi(\z|\x)-\log\frac{P(\x|\z)p(\z)}{P(\x)}\right]
 \end{equation}
  using the Bayes' rule for $P(\z|\x)$. This expression can be rearranged as follows
 \begin{equation}
     \log P(\x) = \mathbbm{E}_{\z\sim q_\phi}\left[\log\frac{P_\theta(\x|\z)p(\z)}{q_\phi(\z|\x)}\right]+ D_{KL}(q_\phi(\z|\x)||P(\z|\x))\; ,
     \label{eq:lik-vae}
 \end{equation}
 where $\mathbbm{E}_{\z\sim q_\phi}[\log P(\x)]=\log P(\x)$ once $P(\x)$ does not depend on $\z$.  
 
 The first term on the right side of this expression is called the Evidence Lower Bound (ELBO), ${\cal L}(\x;\theta,\phi)$. Because KL divergence is always non-negative, $\log P(\x) \geq {\cal L}(\x;\theta,\phi)$. This inequality is very convenient for obtaining an objective function for the learning process. The posterior distribution $P(\z|\x)$ is probably a too difficult multidimensional distribution to be learned, but $P_\theta(\x|\z)$ is the deterministic neural network decoder while $p(\z)={\cal N}(0,1)$ is a prior distribution that can be taken as a simple normal distribution, for example. Thus, the first term of Eq.~\eqref{eq:lik-vae} can be modeled. 

 All this leads us to carry the inference process via a Maximum Likelihood Estimation (MLE). The goal is to maximize $\log P(\x)$, which is the same as maximizing the ELBO with respect to the neural net parameters $\theta$ and $\phi$, 
 \begin{eqnarray}
     \underset{\theta,\phi}{\mathrm{argmax}}\; \frac{1}{|{\cal D}|}\sum_{\x\in{\cal D}} \log P(\x) &=& \underset{\theta,\phi}{\mathrm{argmax}} \; \mathbbm{E}_{\x\in{\cal D}}[\log P(\x)] = \underset{\theta,\phi}{\mathrm{argmin}}\; \mathbbm{E}_{\x\in{\cal D}}[-{\cal L}(\x;\theta,\phi)] \nonumber \\
     & = & \underset{\theta,\phi}{\mathrm{argmin}}\; \mathbbm{E}_{\x\in{\cal D}}[-\mathbbm{E}_{q_\phi}[\log P_\theta(\x|\z)]+D_{KL}(q_\phi(\z|\x)||p(\z))]\; .
 \end{eqnarray}
 This is valid as long as $q_\phi(\z|\x)$ approaches the true posterior distribution $P(\z|\x)$ and saturates the lower bound as $D_{KL}(q_\phi(\z|\x)||P(\z|\x))\to 0$. 
 
 Now, we are ready to answer the question made previously: how to learn $\mu_\phi$ and $\sigma^2_\phi$? The MLE posed above can be solved by minimizing the loss function
 \begin{eqnarray}
     \text{Loss}(\x;\theta,\phi) & = & L_R+L_{KL} \nonumber \\
     &=& ||\x-\x^\prime(\theta)||+D_{KL}(q_\phi(\z|\x)||p(\z)) \nonumber \\
     &=& ||\x-\x^\prime(\theta)||-\frac{1}{2}[1+\log\sigma^2_\phi(\x)-\mu_\phi^2(\x)+\exp(\log\sigma^2_\phi(\x))]\; ,
 \end{eqnarray}
 where $\x^\prime(\theta) = P_\theta(\x|\z=\mu_\phi(\x)+\sigma_\phi(\x)\odot\epsilon),\; \epsilon\sim{\cal N}(0,1)$. $L_R=||\x-\x^\prime||$ is the reconstruction loss, and the distance measure between $\x$ and $\x^\prime$ can be chosen as the mean absolute error, the mean square error, or a cross-entropy measure, for example. The KL divergence can be calculated analytically when  $q_\phi(\z|\x)$ and $p(\z)$ are Gaussian functions as discussed earlier, resulting in the KL-loss, the $L_{KL}$ term. This is the standard VAE loss. 


 How can this algorithm be used for a regression task? The key ingredient is to build an orthogonal dimension in the latent space that is sensitive to variations of the target. Embedding this dimension into the latent space, hopefully, correlates the target variable to the data representation. The latent representation is then said to be disentangled. 
 
 In practice, VAER\footnote{The source code can be found in this address: https://github.com/QingyuZhao/VAE-for-Regression.} works via the variational inference of a {\it probabilistic regressor} for the target vector, $\tgt$. The likelihood distribution is now given by
 \begin{equation}
     P(\x) = \int_{{\cal Z},{\cal R}} P(\x,\z,\tgt)d\z d\tgt\; ,
 \end{equation}
 and taking the same steps that led us to Eq.~\eqref{eq:lik-vae}, gives us the ELBO for VAER
 \begin{equation}
     {\cal L}(\x;\{\theta\}) = \mathbbm{E}_{(\z,\tgt)\sim Q_\phi(\z,\tgt|x)}\left[\log\frac{P_\theta(\x,\z,\tgt)}{Q_\phi(\z,\tgt|\x)}\right]\; .
 \end{equation}
 
 The novelty is that the variables are now conditioned to $\tgt$. Assuming that $\z$ and $\tgt$ are independent variables, we have $Q_\phi(\z,\tgt|\x)=q_\phi(\z|\x)q_\varphi(\tgt|\x)$, where $q_\varphi(\tgt|\x)$ is a neural network regressor. Working on the ELBO expression above, we have (denoting parameters collectively as $\{\theta\})$ 
 \begin{equation}
     {\cal L}(\x;\{\theta\}) = \mathbbm{E}_{\z\sim q_\phi(\z|\x)}[\log P_\theta(\x|\z)]-\mathbbm{E}_{\tgt\sim q_\varphi(\tgt|\x)}[D_{KL}(q_\phi(\z|\x)||P_\vartheta(\z|\tgt))]-D_{KL}(q_\varphi(\tgt|\x)||p(\tgt))\; .
     \label{eq:elbo-vaer}
 \end{equation}

 $P_\vartheta(\z|\tgt)$ is the {\it latent generator}~\cite{DBLP:journals/corr/abs-1904-05948}, an essential component to correlate the latent vector to the regression target through $P_\vartheta(\z|\tgt)={\cal N}(\z;\mathbf{u}^T\odot\tgt,\sigma^2\mathbf{1})$ where $\mathbf{u}$ is a normalized vector. Note that the mean is a linear model of $\tgt$: $\mu_\vartheta(\tgt)=\mathbf{u}^T\odot\tgt$. This is sufficient to correlate $\tgt$ to a disentangled dimension from $\z$ such that traversing $\mathbf{u}$ yields $\tgt$-specific latent representations. Just like VAEs, here $q_\phi(\z|\x)$ is a Gaussian whose mean, $\mu_\phi(\x)$, and variance, $\sigma_\phi^2(\x)$, are neural net models while $P_\theta(\x|\z)$ is a neural net decoder. The regressor $q_\varphi(\tgt|\x)$ is actually a {\it probabilistic regressor} within this variational inference approach, and it is also modeled as a Gaussian distribution: $q_\varphi(\tgt|\x)={\cal N}(\tgt;\mu_\varphi(\x),\sigma^2_\varphi(\x)\mathbf{1})$ where $\mu_\varphi$ and $\sigma^2_\varphi$ are neural nets. The prior on $\tgt$ is assumed to be a simple standard Gaussian distribution, $p(\tgt)={\cal N}(\tgt;0,1)$.
 \begin{figure}[t]
    \centering
    \includegraphics[width=0.7\linewidth]{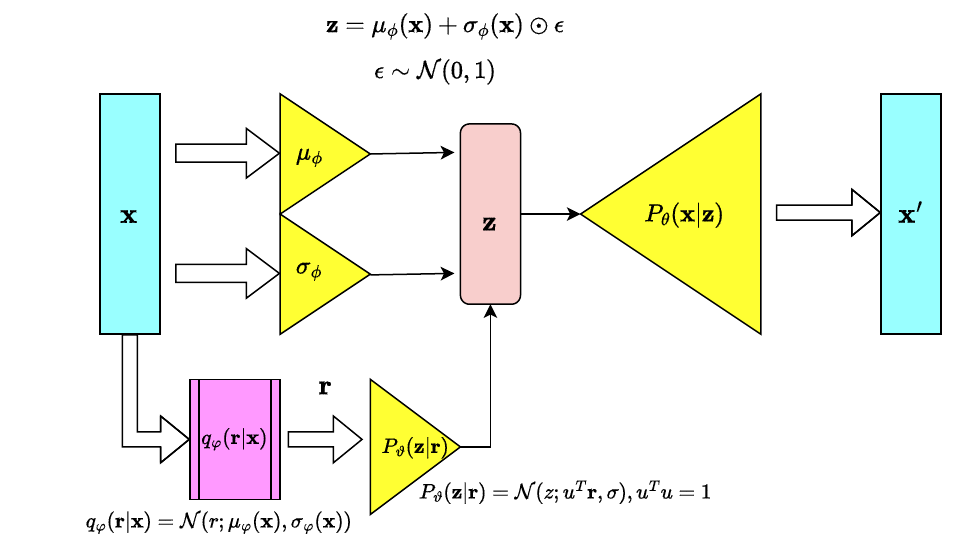}
    \caption{The graphical diagram of the Variational Autoencoder for Regression (VAER). Yellow triangles represent neural networks. The central rectangle represents the latent space, while the external ones, in cyan, are the input and output spaces. The purple rectangle, $q_\varphi(\tgt|\x)$, is the probabilistic regressor from which we make predictions. }
    \label{fig:vaer}
\end{figure}
 The loss function of VAER can now be derived,
 \begin{eqnarray}
     \text{Loss}(\x,\tgt;\{\theta\}) & = & L_R+L_{KL}+L_{reg} \nonumber \\
     L_R &=& ||\x-\x^\prime(\theta)|| \nonumber \\
     L_{KL} &=& -\frac{1}{2}\left[1+\log\sigma^2_\phi(\x)-\log\sigma^2_\vartheta(\tgt) -\frac{(\mu_\phi(\x)-\mu_\vartheta(\tgt))^2}{\sigma^2_\vartheta(\tgt)}-\frac{\sigma^2_\phi(\x)}{\sigma^2_\vartheta(\tgt)}\right] \nonumber \\
     L_{reg} &=& \frac{1}{2}\left[\log\sigma^2_\varphi(\x)+\frac{(\tgt-\mu_\varphi(\x))^2}{\sigma^2_\varphi(\x)}\right], \nonumber \\ 
     \label{eq:vaer-loss}
 \end{eqnarray}
where, again, $\x^\prime(\theta) = P_\theta(\x|\z=\mu_\phi(\x)+\sigma_\phi(\x)\odot\epsilon),\; \epsilon\sim{\cal N}(0,1)$. We depict a graphical diagram of VAER in Figure~\ref{fig:vaer}. The predicted target can be taken from $\hat{\tgt}=\mu_\varphi(\x)$ or, when convenient, as $\hat{\tgt}\sim {\cal N}(\tgt;\mu_\varphi(\x),\sigma^2_\varphi(\x)\mathbf{1})$. Let us now discuss the practical application of VAER to our problem.

\section{Simulation Details}
\label{sec:sim}

 We simulate partonic level events with \texttt{MadGraph5}~\cite{Alwall:2014hca} at the 14 TeV LHC for two types of process: 
 \begin{enumerate}
     \item Double Higgs production and decay
     \begin{equation}
         pp\to hh(j)\to b\bar{b}W^+W^-(j)\to b\bar{b}\ell^+\ell^{\prime -}\nu_\ell\bar{\nu}_{\ell^\prime}(j)
     \end{equation}
     up to one extra jet. The $W$ boson's leptonic decays comprise electrons and muons, $\ell=e,\mu$. The trilinear coupling is treated as a free parameter that controls the interference between the triangle and the box diagrams, and the Yukawa couplings are kept fixed at their SM values. We simulate 100k events for each $\lambda = \kappa \lambda_{SM}$, $\kappa$ from $-3$ to $3$ with steps of 0.5.

    \item The main background source, the top quark pair production
    \begin{equation}
        pp\to t\bar{t}\to b\bar{b}\ell^+\ell^{\prime -}\nu_\ell\bar{\nu}_{\ell^\prime}
    \end{equation}
    at the next-to-leading order QCD.
     
 \end{enumerate}

  Hadronization of jets was performed with \texttt{Pythia8}~\cite{Sjostrand:2007gs}, and detector effects were simulated with \texttt{Delphes3} with default settings, while jet reconstruction and clustering were performed with \texttt{Fastjet}~\cite{Cacciari:2011ma}.
  The MLM merging scheme~\cite{Mangano:2006rw} was adopted to merge hard and soft radiation from \texttt{MadGraph5} and \texttt{Pythia8}, respectively. The following basic selection criteria were imposed to generate the events
  \begin{eqnarray}
    && 2\; b\hbox{-tagged jets, 2 opposite charged leptons} \nonumber \\
    && p_T(\ell) > 15\hbox{ GeV},\; |\eta_\ell| < 2.5 \nonumber \\
    && p_T(b) > 30\hbox{ GeV},\; |\eta_b| < 3.0 \nonumber \\
    && \met > 20\hbox{ GeV}\;\; .
    \label{eq:cuts}
  \end{eqnarray}
  We also recorded the four-momenta of up to two leading non-$b$ jets ($j$) of the events with $p_T(j)>20\hbox{ GeV}$, and $|\eta_j|<3$.

\subsection{Kinematic Variables and Representation of Events}

   The target of the reconstruction is the double Higgs and the top pair invariant masses so besides the two $b$-jets, the two hardest non-$b$ jets, the two opposite charged leptons, and the missing transverse momentum at the detector level, we also kept the four-momenta, in the laboratory frame, of the intermediate Higgs bosons and top quarks of the event as generated at the parton level. Note that NLO QCD radiation effects are taken into account in these four-momenta. It would be possible to reconstruct the partonic center-of-mass energy, $\sqrt{\hat{s}}$, of the collision once we have the four-momenta of the initial state partons at our disposal. This variable also accounts for the energy of all the radiation emitted alongside $hh$ or $t\bar{t}$, which would require a more careful simulation of high-order effects.
  
  The basic representation of the events thus comprises 34 low-level features. This low-level representation is augmented by high-level features described below.
  \begin{itemize}
      \item the transverse momentum, $p_T$, and rapidity, $\eta$, of the two $b$-jets and the two leptons,
      \item the transverse momentum of the pairs $b\bar{b}$, $\ell^+\ell^{\prime -}$, $jj$. In events where only one non-$b$ jet is identified, the transverse momentum is just $p_T(j)$. When the event contains no jets besides the bottom jets, the entries corresponding to those jets are filled with zeroes,
      \item the rapidity of the pairs $b\bar{b}$ and $\ell^+\ell^{\prime -}$,
      \item the energy and $z$-component of the three-momentum of the pairs $b\bar{b}$ and $\ell^+\ell^{\prime -}$, and of the combination $b\bar{b}\ell^+\ell^{\prime -}$,
      \item the invariant masses of the combinations $b\bar{b}$, $\ell^+\ell^{\prime -}$, $b\bar{b}\ell^+\ell^{\prime -}$, $b\bar{b}jj$, $jj\ell^+\ell^{\prime -}$. Again, when just one or no jet $j$ is present, the invariant masses are calculated accordingly. In events where no jets appear, some redundancy between these variables occurs,
      \item the distance in the $\eta\times\phi$ plane: $\Delta R_{ij}=\sqrt{(\Delta\eta_{ij})^2+(\Delta\phi_{ij})^2}$, between the pairs $b\bar{b}$, $\ell^+\ell^{\prime -}$, $b\ell$, $bj$, and $jj$,
      \item the azimuth angle difference, $\Delta\phi_{bb}$, between $b$ and $\bar{b}$, and, $\Delta\phi_{\ell\ell}$, between $\ell^+$ and $\ell^{\prime -}$,
      \item the Barr variable~\cite{Barr:2005dz}: $\cos\theta^*_{bb\ell\ell} = \tanh\left(\frac{1}{2}\Delta\eta(bb,\ell\ell)\right)$ between the $b\bar{b}$ and $\ell^+\ell^{\prime -}$ systems,
      \item the missing transverse momentum, $\ptm$,
      \item $M_T = \sqrt{2|\vec{p}_{T,O}|\ptm-\vec{p}_{T,O}\cdot\not\! \vec{p}_T}$ where $p_O = p_\ell+p_{\ell'}+p_b+p_{\bar{b}}$ 
      \item $\sqrt{\hat{s}_O}=\left[M^2_{bb\ell\ell}+2\not\! p_T\sqrt{M^2_{bb\ell\ell}+p_{T,O}^2} -2\vec{p}_{T,O}\cdot\vec{\not\! p_T}\right]^{1/2}$~\cite{Kim:2018cxf}
  \end{itemize}

   Besides all these kinematic variables, we also compute the {\it Higgsness}, $H$, and the {\it Topness}, $T$, of the events~\cite{Kim:2018cxf}. Higgsness is an adimensional variable defined as
 \begin{eqnarray}
     H &\equiv& \underset{p_\nu, p_{\bar{\nu}}}{\mathrm{argmin}} \left[\frac{(M_{\ell^+\ell^-\nu\bar{\nu}}^2 - m_h^2)^2}{\sigma_h^4} +
     \frac{(M^2_{\nu\bar{\nu}}-M^2_{\nu\bar{\nu},peak})^2}{\sigma_\nu^4} \right.\nonumber\\
     &+& \left. \min\left(\frac{(M^2_{\ell^+\nu}-m_W^2)^2}{\sigma_W^4}+\frac{(M^2_{\ell^-\bar{\nu}}-m_{W^*,peak}^2)^2}{\sigma_{W^*}^4},\; \frac{(M^2_{\ell^-\bar{\nu}}-m_W^2)^2}{\sigma_W^4}+\frac{(M^2_{\ell^+\nu}-m_{W,peak}^2)^2}{\sigma_{W^*}^4}\right)\right], \nonumber \\
     &&
     \label{eq:higgsness}
 \end{eqnarray}
 where $\sigma_h$, $\sigma_W$, and $\sigma_\nu$ might represent experimental uncertainties (in GeV), but for our purposes, they can be treated as free parameters. In the process of construing {\it Higgsness}, the four-momentum of the neutrino and the anti-neutrino must be searched to achieve the maximum compatibility with the decay chain $h\to W^+W^-, W^+\to \ell^+\nu_\ell, W^-\to \ell^{\prime -}\bar{\nu}_{\ell^\prime}$ where one of the $W$ bosons is off its mass shell. The peak of the $M_{\nu\bar{\nu}}$ and $M_{W^*}$ distributions occur approximately at 37 and 31 GeV, respectively. We fixed $\sigma_h=2$ GeV, $\sigma_W=\sigma_{W^*}=5$ GeV, and $\sigma_\nu=10$ GeV as in Ref.~\cite{Kim:2018cxf}.

 By its turn, we define {\it Topness} as follows
 \begin{eqnarray}
     T & \equiv & \min(\chi^2_{12},\chi^2_{21}) \nonumber \\
     \chi^2_{ij} &=& \underset{p_\nu, p_{\bar{\nu}}}{\mathrm{argmin}}\left[\frac{(M^2_{b_i\ell^+\nu}-m^2_t)^2}{\sigma_t^4}+\frac{(M^2_{\ell^+\nu}-m^2_W)^2}{\sigma_W^4}+\frac{(M^2_{b_j\ell^+\bar{\nu}}-m^2_t)^2}{\sigma_t^4}+\frac{(M^2_{\ell^-\bar{\nu}}-m^2_W)^2}{\sigma_W^4}\right], 
 \end{eqnarray}
 where $\sigma_t=5$ GeV, as in Ref.~\cite{Kim:2018cxf}. In this case, as we do not know the $b$-jet charge, we have to test between two options to get the better consistency of the event with the $t\bar{t}$ production and decay chain $t(\bar{t})\to W^+ b(W^-\bar{b})\to b\ell^+\nu(\bar{b}\ell^-\bar{\nu})$. The minimization process was performed with a simplex method from \texttt{Scipy}~\cite{2020SciPy-NMeth}.

\begin{figure}[t]
    \centering
    \includegraphics[width=1\linewidth]{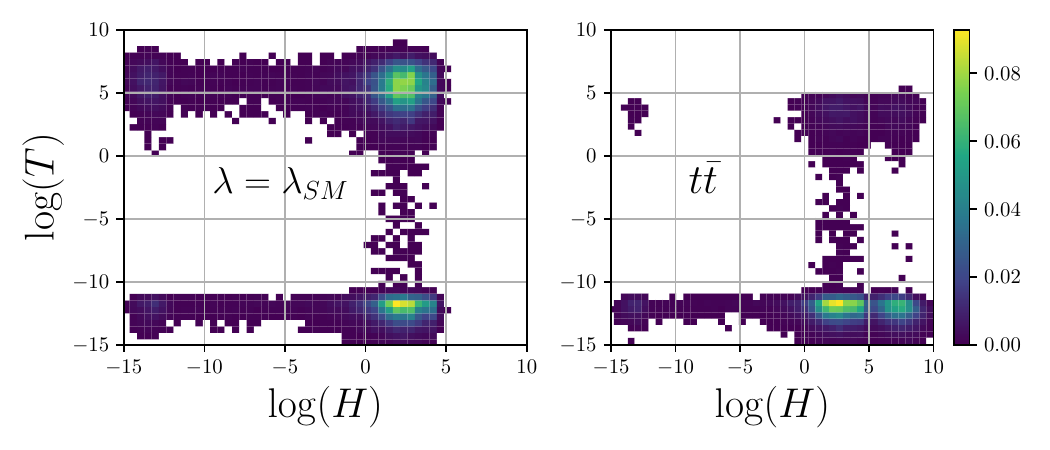}
    \caption{The joint distribution of the logarithm of {\it Higgsness} and {\it Topness} for Higgs pairs events (left panel), and $t\bar{t}$ events (right panel).}
    \label{fig:HT}
\end{figure}

 We show, in Figure~\ref{fig:HT}, the joint {\it Higgsness} and {\it Topness} distributions for the SM double Higgs production and the $t\bar{t}$. We see a clear distinction between the two kinds of events with Higgs pairs concentrating in the region $\log(T)>5$ and $\log(H)<5$. This behavior is largely independent of the strength of the trilinear Higgs self-coupling and also shows a similar pattern for resonant $hh$ production.

\section{Reconstruction of fully leptonic $b\bar{b}W^+W^-$ events: Non-resonant case}
\label{sec:non-res}

 The double Higgs invariant mass is sensitive to the Higgs self-coupling. Besides the total cross section expected at the collider, the shape of $M_{hh}$ might help to measure $\lambda$ and possible deviations from the SM. As discussed before, with the help of powerful discerning variables, like {\it Higgsness} and {\it Topness}, the fully leptonic $b\bar{b}W^+W^-$ mode becomes an interesting option to measure the trilinear Higgs coupling at the LHC. If not used for fits, $M_{hh}$ and $M_{t\bar{t}}$ can be used to further discern between Higgs pairs and top pairs in a cut-based or multivariate analysis.

 The challenge, however, is to recover the information carried away by the neutrinos from $W$s. Neural networks offer the possibility to fit a parametrized function of the observable information brought by leptons, jets, and $b$-jets from data. Our solution is to train a probabilistic neural net regressor from a variational inference process as described in Section~\ref{sec:vaer}. 

 We tested two types of target: (1) a single-valued one, the $hh$ or $t\bar{t}$ mass, denoted collectively as $\mbbllvv$; (2) a 2-component vector, $(\mbbllvv,\ptllvv)$, where $\ptllvv$ denotes the transverse momentum of hardest leptonic $W$ boson. We observed better performance of the vector target across our experiments and tuning, so from now on, we will present the results and analysis for this target. Because $p_T(W)$ is strongly correlated to $p_T(b\bar{b})$, especially in the case of double Higgs, we conjecture that including $\ptllvv$ in the target of the regression task helps to create ties with the vector feature of the events what could explain the better performance of the algorithm. Our focus, however, is the $\mbbllvv$ mass of the event. Let us discuss the preparation of the data to feed the neural networks.

\subsection{Data Preparation, Training and Validation, and Algorithm Structure}

 We generated around $10^6$ events to train and test VAER. The dataset was split into 75\% for training and 25\% for testing. A 5-fold cross-validation was performed to evaluate the error in prediction caused by statistically independent test sets. The training set comprises $t\bar{t}$ and $hh$ for $\kappa_\lambda=-3,-2,-1,0.1,1,2,3$ trilinear couplings. We will refer to this coupling set as the {\it support couplings}. The test set contains the same types of events and $hh$ events with the addition of intermediate trilinear couplings $\kappa_\lambda=-2.5,-1.5,-0.5,0.5,1.5,2.5$. 
 This is the {\it interpolated couplings} set.
 We also generated events for new heavy Higgs bosons from xSM~\cite{Profumo:2007wc,Profumo:2014opa}, with masses from 300 to 1000 GeV, decaying to $hh\to b\bar{b}W^+W^-$. We will discuss the resonant case in detail ahead. 

 To establish the generalization power of VAER, intermediate couplings and heavy Higgs events are not presented to the algorithm during the training phase. The intermediate coupling events test the interpolation ability of the algorithm, which is supposed to learn the $b\bar{b}\ell^+\ell^{\prime -}\nu_\ell\bar{\nu}_{\ell^\prime}$ mass from the observable information. For that purpose, diversity is essential. The heavy Higgs events test the extrapolation power of the algorithm once they populate regions of the representation space that are poorly populated by training examples. We should expect that extrapolation works significantly worse than interpolation. 
 \begin{table}[t]
 \centering
 \begin{tabular}{c|c|c|c}
 \hline
    Hyperparameter/architecture  &  Encoder & Decoder & Regressor\\
    \hline\hline
    L1 regularization  & $10^{-5}$ & -- & -- \\
    kernel initialization & Glorot uniform & Glorot uniform & Glorot uniform \\
    layer activation & $\tanh$ & $\tanh$ & $\tanh$ \\
    numbers of layers and neurons & (1024,512,256,128) & (128,256,512,1024) & (128) \\
    total of parameters & 759955 & 759107 & 258 \\
    \hline\hline
 \end{tabular}
\caption{Hyperparameters and architecture of the decoder, the encoder, and the regressor neural networks. No dropout layers were needed. The total number of parameters of this VAER configuration is $\sim 1.5\times 10^6$.}
\label{tab:hyper}
 \end{table}
 
 Training a neural network with signal events of different model parameters to help it to generalize across the parameters space was shown to be successful in Ref.~\cite{Baldi:2016fzo}. The parametrized neural networks obtained from this framework are fed with physics parameters and then used to classify events for intermediate points of the parameters space for which the algorithm was not trained, saving time and computational resources. In our case, we do not provide any physics parameters to the algorithm, neither trilinear couplings nor masses. Nonetheless, as we are going to show, VAER learns the target variables across those parameter spaces.

 To reduce the magnitude of the target variables, we took their logarithm for the regression task. The features and target vector were scaled with the \texttt{RobustScaler} from \texttt{scikit-learn}~\cite{scikit-learn}. This scaler removes the median of the data feature-wise and scales them with the interquantile range between the first and third quartiles of the data, making the dataset less sensitive to outliers events. 

 The algorithm is trained for 2000 stochastic gradient descent iterations in batches of 1024 examples. A stopping criterion is adopted, halting the training if no reduction in the loss function is observed after 20 iterations. The learning rate is reduced by half if no improvement is observed after 10 iterations. The initial learning rate is $10^{-3}$. The neural networks were built with \texttt{Keras}~\cite{chollet2015keras} and \texttt{Tensorflow}~\cite{tensorflow2015-whitepaper}. The optmizer adopted was the \texttt{AdamW}~\cite{2017arXiv171105101L} with a weight decay of $10^{-4}$.

 We tested several architectures and hyperparameters, but no extensive tuning was performed. Improvements in the performance of the algorithm can thus be achieved. We display, in Table~\ref{tab:hyper}, the architecture and the hyperparameters of the various components of VAER. The dimension of the latent space was 3. The target loss, $L_{reg}$ in Eq.~\eqref{eq:vaer-loss}, was multiplied by $\beta=10$ to encourage the algorithm to better predict the target variables.

\subsection{$M_{bb\ell\ell\nu\nu}$ in the Standard Model}
\begin{figure}[t]
    \centering
    \includegraphics[width=0.45\linewidth]{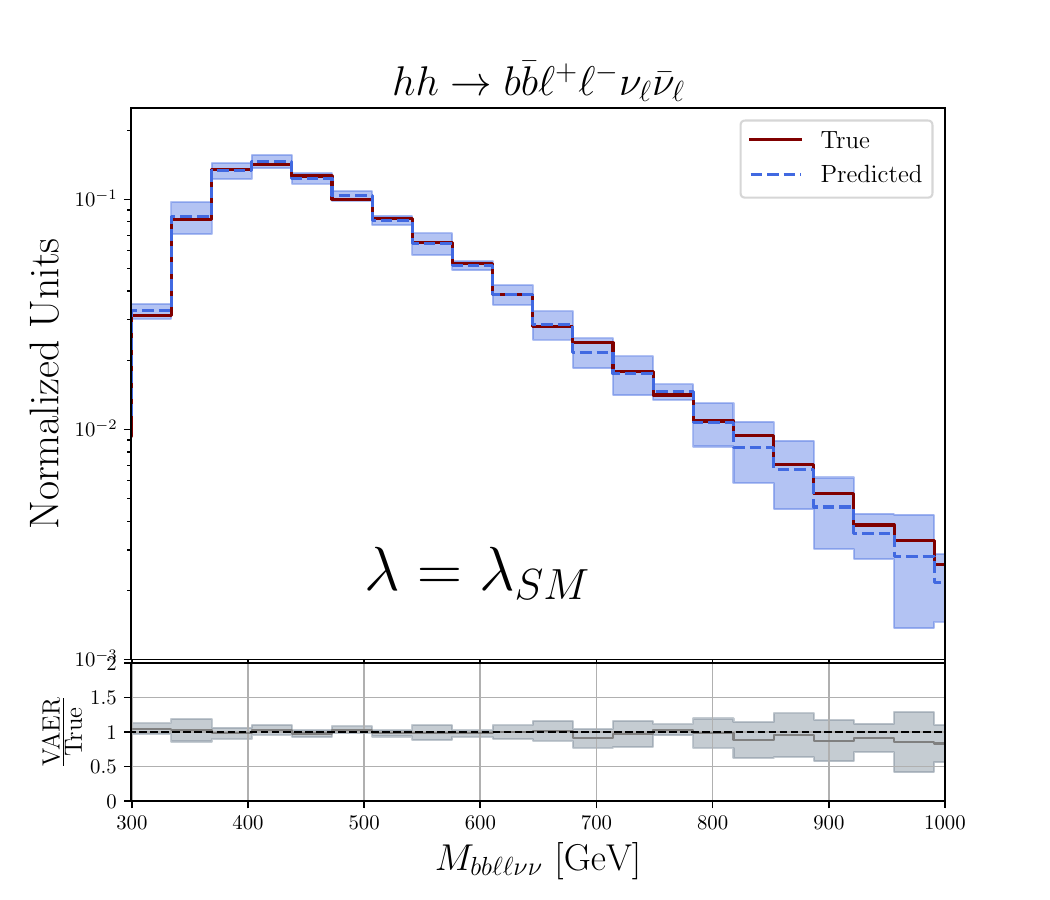}
    \includegraphics[width=0.45\linewidth]{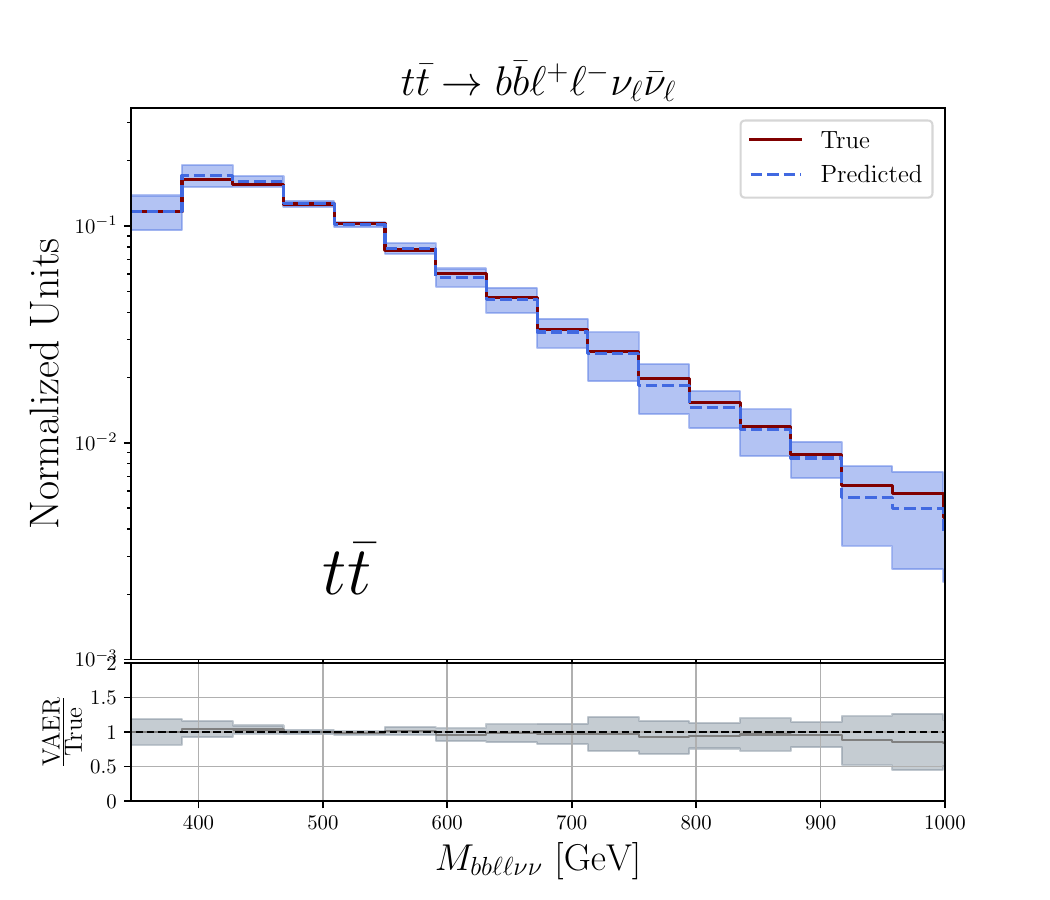} \\
    \includegraphics[width=0.45\linewidth]{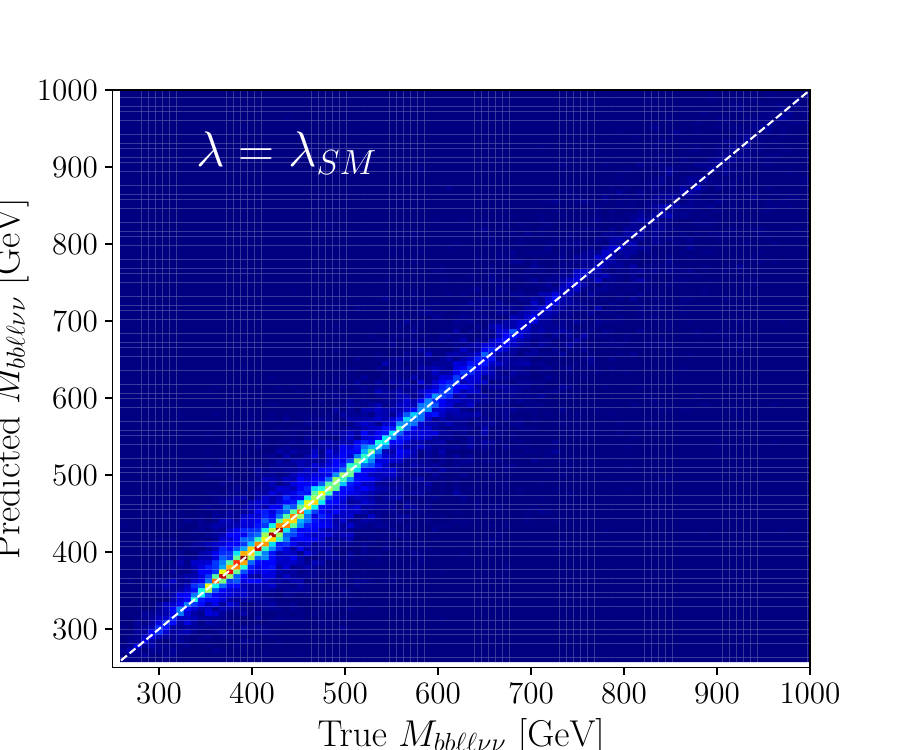}
    \includegraphics[width=0.45\linewidth]{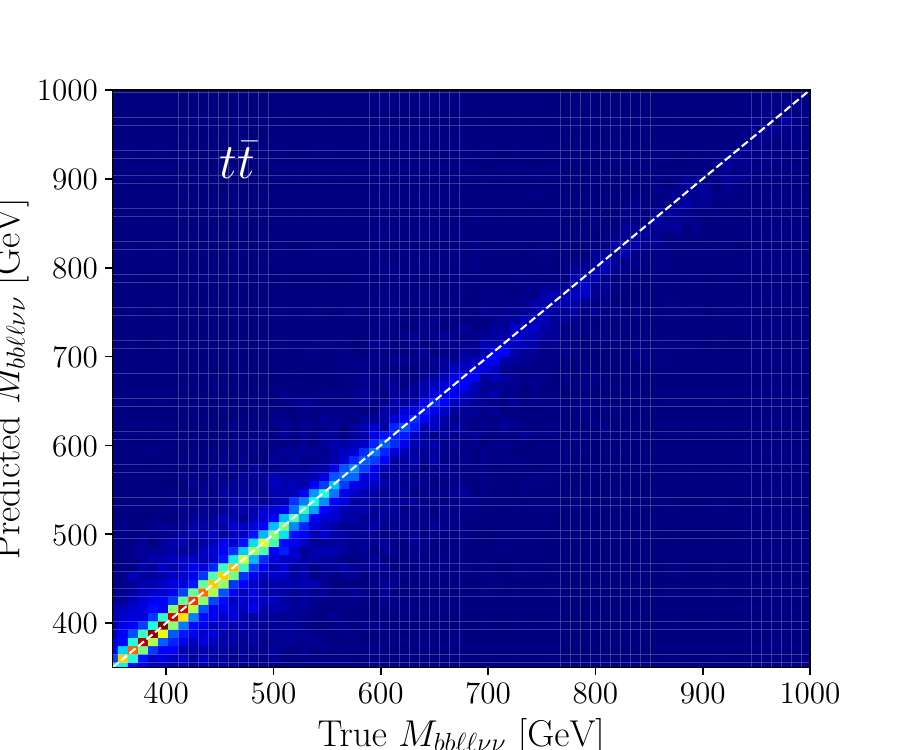} 
    \caption{Upper panels: the true and predicted SM $hh$ and the $t\bar{t}$ invariant masses at left and right, respectively. The blue-shaded regions represent the cross-validation uncertainties in the prediction. The ratio between VAER  prediction and ground truth is also depicted in these plots. Lower panels: scatter plots for true versus predicted masses.}
    \label{fig:m2-sm}
\end{figure}

 We now present the results for the reconstruction of $\mbbllvv$ in the Standard Model. In Figure~\ref{fig:m2-sm}, we depict the $\mbbllvv$ mass for the SM $hh$ production (left panel) and the $t\bar{t}$ background (right panel). The lower panels show the true-to-predicted ratio. The blue shaded area in the histograms represents the variation of predictions from the 5-fold cross-validation where the test set is split into 5 independent sets of events. The dashed blue line is the mean prediction from the five test sets. The agreement between true and predicted invariant masses is within a few percent both for $hh$ and $t\bar{t}$ production up to 1 TeV. The uncertainty increases for higher invariant masses as the number of events in the tail of the distributions drops.
 \begin{table}[t]
     \centering
     \begin{tabular}{c|c|c|c|c|c}
     \hline\hline
        Process & $\sqrt{\hbox{MSE}}$ $(\downarrow)$ & MAE $(\downarrow)$ & $R^2$ $(\uparrow)$& $JS_D$ $(\downarrow)$ & fraction@$\Delta_{tp}\leq10$\% $(\uparrow)$\\
        \hline
        {\color{purple}$t\bar{t}$} & {\color{purple}$66.1\pm 5.7$} & {\color{purple}$35.3\pm 0.4$} & {\color{purple}$0.830\pm 0.010$} &  {\color{purple}$0.011\pm 0.014$} & {\color{purple}79.2\%} \\
         \hline
         $\kappa_\lambda=-3$ & $38.8\pm 4.0$ & $22.3\pm 0.5$ & $0.896\pm 0.003$ & $0.010\pm 0.002$ & 83.3\% \\
        {\color{cyan}$\kappa_\lambda=-2.5$} & {\color{cyan}$56.1\pm 6.4$} &  {\color{cyan}$38.2\pm 0.7$} &  {\color{cyan}$0.79\pm 0.01$} &   
          {\color{cyan}$0.018\pm 0.002$} & {\color{cyan}68.1\%} \\
          $\kappa_\lambda=-2$ & $41.0\pm 5.8$ & $23.3\pm 0.6$ & $0.898\pm 0.004$ & $0.005\pm 0.001$ & 83.5\% \\
        {\color{cyan}$\kappa_\lambda=-1.5$} & {\color{cyan}$54.7\pm 6.6$} & {\color{cyan}$37.6\pm 0.9$} & {\color{cyan}$0.803\pm 0.013$} & {\color{cyan}$0.025\pm 0.006$}  & {\color{cyan}69.4\%} \\
          $\kappa_\lambda=-1$ & $41.2\pm 5.6$ & $23.4\pm 0.3$ & $0.907\pm 0.005$ & $0.005\pm 0.002$ & 83.8\% \\
          {\color{cyan}$\kappa_\lambda=-0.5$} & {\color{cyan}$61.2\pm 10.0$} & {\color{cyan}$40.8\pm 1.0$} & {\color{cyan}$0.799\pm 0.015$} & {\color{cyan}$0.015\pm 0.004$} & {\color{cyan}68.0\%} \\
          $\kappa_\lambda=+0.1$ & $45.2\pm 6.4$ & $24.9\pm 0.4$ & $0.896 \pm 0.010$ & $0.005\pm 0.002$ & 84.2\% \\
          {\color{cyan}$\kappa_\lambda=+0.5$} & {\color{cyan}$64.7\pm 8.8$} & {\color{cyan}$43.0\pm 1.1$} & {\color{cyan}$0.783\pm 0.012$} & {\color{cyan}$0.014\pm 0.003$} & {\color{cyan}68.7\%} \\
         {\color{purple}$\kappa_\lambda=+1$} & {\color{purple}$47.8\pm 7.7$} & {\color{purple}$26.8\pm 0.6$} & {\color{purple}$0.890\pm 0.010$} & {\color{purple}$0.005\pm 0.001$} & {\color{purple}83.7\%} \\
        {\color{cyan}$\kappa_\lambda=+1.5$} & {\color{cyan}$70.5\pm 6.8$} & {\color{cyan}$47.3\pm 0.6$} & {\color{cyan}$0.795\pm 0.008$} & {\color{cyan}$0.015\pm 0.002$} & {\color{cyan}68.0\%} \\
         $\kappa_\lambda=+2$ & $55.0\pm 5.7$ & $28.4\pm 0.6$ & $0.915\pm 0.003$ & $0.011\pm 0.003$ & 83.4\% \\
         {\color{cyan}$\kappa_\lambda=+2.5$} & {\color{cyan}$44.7\pm 5.8$} & {\color{cyan}$50.6\pm 1.5$} & {\color{cyan}$0.82\pm 0.02$} & {\color{cyan}$0.039\pm 0.005$} & {\color{cyan}66.7\%} \\
         $\kappa_\lambda=+3$ & $51.0\pm 7.2$ & $26.6\pm 0.8$ & $0.938\pm 0.003$ & $0.027\pm 0.002$ & 82.1\% \\
         \hline\hline
     \end{tabular}
     \caption{Root MSE (in GeV), MAE (in GeV), $R^2$, the Jensen-Shannon distance, and the fraction of events where the relative difference of true and predicted invariant mass is less than 10\% for the $t\bar{t}$ and $hh$ with various trilinear couplings. In purple, we highlight the SM results, while in black(cyan), we depict the support(interpolated) couplings for which VAER was(was not) trained to make predictions. An up(down) arrow $\uparrow(\downarrow)$ indicates that larger(smaller) is better. Uncertainties were computed from a 5-fold cross-validation.} 
     \label{tab:metrics}
 \end{table}

 A quantitative assessment of our results can be read in Table~\ref{tab:metrics}. Let us call the true $\mbbllvv$ mass of an event, $t$, and the predicted one as $p$. To quantitatively access the performance of the algorithm, we compute the root mean squared error, $\sqrt{\hbox{MSE}}=\sqrt{\overline{(t-p)^2}}$; the mean absolute error, $\hbox{MAE}=\overline{|t-p|}$; the binned Jensen-Shannon divergence, $JS_D$,
 \begin{equation}
     JS_D = \frac{1}{2}\sum_{bins} \left[t_i\log\left(\frac{t_i}{p_i}\right)+p_i\log\left(\frac{p_i}{t_i}\right)\right]\; ;
 \end{equation}
 the fraction of events whose relative difference between true and predicted invariant mass
 \begin{equation}
     \Delta_{tp}=\left|\frac{t-p}{t}\right|\; 
     \label{eq:dtp}
 \end{equation}
 is less than 10\%; and the correlation coefficient, $R^2$, defined as follows
 \begin{equation}
     R^2=1-\frac{\sum_{i}(t_i-p_i))^2}{\sum_{i}(t_i-\overline{t})^2}\; ,
 \end{equation}
 where $\overline{t}$ is the mean of the true target. Except for the $JS_D$, which is computed from binned invariant mass distributions, all the other metrics are evaluated on an event-by-event basis.

 Corroborating the visual agreement we see in Figure~\ref{fig:m2-sm}, the purple entries of Table~\ref{tab:metrics} show the excellent performance of VAER in predicting the $b\bar{b}\ell\ell\nu\nu$ mass for signal and background. The MAE of both SM $hh$ and $t\bar{t}$ events are both comparable to the bin width of the distributions. The MSE of $t\bar{t}$ events are larger than $hh$ ones as the background presents a harder spectrum. In both cases, the correlation coefficient is high, especially for Higgs events. The lower panels of Figure~\ref{fig:m2-sm} show the scatter plots of true versus predicted $\mbbllvv$ and confirm the high correlation coefficient. 

\begin{figure}[t]
    \centering
    \includegraphics[width=0.475\linewidth]{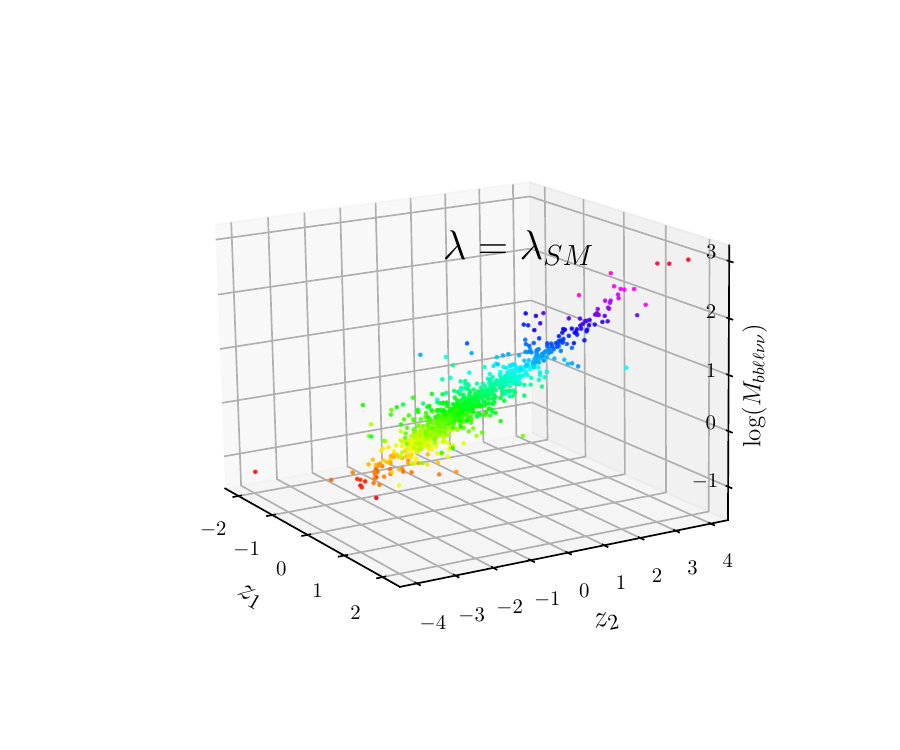}
    \includegraphics[width=0.475\linewidth]{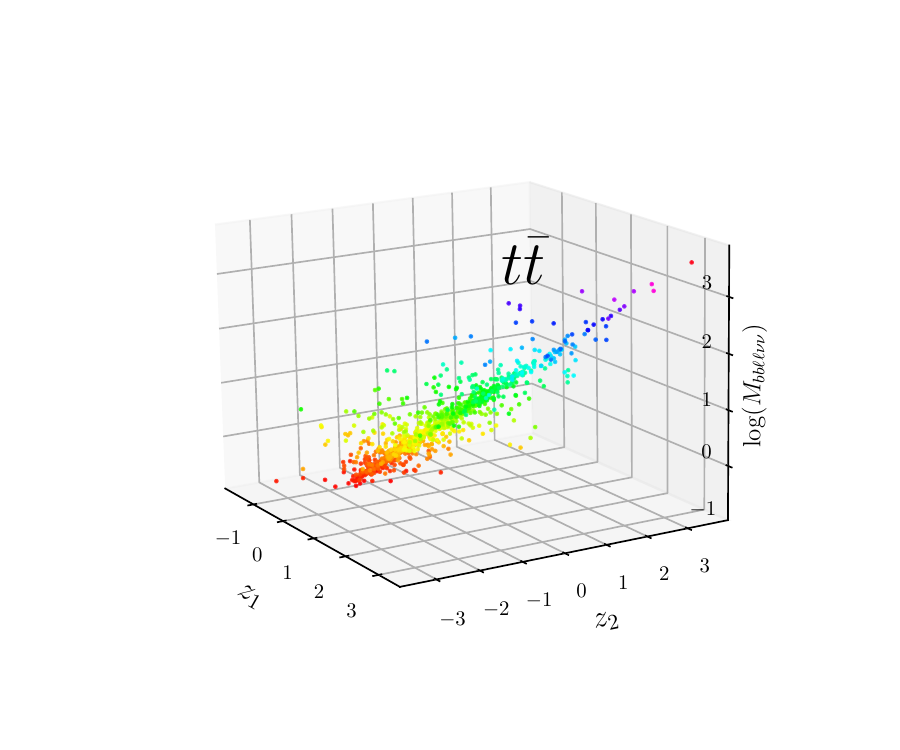} 
    \caption{The disentangled dimension associated with the latent space representations for SM $hh$ and $t\bar{t}$ predictions.}
    \label{fig:m2-disentangled}
\end{figure}

 As discussed in Section~\ref{sec:vaer}, a VAE for regression associates a disentangled dimension to the latent space representation of the events. In Figure~\ref{fig:m2-disentangled} we show $\log(\mbbllvv)$ as a function of two out of the three latent space dimensions. As anticipated, the linear regression model in terms of the latent dimension suffices for a good prediction. 

\subsection{Varying the Trilinear Higgs Coupling: Support Couplings}

 \indent Besides the SM double Higgs production and its main background source, VAER is also able to predict $\mbbllvv$ for non-SM trilinear couplings spanned in the training phase. The $hh$ invariant mass shape changes considerably from $\lambda=-3\lambda_{SM}$ to $\lambda=+3\lambda_{SM}$ due to the relative importance of the triangle amplitude in the interference with the box contribution. 

 In Leading Order, the differential cross section of $hh$ production can be expanded in powers of $1/m_t^2$~\cite{PhysRevD.87.014007}. Ignoring symmetry factors, charges, and couplings, it reads 
 \begin{eqnarray}
     \frac{d\hat{\sigma}}{dt} &\propto& \frac{1}{s^2} \left(\left|\kappa_\lambda F_\triangle+F_\square\right|^2+\left|G_\square\right|^2\right)\label{eq:dsdt-1} \\
     F_\triangle &=& \frac{4 s m_h^2}{s-m_h^2}\left(1+\frac{7s}{120 m_t^2}\right)\times \lambda y_t+{\cal O}(1/m_t^4)\label{eq:dsdt-2}\\
     F_\square &=& -\frac{4}{3}s\left(1+\frac{7 m_h^2}{20 m_t^2}\right)\times y_t^2+{\cal O}(1/m_t^4)\label{eq:dsdt-3}\\
     G_\square &= &-\frac{11}{45}\frac{s p_T^2}{m_t^2}\left(1+\frac{62m_h^2-5s}{154m_t^2}\right)\times y_t^2+{\cal O}(1/m_t^4)\; ,
     \label{eq:dsdt-4}
 \end{eqnarray}
 where $F_\triangle$ is the loop function of the triangle contribution that contains the trilinear coupling, $\lambda$, and the top quark Yukawa coupling, $y_t$, while $F_\square$ and $G_\square$ come from the box diagram and are proportional to $y_t^2$; $p_T$ is the Higgs boson transverse momentum, $s=(p_1+p_2)^2=M_{hh}^2$, and $t=(k_1-p_1)^2$. This is a crude approximation to the partonic invariant mass distribution, and it is shown that including $1/m_t^4$ terms actually worsens the agreement with the exact results~\cite{PhysRevD.87.014007}. However, this approximation captures the main features of the $hh$ mass for $m_{hh}\lesssim 1$ TeV. 
 
 Taking into account the gluon luminosity, the differential $M_{hh}$ distribution is
 \begin{equation}
     \frac{d\sigma}{dM_{hh}}=\frac{2M_{hh}}{S}\times  \int_{\tau}^1 \hat{\sigma}(gg\to hh)\times g(x,\mu_F)g(\tau/x,\mu_F)\frac{dx}{x},
 \end{equation}
 where $\tau=M_{hh}^2/S$, $\sqrt{S}=14$ TeV, $x$ and $\tau/x$ are the fractions of the protons' momenta brought by the gluons to the hard scattering. For our purposes, a simplified gluon distribution function might be taken as $g(x,\mu_F)\approx 1/x^\delta$ for $x\ll 1$. We took $\delta=2$ to mimic the SM distribution as closely as possible. Again, these approximations are crude but capture the basic dynamics that build the $M_{hh}$ distribution.

 To understand how $\lambda$ affects the distributions, first notice that the triangle contribution is enhanced compared to the box contribution towards the $hh$ production threshold due to the propagator $1/(s-m_h^2)$. 
 Second, the role played by the interference term is dictated by $\kappa_\lambda$ and the relative sign between $F_\triangle$ and $F_\square$. 
 Finally, $G_\square$ effectively contributes only to high $p_T$. 
\begin{figure}[t]
    \centering
    \includegraphics[width=0.31\linewidth]{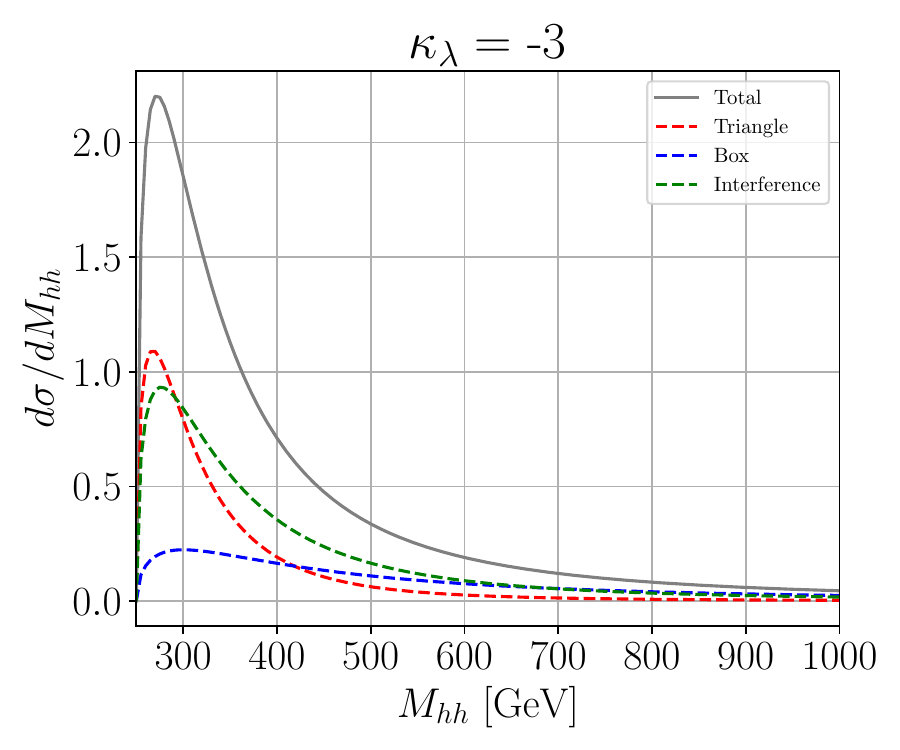}
    \includegraphics[width=0.31\linewidth]{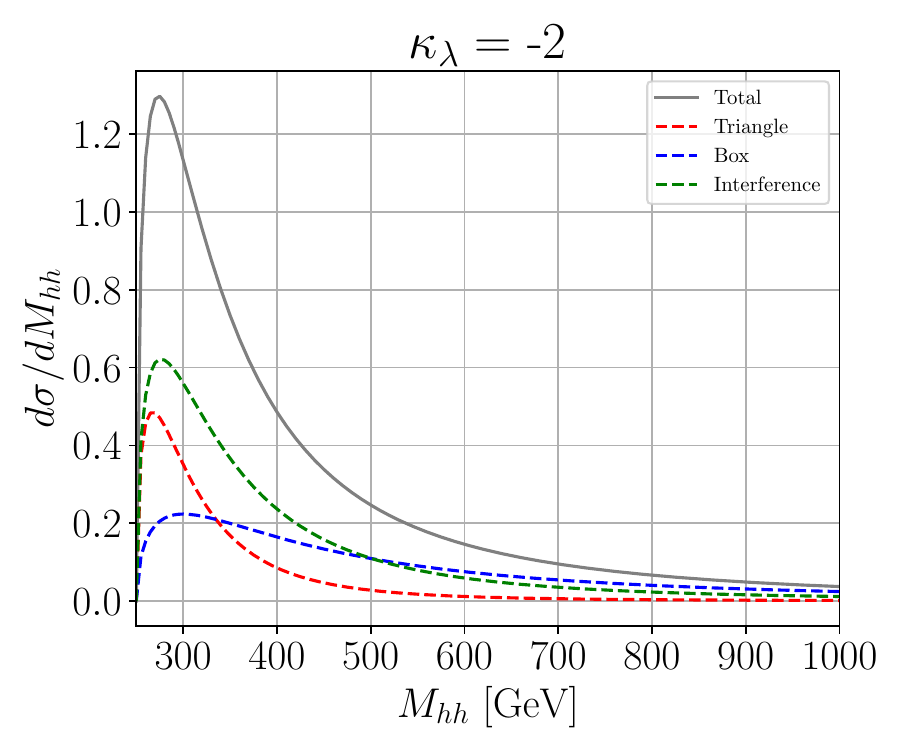} 
    \includegraphics[width=0.31\linewidth]{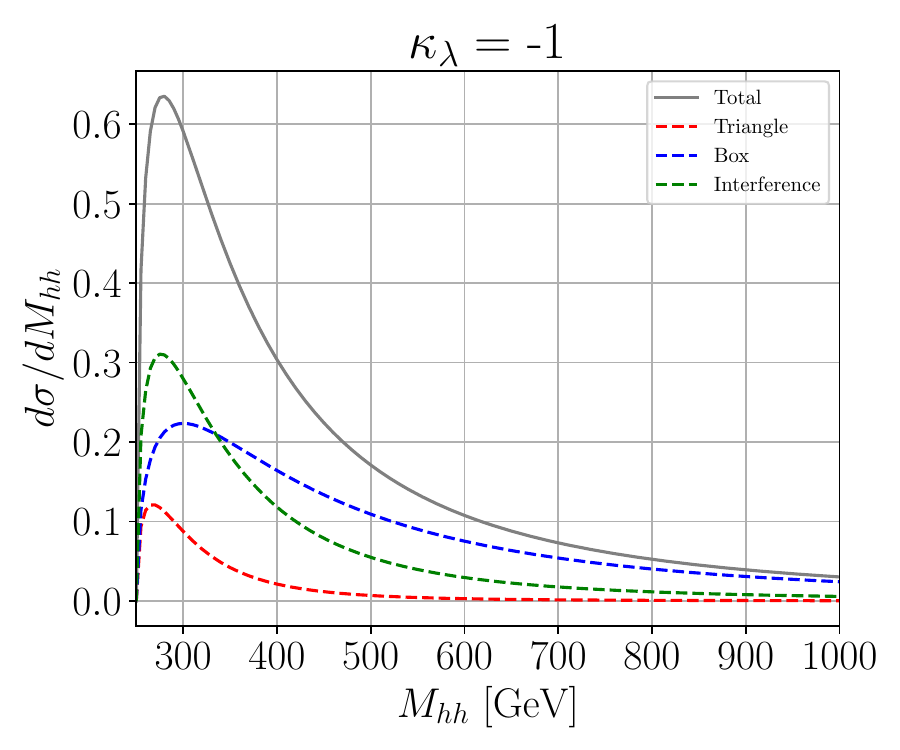} \\ 
    \includegraphics[width=0.31\linewidth]{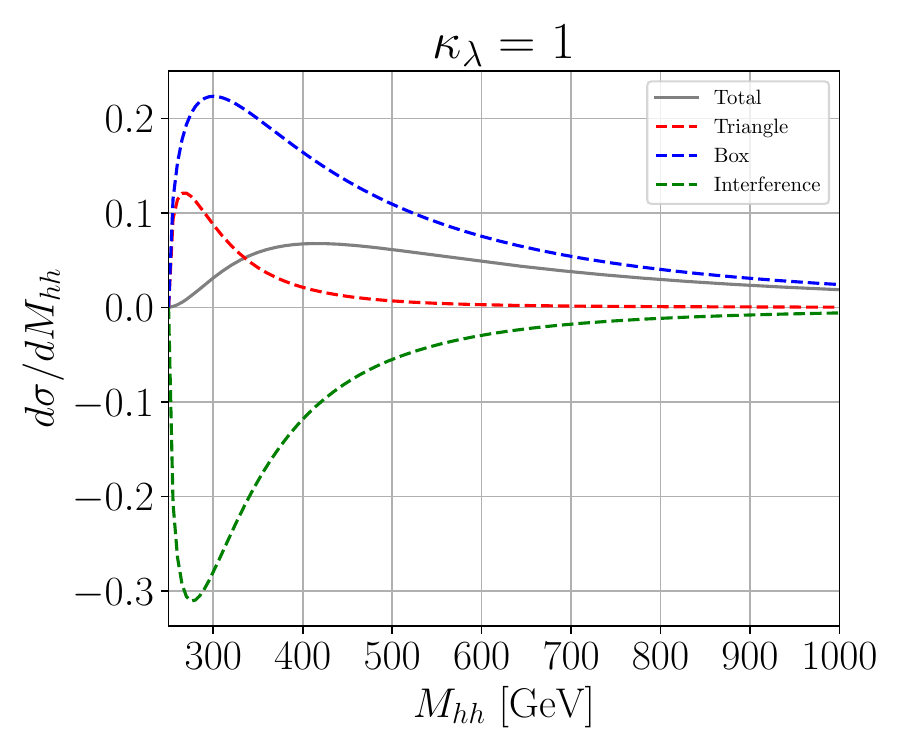}
    \includegraphics[width=0.31\linewidth]{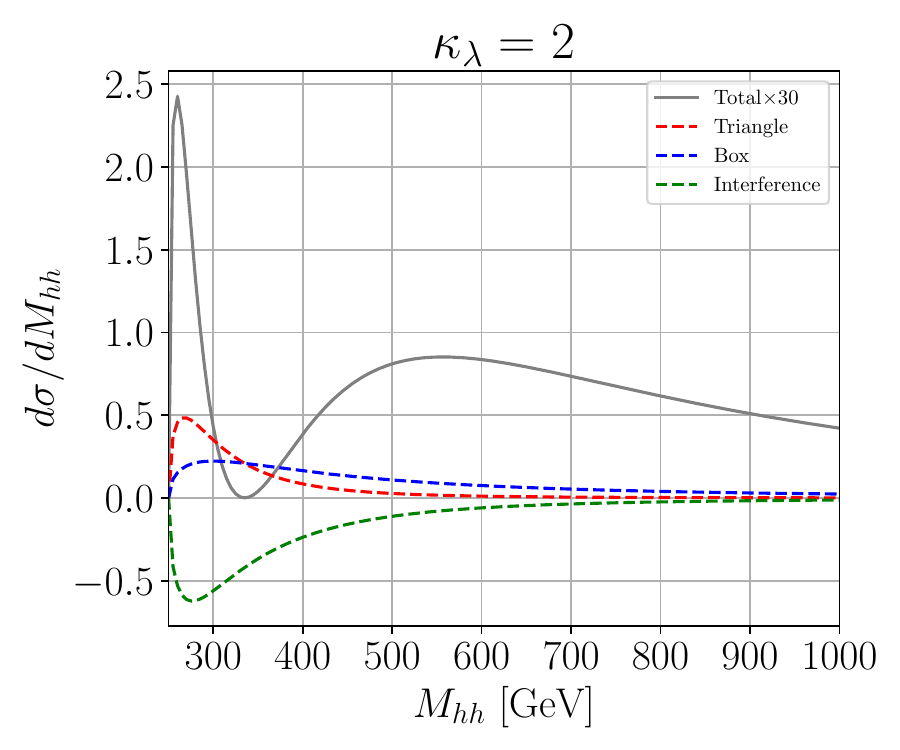}
    \includegraphics[width=0.31\linewidth]{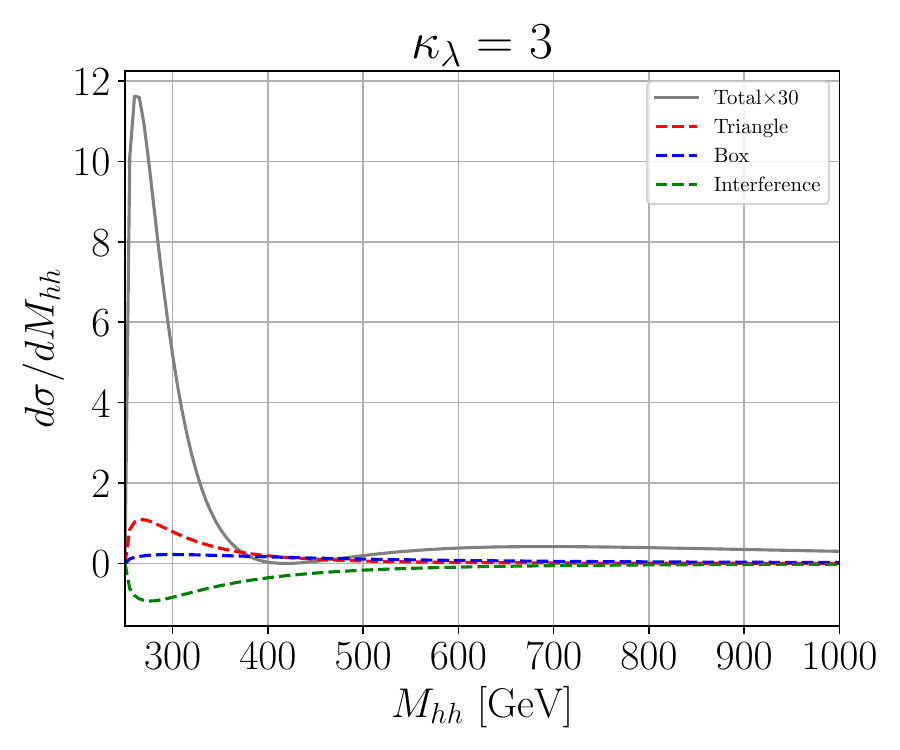}
    \caption{Approximate invariant $hh$ mass distributions for various trilinear couplings, $\lambda=\kappa_\lambda\times \lambda_{SM}$. The expressions for each contribution can be read from Eq.~\eqref{eq:dsdt-1}-\eqref{eq:dsdt-4}. The histograms of $\kappa_\lambda=2$ and $3$ were multiplied by 30 for better visualization.}
    \label{fig:m2-nonsm-approx}
\end{figure}

 In Figure~\ref{fig:m2-nonsm-approx}, we show $d\sigma/dM_{hh}$ (in arbitrary units) as a function of $M_{hh}$ for some trilinear couplings. Negative $\lambda$ turns the interference constructive with all the contributions reinforcing each other once the interference inherits a similar kinematic structure from the triangle and box contributions and, in special, the $1/(s-m_h^2)$ propagator that makes it also peak towards $M_{hh}\sim 2m_h$. When the trilinear coupling is positive, however, the interference term is negative and contributes destructively to $d\sigma/dM_{hh}$. For the SM production, the cancellation of the peak near the threshold is almost exact, and $M_{hh}$ increases, reaching a peak right after $400$ GeV.
 For larger $\lambda$, on the other hand, the interference term cancels the other contributions for larger $M_{hh}$, carving a dip in the distribution, causing a kind of amplitude-zero situation where no events are expected for certain $M_{hh}$ values.  Large $\lambda$ of both signs tend to resemble each other once the triangle contribution dominates.
 
The behavior of the contributions is also important to predict the impact of cuts. For example, large transverse momentum cuts favor the box contribution because the triangle amplitude is an $s$-channel diagram where Higgs bosons are mainly produced near the production threshold. 

\begin{figure}[t]
    \centering
    \includegraphics[width=0.32\linewidth]{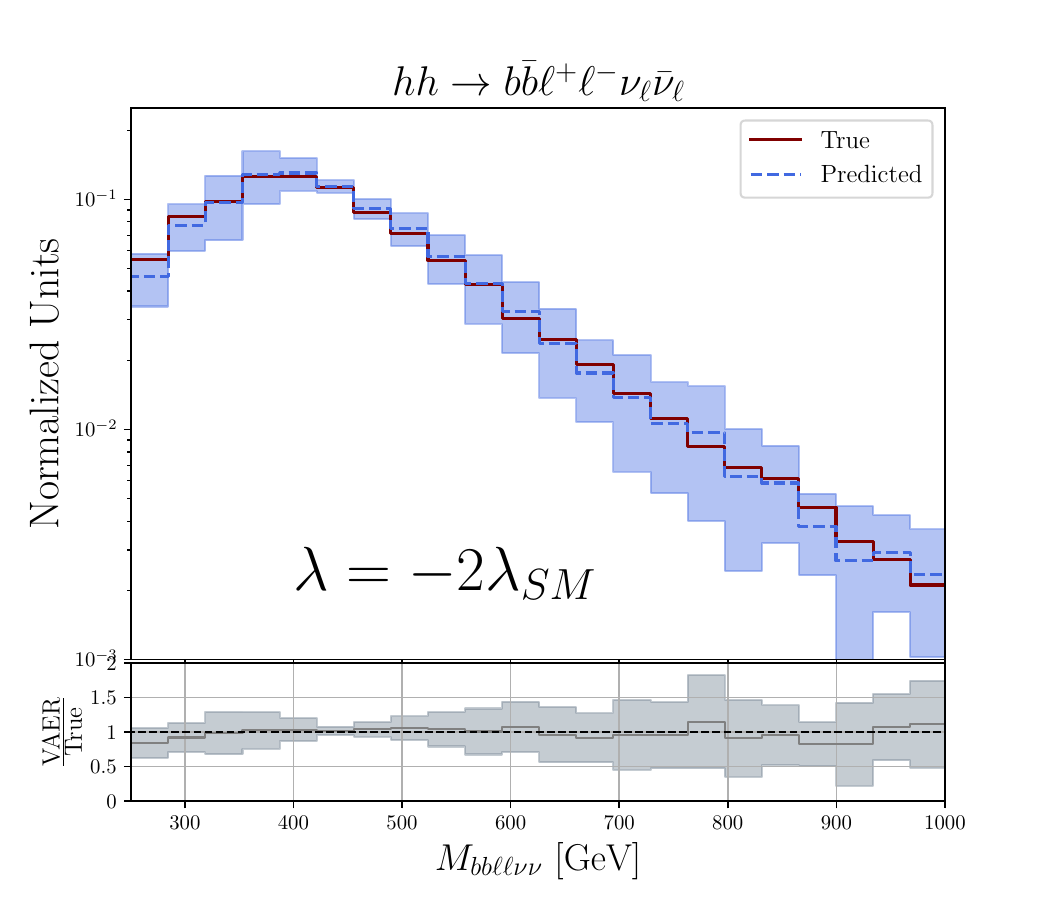}
    \includegraphics[width=0.32\linewidth]{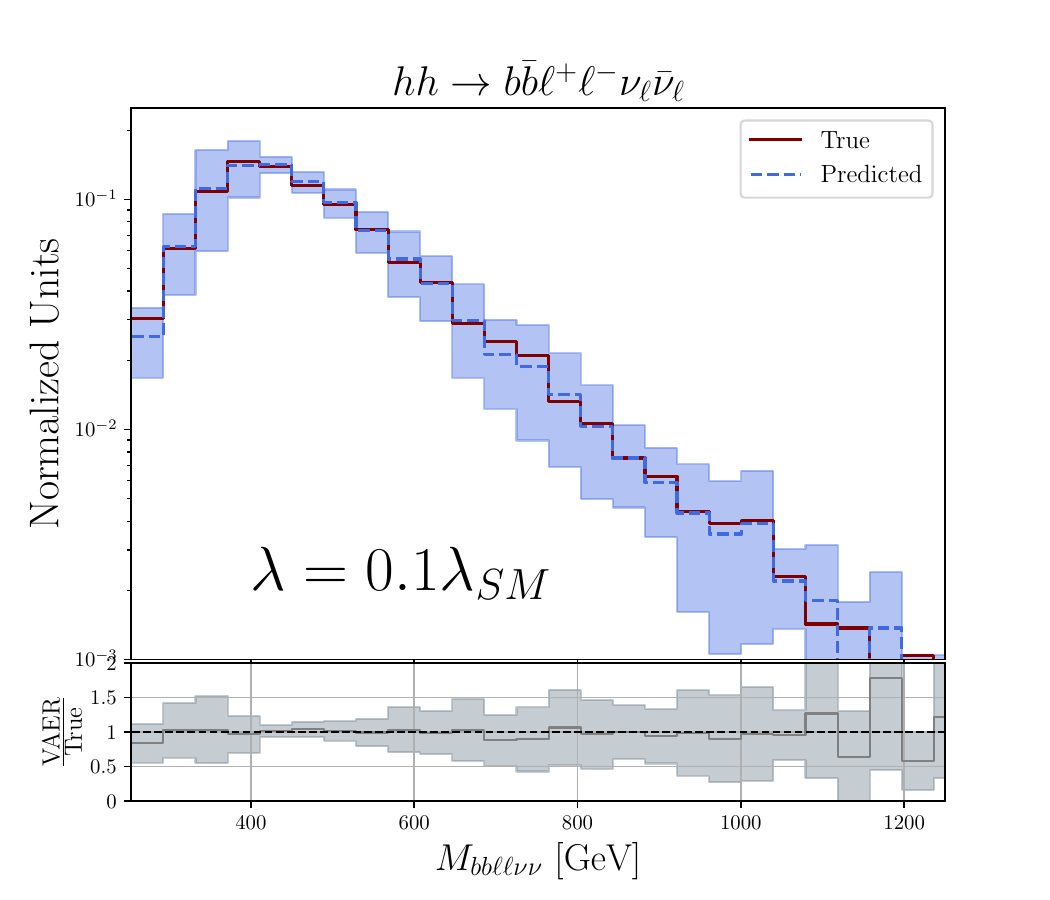} 
    \includegraphics[width=0.32\linewidth]{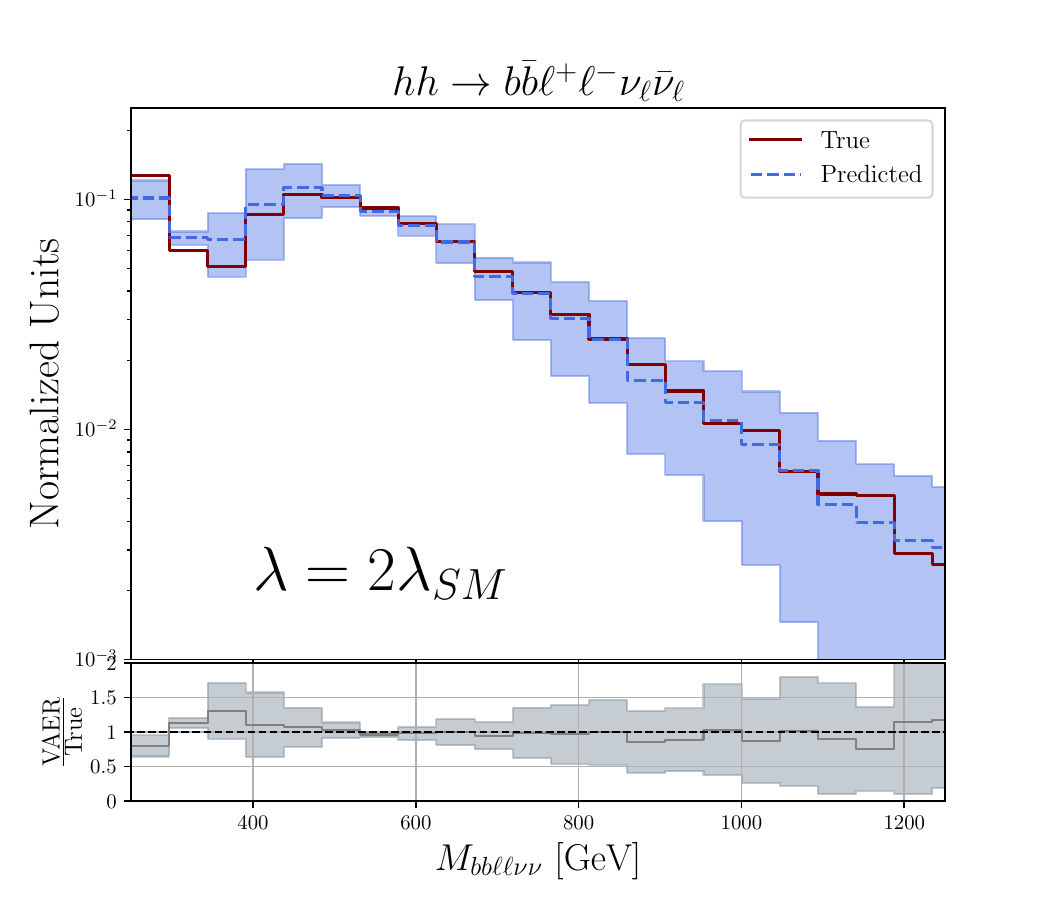} \\
    \includegraphics[width=0.32\linewidth]{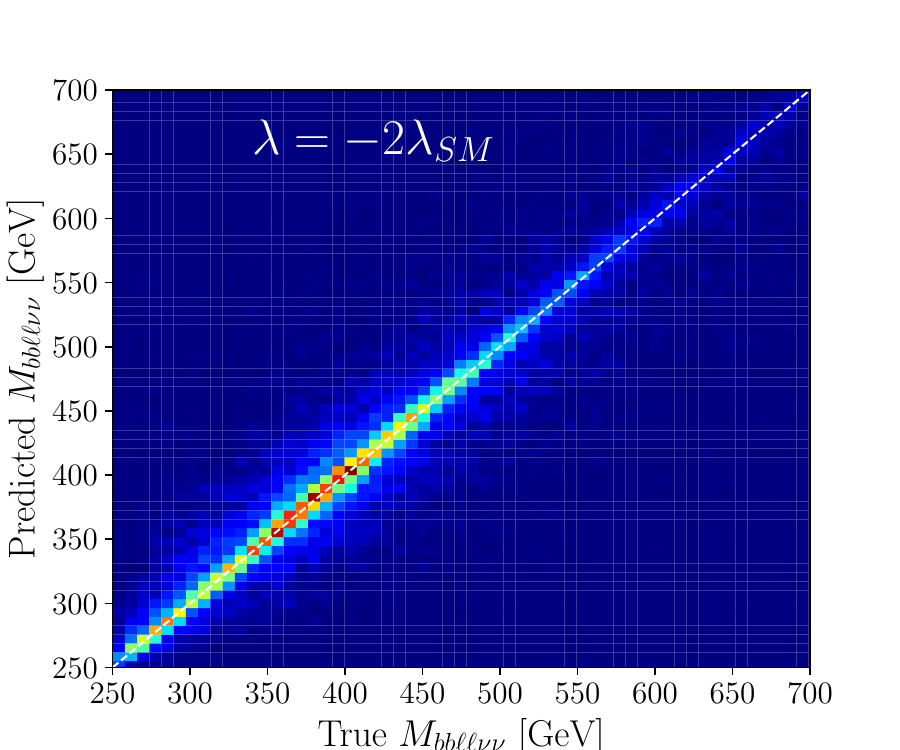} 
    \includegraphics[width=0.32\linewidth]{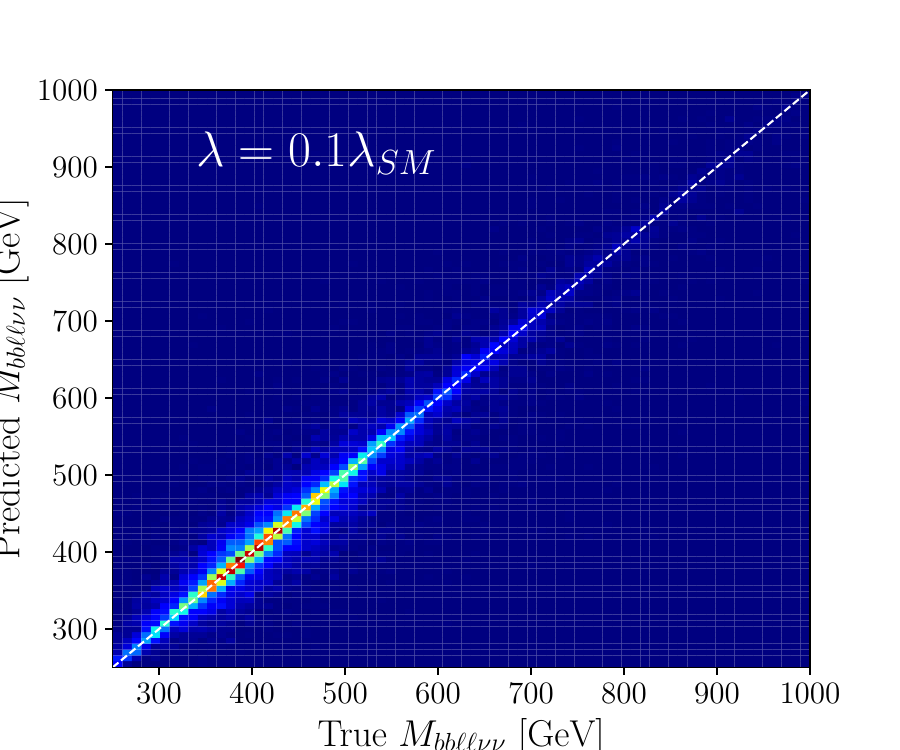}
    \includegraphics[width=0.32\linewidth]{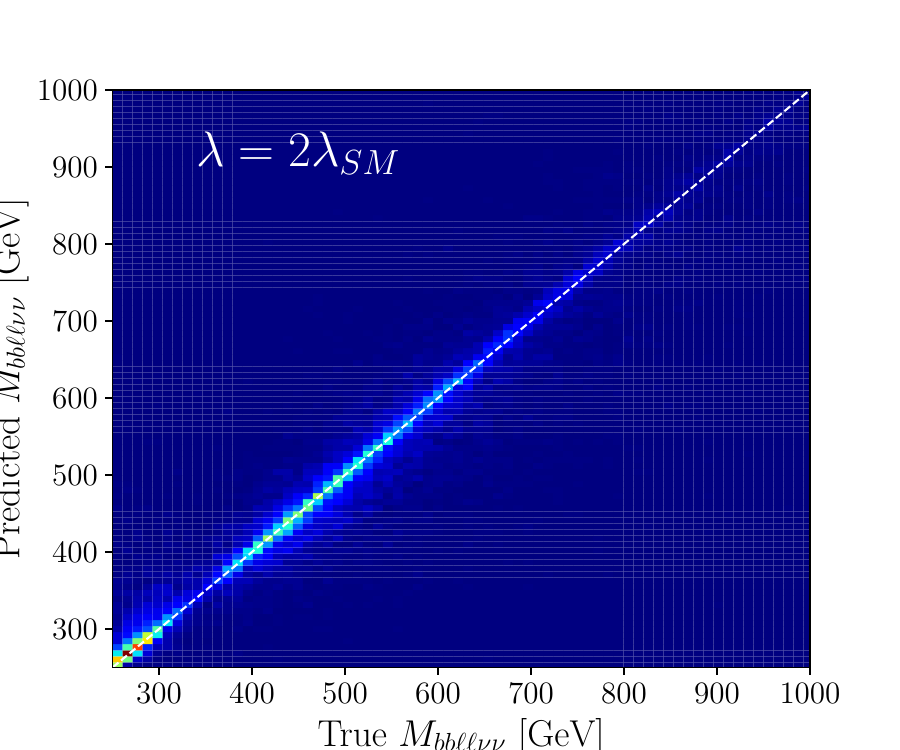}
    \caption{Upper panels: the true and predicted $hh$ invariant masses for the support couplings $\kappa_\lambda=-2$, $0.1$, and $2$. The blue-shaded regions represent the cross-validation uncertainties in the prediction. The ratio between VAER  prediction and ground truth is also depicted in these plots. Middle panels: scatter plots for true versus predicted masses. }
    \label{fig:m2-support}
\end{figure}
 Figure~\ref{fig:m2-support} shows the true and the predicted $d\sigma/dM_{hh}$ for some support couplings.  The agreement is good in all cases. A quantitative assessment of the predictions is given in the black entries of Table~\ref{tab:metrics}. For $\kappa_\lambda=2$, VAER predicts the shape of the dip in the distribution with good accuracy, as we see in the rightmost panel of Figure~\ref{fig:m2-support}. The bin right at the local minimum of the distribution, where the disagreement is the largest, differs by $\sim 20$\%. The disagreement increases for $\kappa_\lambda=3$, reaching an excess of 50\% compared to truth. Interestingly, this prediction is expected if we take detector smearing and higher-order corrections into account.
 
\subsection{Varying the Trilinear Higgs Couplings: Interpolated Couplings}

 Collision events associated with the true trilinear coupling might not pertain to the training set of the algorithm. The solution is to cover a finite grid of couplings during the learning process and expect the neural networks to generalize for intermediate couplings that were not presented to the regressor.  In principle, this can be achieved with parametrized neural networks~\cite{Baldi:2016fzo}  where the value of the coupling is concatenated with the features matrix. This approach is very useful for training a classifier that depends on theory parameters that affect the kinematic distributions and change the label prediction of the algorithm. It saves an enormous time in generating events during the training phase but it cannot be used, of course, in predicting the labels of data without knowing the true theory parameters. The same caveat applies to a regression problem.

 Contrary to unsupervised classification algorithms, like anomaly detection, for example, predicting a real-valued target function that depends on unknown theory parameters is a hard task. In our case, there is also the issue related to the missing neutrinos that carry information away. The target we need to predict is a function $\mbbllvv(\mathbf{x}_{obs}|\mathbf{\theta})$ where $\mathbf{x}_{obs}$ comprises only observable information and $\mathbf{\theta}$, the model parameters, are unknown. Moreover, the background must be taken into account in a joint learning process, that is it, we also want a single regressor to be able to correctly predict the background and the signal irrespective of the unknown theory parameters.
 
 In Ref.~\cite{Alves:2022gnw}, a combination of neural networks for signal {\it versus} background separation and $k$-nearest neighbors regressor are used for a post-discovery regression. In $k$NNNN, a pre-classification step to separate signal and background precedes the regression of the invariant mass. Once the event is classified, the algorithm uses a dedicated regressor for that specific class. The regressor is a simple $k$NN that precludes a training phase. As a clustering algorithm, it is unsupervised, which is a good feature but its weakness is needing a classification step to guide the regression.
\begin{figure}[t!]
    \centering
    \includegraphics[width=0.32\linewidth]{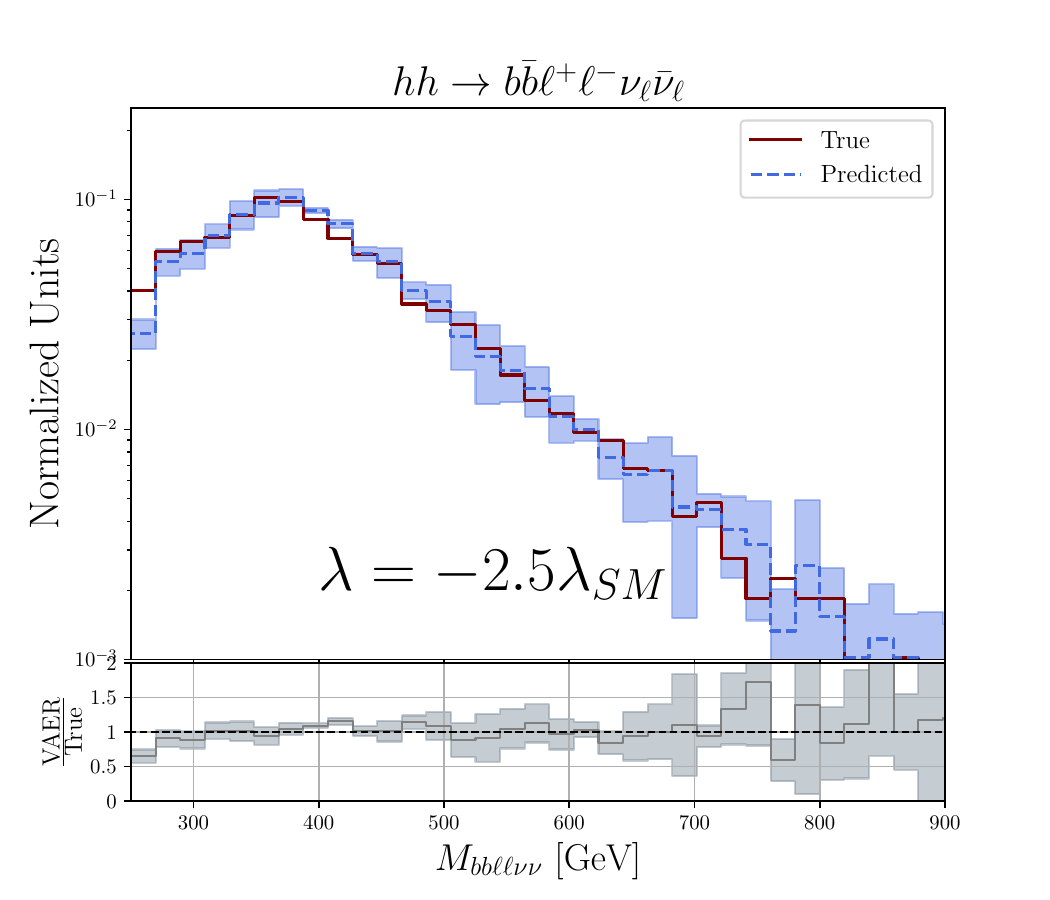}
    \includegraphics[width=0.32\linewidth]{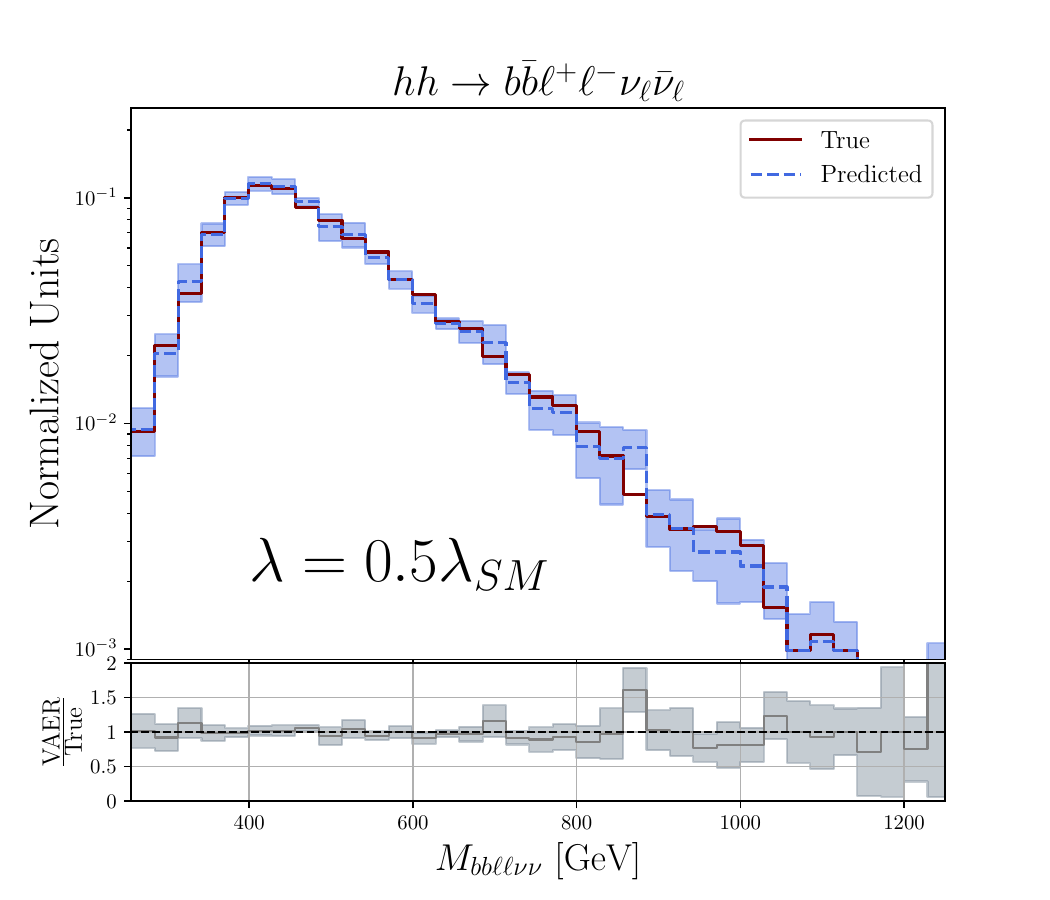}
    \includegraphics[width=0.32\linewidth]{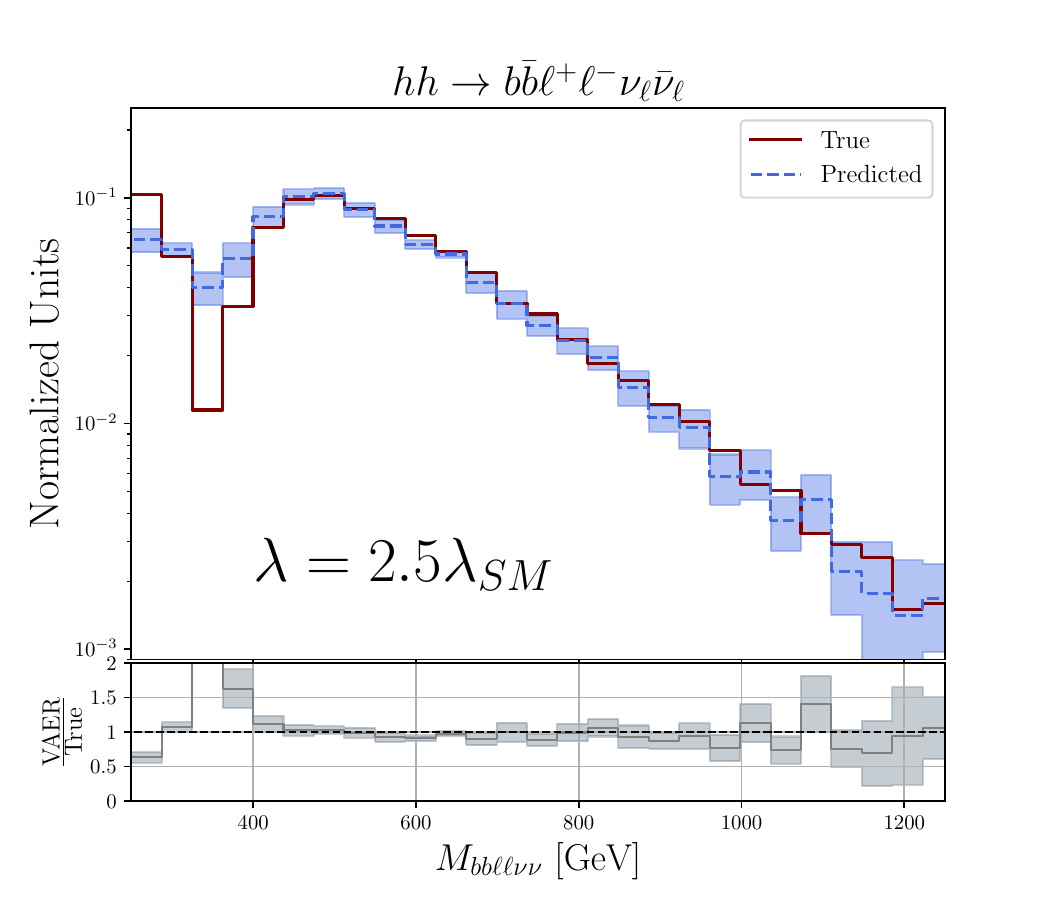} \\
    \includegraphics[width=0.32\linewidth]{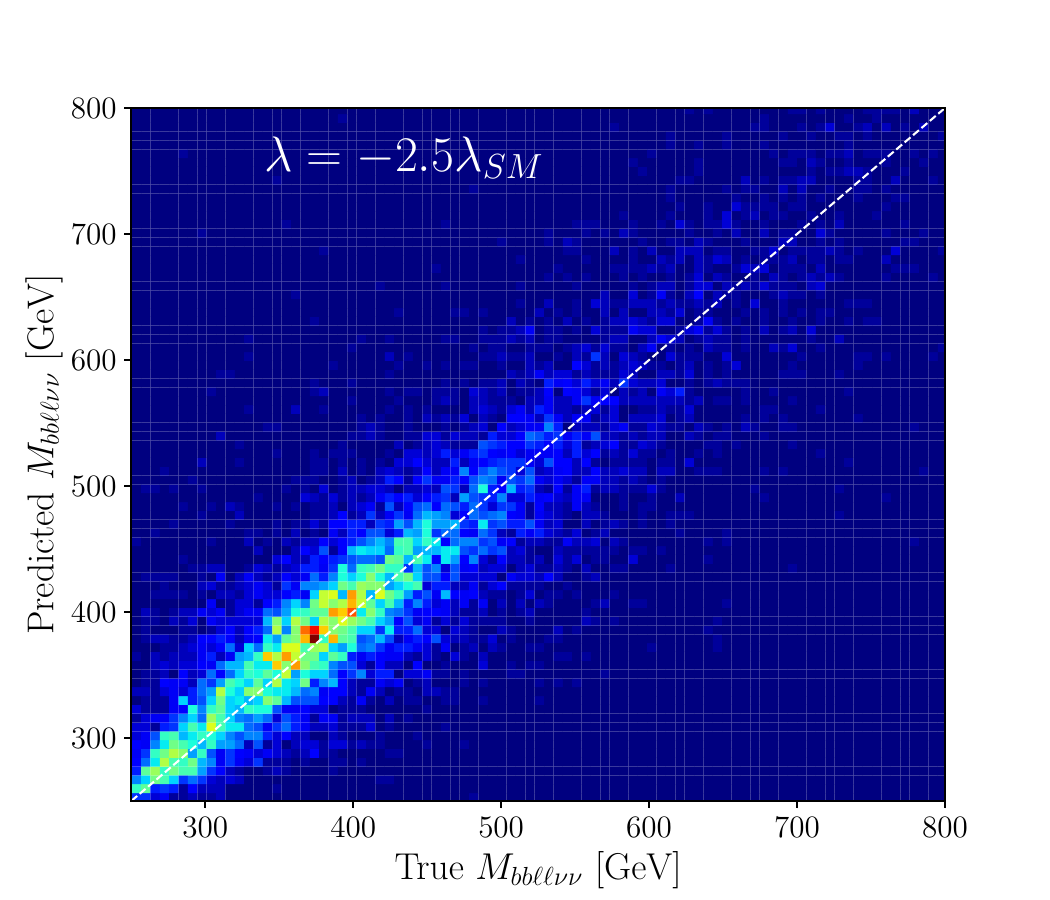}
    \includegraphics[width=0.32\linewidth]{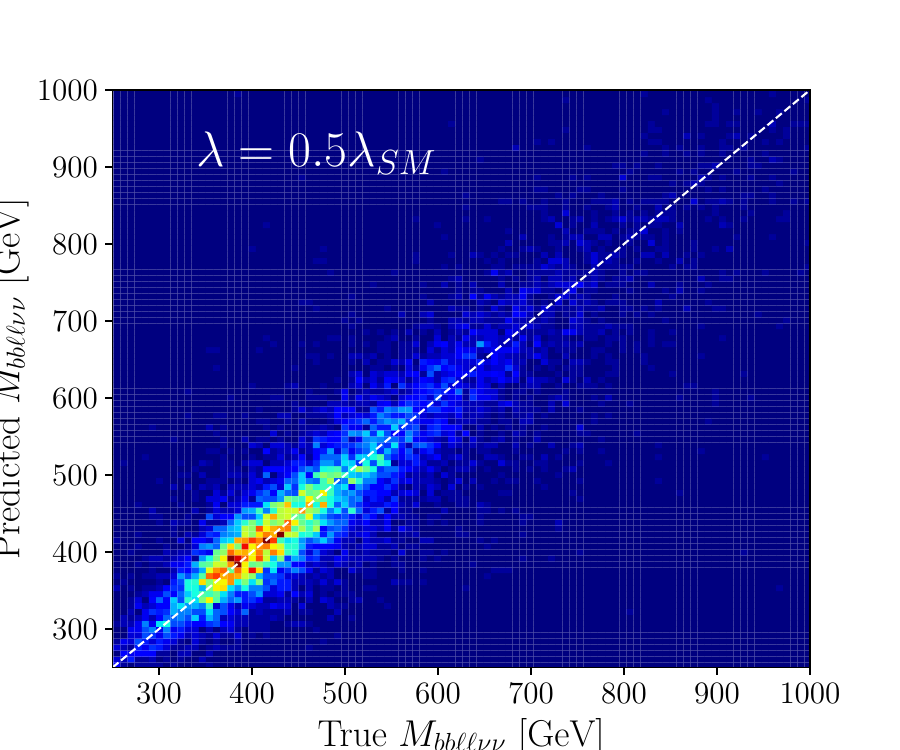}
    \includegraphics[width=0.32\linewidth]{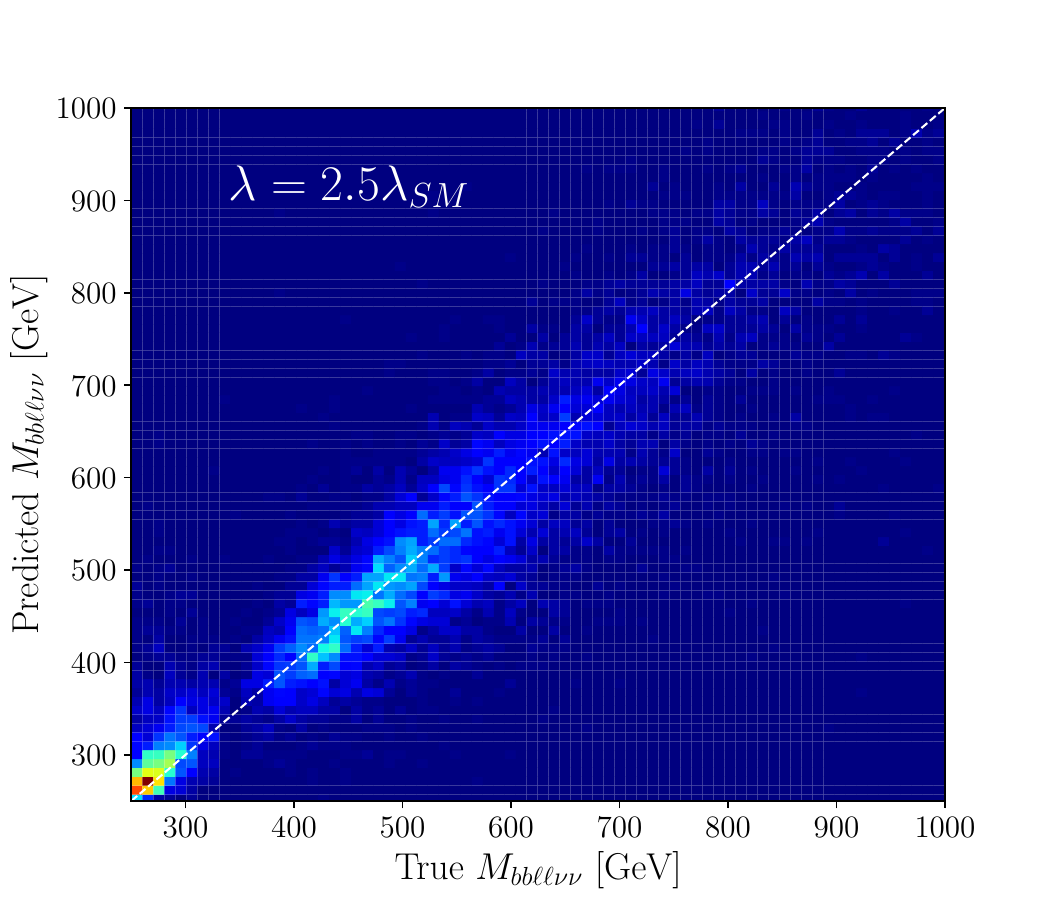} 
    \caption{Upper panels: the true and predicted $hh$ invariant masses for the interpolated couplings $\kappa_\lambda=-2.5$, $0.5$, and $2.5$. The blue-shaded regions represent the cross-validation uncertainties in the prediction. The ratio between VAER  prediction and ground truth is also depicted in these plots. Lower panels: scatter plots for true versus predicted masses. }
    \label{fig:m2-bsm-1}
\end{figure}
 Deep learning regressors were used in the reconstruction of tops from semi-leptonic $t\bar{t}$ events~\cite{Erdmann_2019} and $t\bar{t}h$ reconstruction~\cite{Erdmann_2017}. In those cases, the jet combinatorics have to be solved to correctly assign the jets to top quarks for their reconstruction, enabling mass and top-Higgs coupling measurements, respectively. Contrary to our task, that reconstruction assumes a pure and unambiguous identification of samples. If some other type of events other than those the neural networks were trained to recognize are present, there is no guarantee that they will generalize properly.
%
%
 The examples cited above show some of the difficulties in the task of machine learning-assisted regression of kinematic variables without some previous knowledge of the events. What VAER tries to emulate is a function of observable information that predicts the $b\bar{b}\ell^+\ell^{\prime -}\nu_\ell\bar{\nu}_{\ell^\prime}$ mass with less previous information about the nature of the events. The framework is not completely unsupervised, though. We trained the algorithm with some of the types of events we guess that might appear in that channel, signals, and background.
 However, the regressor training occurs in a single stage, and no previous classification or label assignment is needed. In this respect, VAER offers a step ahead in the solution of reconstructing events with missing information.

 Concerning the signals, a useful algorithm recognizes events associated with new trilinear couplings, and possibly other model parameters, that did show up in the training phase. It must generalize the reconstruction to other parameters never seen, at least for parameters inside the range of the support grid couplings. In Figure~\ref{fig:m2-bsm-1},  we depict $\mbbllvv$ for some intermediate $\lambda$. None of them were previously presented to VAER. In the upper plots of Figure~\ref {fig:m2-bsm-1}, we show the distributions for $\kappa_\lambda=-2.5$, $0.5$, and $+2.5$. The lower plots show the true {\it versus} predicted masses. The visual agreement is again corroborated by the qualitative assessment of the performance displayed in the cyan entries of Table~\ref{tab:metrics}.  
%
%
 A general feature that might be improved is that the prediction deteriorates at the extremes of the distribution, in the first bin, at the onset of the distribution, and in the last bins, in the tail. This might be mitigated by choosing larger bins and possibly by increasing the number of examples at the training phase. A coarser grid of support couplings can also help to bring the predictions closer to the ground truth in those bins.
 As in the case of $\kappa_\lambda=+3$, the algorithm correctly identifies the dip in the distribution caused by the destructive interference when $\kappa_\lambda=+2.5$ as we see in the rightmost panel of Figure~\ref{fig:m2-bsm-1}, but it is shallower than the true distribution. In all cases, the bulk of the distribution around the peak value is very well predicted. Overall, however, the interpolated predictions present a diminished quality compared to the support ones, although they are still good.
 

\subsection{Robustness against Kinematic Cuts}

 If VAER truly emulates $\mbbllvv$ as a function of observable kinematics, it should reconstruct the event in the whole of the phase space, just like any parametric function. 
\begin{figure}[t!]
    \centering
    \includegraphics[width=0.475\linewidth]{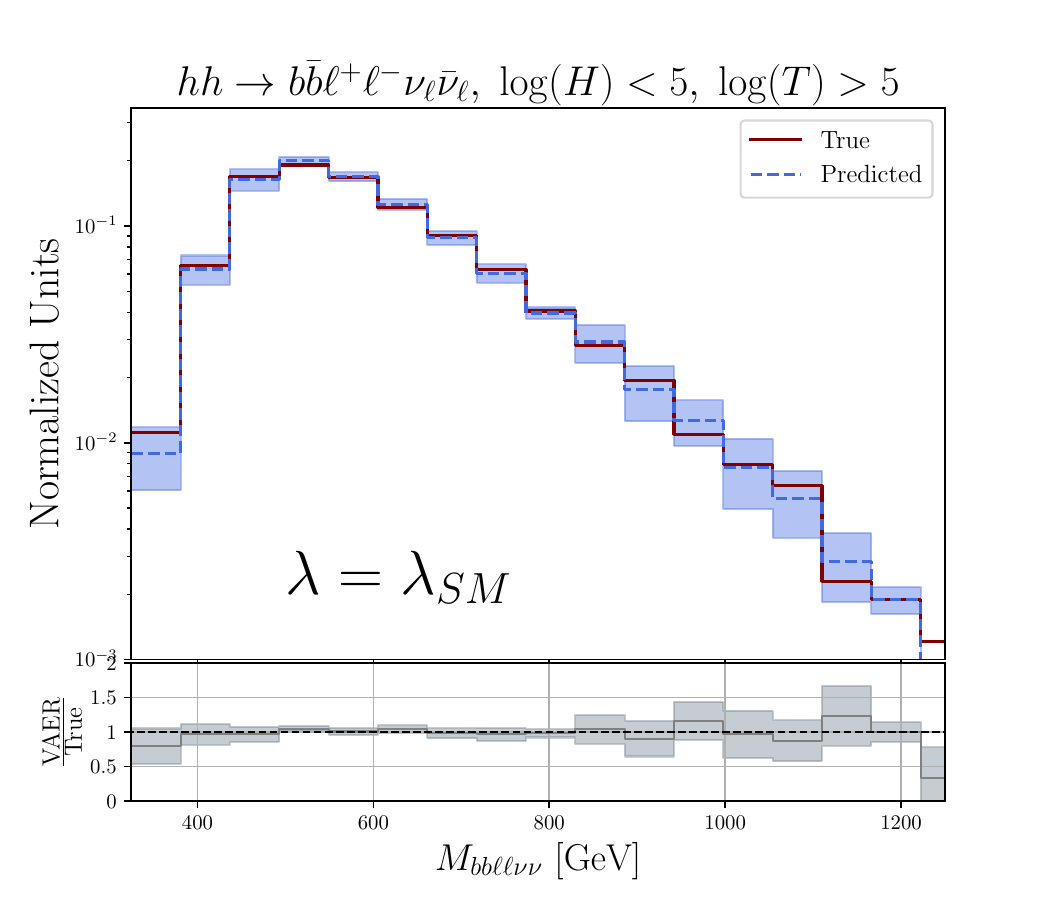}
    \includegraphics[width=0.475\linewidth]{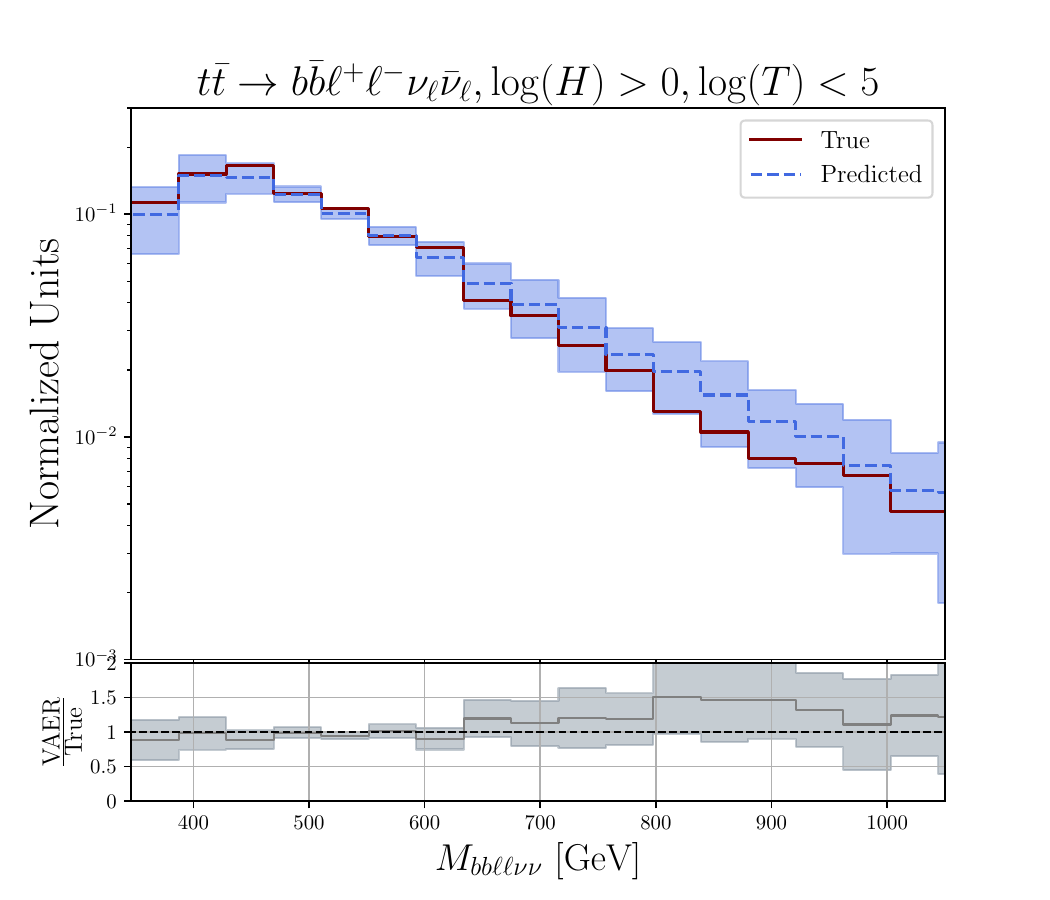} \\
    \includegraphics[width=0.475\linewidth]{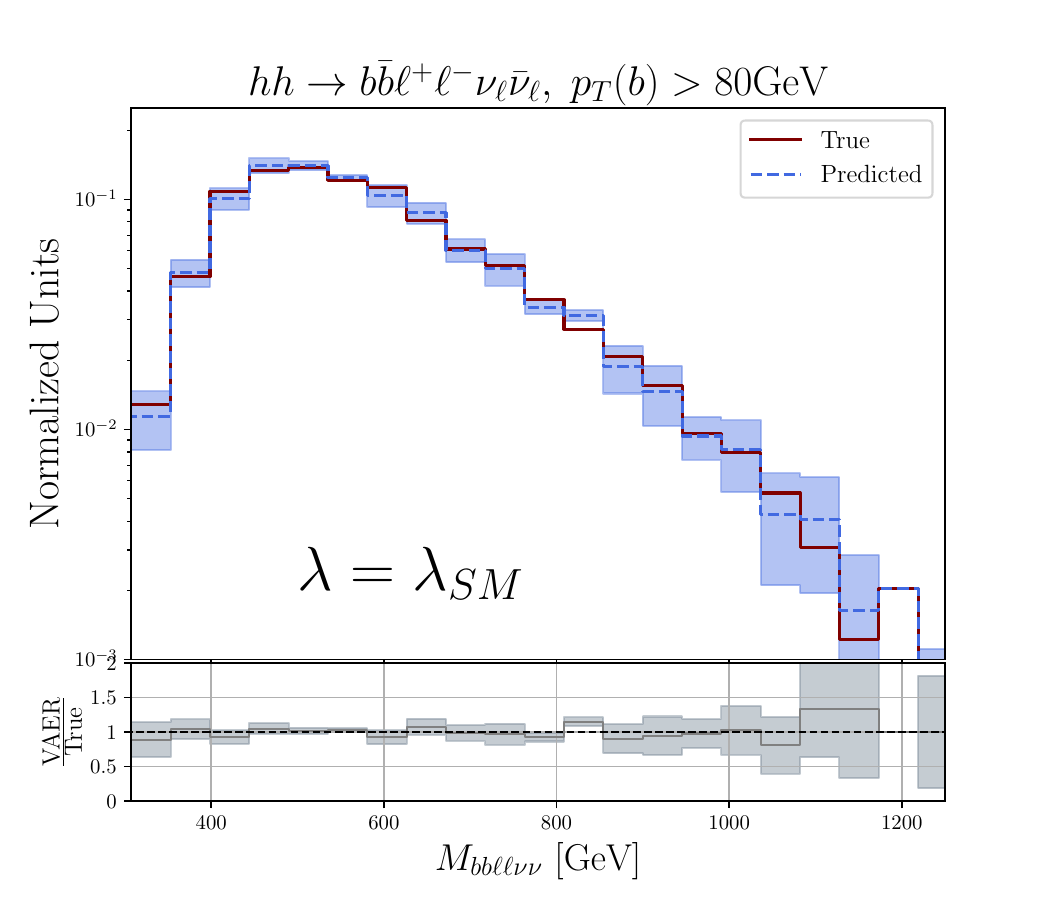}
    \includegraphics[width=0.475\linewidth]{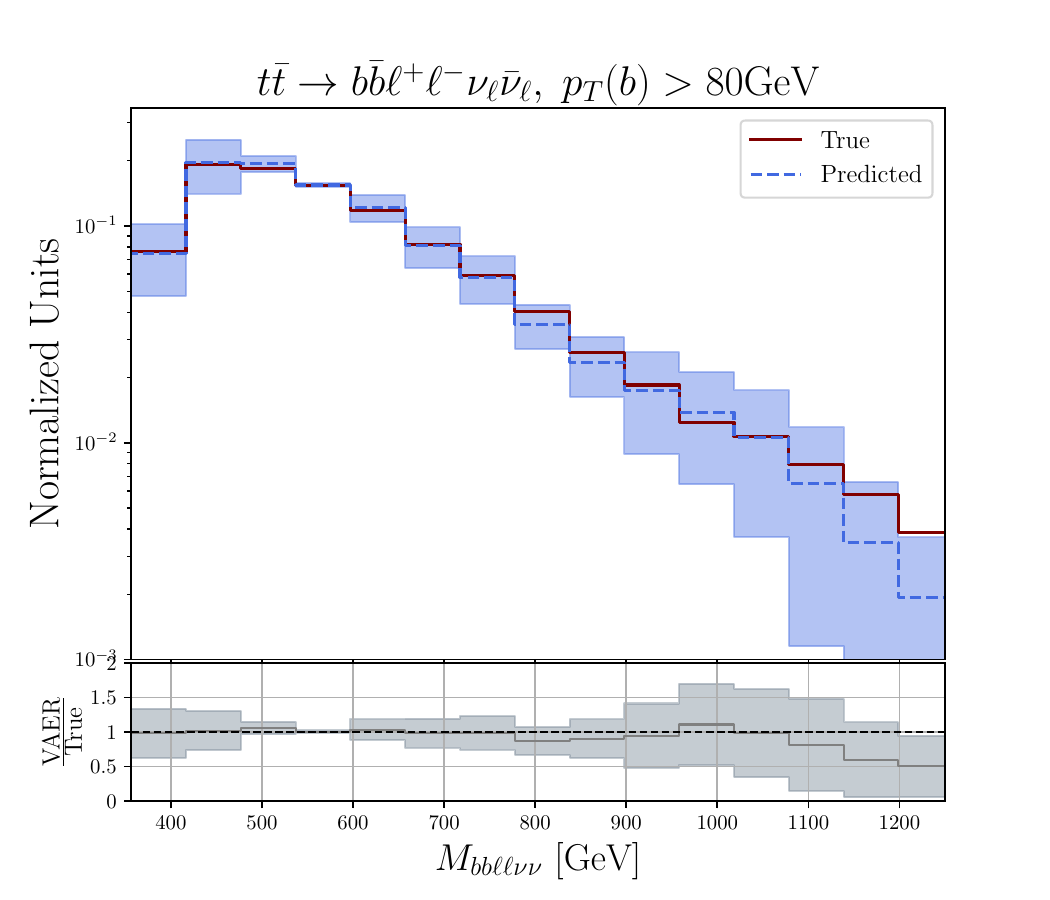} 
    
    \caption{Upper panels: the true and predicted SM $hh$ and $t\bar{t}$ after cuts on Higgsness and Topness variables as described in Eq.~\eqref{eq:more-cuts}. Lower panels: distributions of events selected after imposing $p_T(b)>80$ GeV besides the basic cuts. The blue-shaded regions represent the cross-validation uncertainties in the prediction. The ratio between VAER  prediction and ground truth is also depicted in these plots.  }
    \label{fig:m2-cuts}
\end{figure}
 We tested VAER in predicting the $hh$ and $t\bar{t}$ invariant masses with the following harder cuts besides the basic ones
 \begin{eqnarray}
     && \log(H) < 5,\;\; \log(T)>5 \; \hbox{for}\; hh\\
     && \log(H) > 0,\;\; \log(T)<5 \; \hbox{for}\; t\bar{t}\\
     && p_T(b)>80\; \hbox{GeV}\; \hbox{for both}\; hh,t\bar{t}\; .
     \label{eq:more-cuts}
 \end{eqnarray}

 {\it Higgsness} and {\it Topness} are two very distinctive variables to separate signal and background. Double Higgs events tend to have smaller {\it Higgsness} and larger {\it Topness} compared to $t\bar{t}$, which motivates the cuts on those variables to isolate data from $hh$ and $t\bar{t}$. A way to increase $b$-jet tagging is to impose a harder bottom-jet transverse momentum cut. For $p_T(b)>80$ GeV, for example, \texttt{Delphes3}~\cite{deFavereau:2013fsa} reaches a higher $b$-jet tagging efficiency of around 70\% mimicking the detector's true efficiency. In Figure~\ref{fig:m2-cuts}, we show the SM $hh$ and $t\bar{t}$ mass distributions for the cuts of Eq.~\eqref{eq:more-cuts}. The agreement remains good, especially for $hh$ events. In the case of harder cuts in $\log(H)$ and $\log(T)$ to isolate $t\bar{t}$ events, we observe a somewhat harder predicted spectrum compared to truth. In all cases, though, a very good agreement is achieved for masses up to 800 GeV. Moreover, the true distribution always lies within the error band of the cross-validation.

 These experiments give us confidence that the VAER prediction can be useful in helping the phenomenological analysis of these types of events by providing another distinctive kinematic variable to isolate the signal events. We reinforce that the training dataset just contains events with the basic selection requirements of Eq.~\eqref{eq:cuts}.

\subsection{Chi-Square Computation with VAER distributions}

 The sensitivity of $M_{hh}$ to $\lambda$ makes it a good target for inferring the Higgs trilinear self-coupling offering its shape along with the number of events of its normalization to measure that theory parameter. In the case where $hh\to b\bar{b}W^+W^-\to b\bar{b}\ell^+\ell^{\prime -}\nu_\ell\bar{\nu}_{\ell^\prime}$, the $\mbbllvv$ distribution inherits that sensitivity but, of course, it must be reconstructed despite the missing neutrinos components. VAER, as we have shown, provides accurate histograms of $\mbbllvv$ that can be used for statistical inference of $\lambda$. 
 
\begin{figure}[t]
    \centering
    \includegraphics[width=0.8\linewidth]{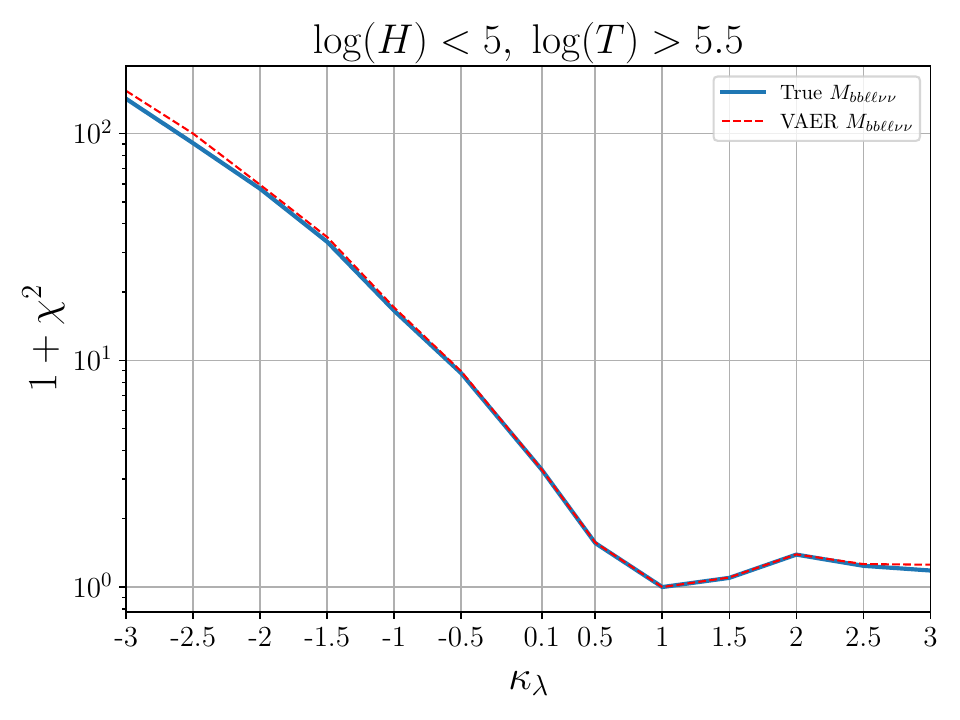}
    
    \caption{The $\chi^2$ between $S(\kappa_\lambda\neq 1)$ and the SM $S$, the number of signal events for each $\lambda$ hypothesis. We depict both the true and the predicted curves, adding 1 to $\chi^2$ just to enable us to show them in log scale.}
    \label{fig:chi2}
\end{figure}

 To demonstrate its usefulness for practical purposes, we show, in Figure~\ref{fig:chi2}, a simplified $\chi^2$ computation ignoring backgrounds after imposing a hard cut on {\it Higgsness} and {\it Topness} variables of $\log(H)<5,\; \log(T)>5.5$. We checked that no $t\bar{t}$ survives to those cuts. The number of signal events, however, is also small, a few tens at most, and other cuts might be needed to surely ignore the backgrounds~\cite{Kim:2018cxf}. We do not intend to calculate bounds to $\lambda$ in this work but just to demonstrate that the VAER prediction can be used for that purpose. The computation was performed using a 10-bins histogram of $\mbbllvv$.
 
 The $\chi^2$ is thus computed as
 \begin{equation}
     \chi^2 = \sum_{i=1}^{\hbox{\# bins}}\frac{\left[S_i(\kappa_\lambda\neq 1)-S_i(\kappa_\lambda=1)\right]^2}{S_i(\kappa_\lambda=1)}
 \end{equation}
 to test an alternative $\lambda$ hypothesis against the SM one. In this formula, $S(\kappa_\lambda)$ is the number of signal events for a given $\kappa_\lambda$, after the hard {\it Higgsness} and {\it Topness} cuts mentioned in the previous paragraph.

 As we see in Figure~\ref{fig:chi2}, the agreement between the $\chi^2$ computed from the true and the VAER predicted $\mbbllvv$ distributions is very good. The VAER $\chi^2$ curve is slightly above the true curve, making the inference a bit conservative.

\section{Reconstruction of fully leptonic $b\bar{b}W^+W^-$ events: Heavy Higgs decay}
\label{sec:res}
\begin{figure}[t]
    \centering
    \includegraphics[width=0.475\linewidth]{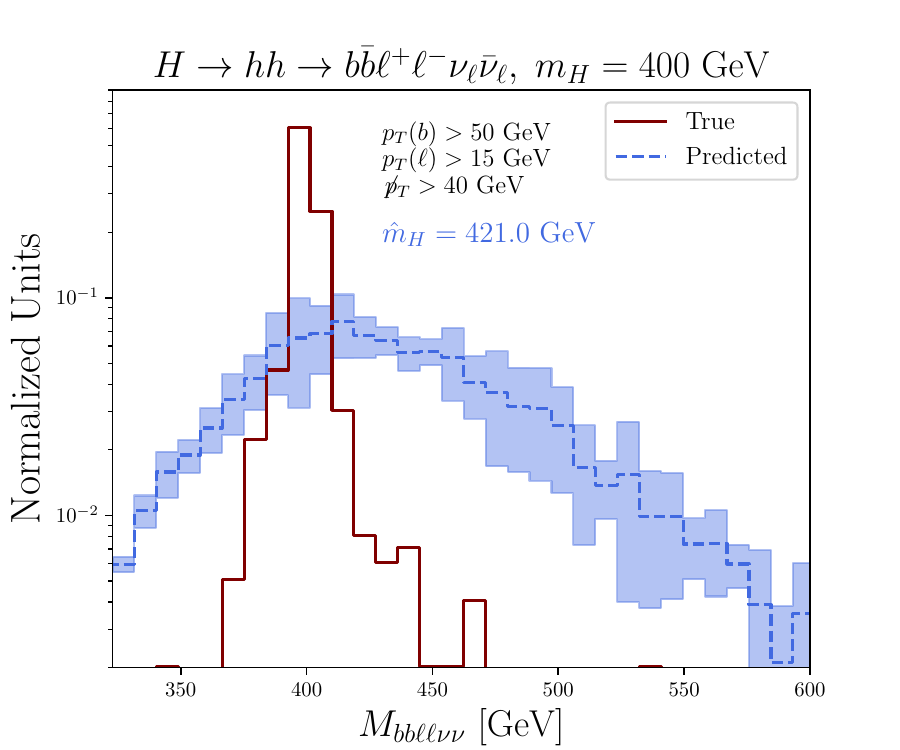}
    \includegraphics[width=0.475\linewidth]{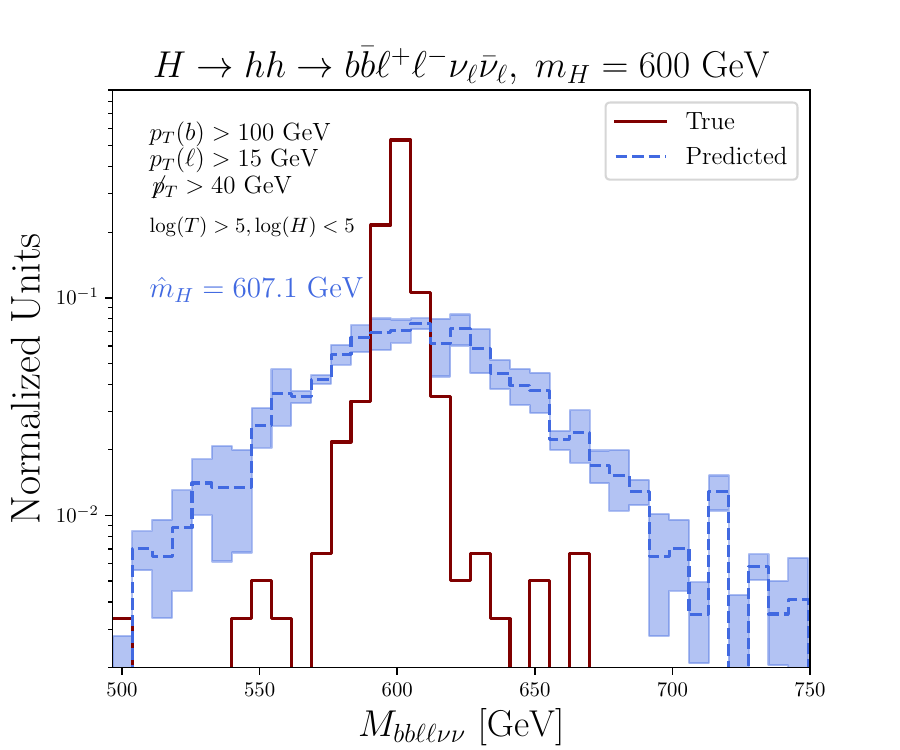} \\
    \includegraphics[width=0.475\linewidth]{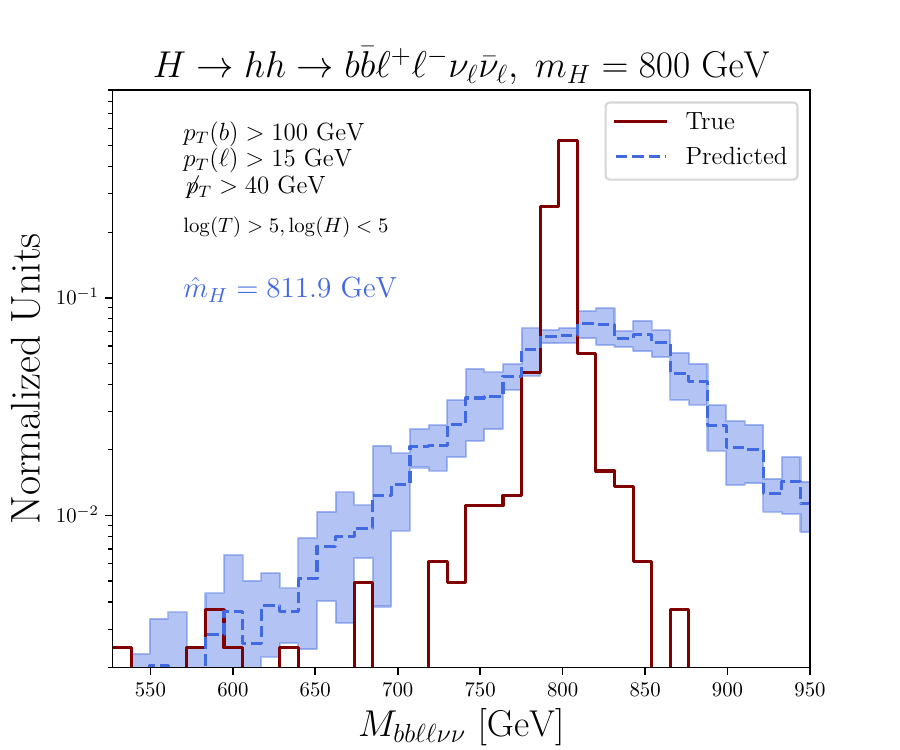}
    \includegraphics[width=0.475\linewidth]{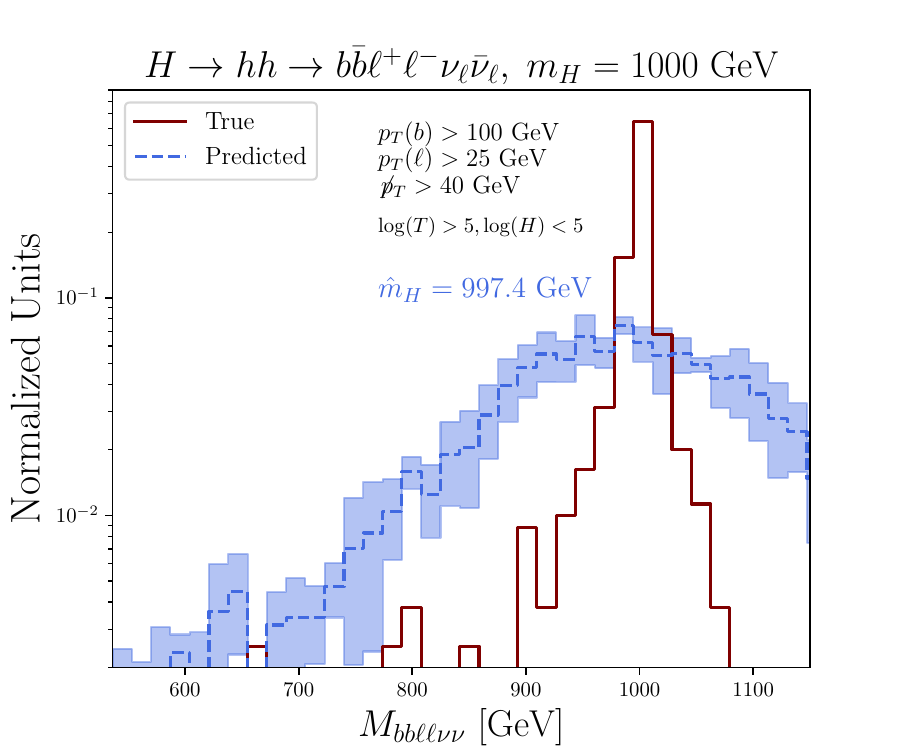} 
    
    \caption{True and predicted $\mbbllvv$ for a heavy Higgs boson of 400, 600, 800, and 1000 GeV masses decaying to $hh\to b\bar{b}\ell^+\ell^{\prime -}\nu_\ell\bar{\nu}_{\ell^\prime}$. As in the previous plots, the blue-shaded regions represent the cross-validation uncertainties in the prediction. The total width of the new scalar is fixed at $m_H/10$. Different cuts were applied besides the basic ones. The location of the peak of the predicted distributions is $\hat{m}_H$.}
    \label{fig:m2-bsm-resonant}
\end{figure}

 In extended scalar models, like xSM~\cite{Profumo:2007wc,Profumo:2014opa,Huang:2017jws}, besides shifts in trilinear couplings, new heavy Higgs bosons, $H$, might appear in the particle spectrum. If the new scalar has a sizeable decay into SM Higgs bosons, a resonance in $hh$ mass would be a smoking gun signature. Of course, the resonance is missing if the $W$ bosons of $hh\to b\bar{b}W^+W^-$ decay leptonically so VAER can be used to reconstruct the peak of the $H$ decay. 

 To test VAER in resonant $hh$ production, we generated events for $gg\to H\to hh\to b\bar{b}\ell^+\ell^{\prime -}\nu_\ell\bar{\nu}_{\ell^\prime}$ with \texttt{MadGraph5}, \texttt{Delphes3}, and \texttt{Pythia8}. We tested four hypothetical masses, $m_H$: 400, 600, 800, and 1000 GeV. In all cases, we fixed the total width of the new boson to $m_H/10$. All simulation parameters were fixed as the non-resonant cases.

 We also hardened the selection cuts to mimic the possible experimental searches in that channel. We display the true and predicted $\mbbllvv$ masses in Figure~\ref{fig:m2-bsm-resonant}. The selection cuts are shown in the plots. We compute the mode of the binned distributions and found around 5\%, 1\%, 1\%, and 0.3\% discrepancies to the true mass for 400, 600, 800, and 1000 GeV masses, respectively. As we see in Figure~\ref{fig:m2-bsm-resonant}, despite VAER predicting the peak of the distributions accurately, it does not capture its width, predicting a much broader distribution compared to the true case. The mass prediction improves for larger masses and harder cuts but the effect on width remains. We point out that VAER has not been trained to predict a resonance signal. The prediction can be considered unsupervised in this sense. 

 In Ref.~\cite{Huang:2017jws}, xSM new Higgs bosons decaying to $hh\to b\bar{b}W^+W^-$, and leptonic $W$ decays, are reconstructed using the Heavy Mass Estimator (HME) technique~\cite{Elagin:2010aw}. The HME technique resembles the {\it Higgsnes} calculation but it keeps the solutions to the neutrinos' momenta and uses them to calculate the mass of $H$. The results from Ref.~\cite{Huang:2017jws} for heavy Higgses of masses comparable to those we simulated in this work show a similar accuracy and peak resolution, however, they do not generalize to background events. 

 We postpone to a future investigation a detailed statistical estimate of the signal significance that can be achieved by searching for such a resonance in the tail of the background $\mbbllvv$, but with the help of {\it Higgsness} and {\it Topness} variables, we believe that a statistical analysis may benefit from the VAER reconstruction of the peaks, possibly enabling an estimate of the mass of the resonance.


\section{Conclusions e Outlook}
\label{sec:conclusions}

 Recovering information leaked in the emission of neutrinos, dark matter, or long-lived particles is a research field of its own. Much effort has been put into reconstruction algorithms and proxy functions that might capture the kinematics of the missing components in collision events at high-energy colliders. In this work, we proposed a parametrized function of the observable momenta in the reconstruction of the final state $b\bar{b}\ell^+\ell^{\prime -}\nu_\ell\bar{\nu}_{\ell^\prime}$ from double Higgs production and its leading background source, $t\bar{t}$ pairs. The parameterization is provided by neural networks in a variational autoencoder algorithm designed for regression tasks and trained with a dataset comprising detector-level events generated from a grid of trilinear couplings for $hh$ simulated data besides $t\bar{t}$ data. 

 We showed that VAER presents a very good generalization power, accurately predicting the partonic invariant mass $\mbbllvv$ of the $t\bar{t}$ background and across events associated with the various trilinear coupling of the support grid of the training set. Moreover, it also provides good predictions of $\mbbllvv$ in events of trilinear couplings and resonant new Higgs production and decay into $hh$ which were not present in the training phase corroborating its generalization performance.  

 Its usefulness was tested against harder selection cuts beyond those used to select the training dataset and, once more, confirmed that VAER is capable of learning a function of the observable kinematics to output a variable that encompasses missing momenta. The algorithm is easy to train, not requiring extensive tuning or a large amount of data. All our predictions were validated through statistically independent cross-validation sets and showed a good degree of robustness.

 Reconstructing $\mbbllvv$ opens the possibility of using the shape of a distribution that is very sensitive to $\lambda$ in measuring the trilinear coupling, besides the cross section measurement, in the $b\bar{b}W^+W^-$ channel that has been recently rehabilitated as a competitive channel for double Higgs studies~\cite{Kim:2018cxf}. In conjunction with powerful variables like {\it Higgsness} and {\it Topness}~\cite{Kim:2018cxf}, for example, VAER could provide a variable to compare data and theory to measure the trilinear coupling of the SM scalar potential. As a practical evaluation of the algorithm, we showed that a $\chi^2$ computation based on VAER $\mbbllvv$ histograms can be used as a reliable estimate of the statistic.

 We envisage other applications, though. For example, VAER can be used to reconstruct final states with dark matter particles, including intermediate particles that decay into them. Recovering partonic distributions from detector-level events should also be easy, making the algorithm an option for unfolding. Mass and spin measurements could also benefit from fully available kinematic variables. Of course, without mentioning the original application that motivated us, the regression of a variable from images, as in Ref.~\cite{DBLP:journals/corr/abs-1904-05948}, where VAER could be adapted to infer properties of jets from their images, for example. 

 \vskip0.5cm
\textbf{Acknowledgments}: This study was supported by Conselho Nacional de Desenvolvimento Científico e Tecnológico (CNPq), grants 307317/2021-8 (A.A.), 305802/2019-4 (I.N.M.). A. Alves also acknowledges support from FAPESP 2021/01089-1 grant. 

\bibliography{refs}

\begin{thebibliography}{34}%
\makeatletter
\providecommand \@ifxundefined [1]{%
 \@ifx{#1\undefined}
}%
\providecommand \@ifnum [1]{%
 \ifnum #1\expandafter \@firstoftwo
 \else \expandafter \@secondoftwo
 \fi
}%
\providecommand \@ifx [1]{%
 \ifx #1\expandafter \@firstoftwo
 \else \expandafter \@secondoftwo
 \fi
}%
\providecommand \natexlab [1]{#1}%
\providecommand \enquote  [1]{``#1''}%
\providecommand \bibnamefont  [1]{#1}%
\providecommand \bibfnamefont [1]{#1}%
\providecommand \citenamefont [1]{#1}%
\providecommand \href@noop [0]{\@secondoftwo}%
\providecommand \href [0]{\begingroup \@sanitize@url \@href}%
\providecommand \@href[1]{\@@startlink{#1}\@@href}%
\providecommand \@@href[1]{\endgroup#1\@@endlink}%
\providecommand \@sanitize@url [0]{\catcode `\\12\catcode `\$12\catcode
  `\&12\catcode `\#12\catcode `\^12\catcode `\_12\catcode `\%12\relax}%
\providecommand \@@startlink[1]{}%
\providecommand \@@endlink[0]{}%
\providecommand \url  [0]{\begingroup\@sanitize@url \@url }%
\providecommand \@url [1]{\endgroup\@href {#1}{\urlprefix }}%
\providecommand \urlprefix  [0]{URL }%
\providecommand \Eprint [0]{\href }%
\providecommand \doibase [0]{http://dx.doi.org/}%
\providecommand \selectlanguage [0]{\@gobble}%
\providecommand \bibinfo  [0]{\@secondoftwo}%
\providecommand \bibfield  [0]{\@secondoftwo}%
\providecommand \translation [1]{[#1]}%
\providecommand \BibitemOpen [0]{}%
\providecommand \bibitemStop [0]{}%
\providecommand \bibitemNoStop [0]{.\EOS\space}%
\providecommand \EOS [0]{\spacefactor3000\relax}%
\providecommand \BibitemShut  [1]{\csname bibitem#1\endcsname}%
\let\auto@bib@innerbib\@empty
\bibitem [{\citenamefont {Group}\ \emph {et~al.}(2022)\citenamefont {Group},
  \citenamefont {Workman}, \citenamefont {Burkert}, \citenamefont {Crede},
  \citenamefont {Klempt}, \citenamefont {Thoma}, \citenamefont {Tiator},
  \citenamefont {Agashe}, \citenamefont {Aielli}, \citenamefont {Allanach},
  \citenamefont {Amsler}, \citenamefont {Antonelli}, \citenamefont
  {Aschenauer}, \citenamefont {Asner}, \citenamefont {Baer}, \citenamefont
  {Banerjee}, \citenamefont {Barnett}, \citenamefont {Baudis}, \citenamefont
  {Bauer}, \citenamefont {Beatty}, \citenamefont {Belousov}, \citenamefont
  {Beringer}, \citenamefont {Bettini}, \citenamefont {Biebel}, \citenamefont
  {Black}, \citenamefont {Blucher}, \citenamefont {Bonventre}, \citenamefont
  {Bryzgalov}, \citenamefont {Buchmuller}, \citenamefont {Bychkov},
  \citenamefont {Cahn}, \citenamefont {Carena}, \citenamefont {Ceccucci},
  \citenamefont {Cerri}, \citenamefont {Chivukula}, \citenamefont {Cowan},
  \citenamefont {Cranmer}, \citenamefont {Cremonesi}, \citenamefont
  {D'Ambrosio}, \citenamefont {Damour}, \citenamefont {de~Florian},
  \citenamefont {de~Gouvêa}, \citenamefont {DeGrand}, \citenamefont {de~Jong},
  \citenamefont {Demers}, \citenamefont {Dobrescu}, \citenamefont {D'Onofrio},
  \citenamefont {Doser}, \citenamefont {Dreiner}, \citenamefont {Eerola},
  \citenamefont {Egede}, \citenamefont {Eidelman}, \citenamefont {El-Khadra},
  \citenamefont {Ellis}, \citenamefont {Eno}, \citenamefont {Erler},
  \citenamefont {Ezhela}, \citenamefont {Fetscher}, \citenamefont {Fields},
  \citenamefont {Freitas}, \citenamefont {Gallagher}, \citenamefont
  {Gershtein}, \citenamefont {Gherghetta}, \citenamefont {Gonzalez-Garcia},
  \citenamefont {Goodman}, \citenamefont {Grab}, \citenamefont {Gritsan},
  \citenamefont {Grojean}, \citenamefont {Groom}, \citenamefont {Grünewald},
  \citenamefont {Gurtu}, \citenamefont {Gutsche}, \citenamefont {Haber},
  \citenamefont {Hamel}, \citenamefont {Hanhart}, \citenamefont {Hashimoto},
  \citenamefont {Hayato}, \citenamefont {Hebecker}, \citenamefont {Heinemeyer},
  \citenamefont {Hernández-Rey}, \citenamefont {Hikasa}, \citenamefont
  {Hisano}, \citenamefont {Höcker}, \citenamefont {Holder}, \citenamefont
  {Hsu}, \citenamefont {Huston}, \citenamefont {Hyodo}, \citenamefont {Ianni},
  \citenamefont {Kado}, \citenamefont {Karliner}, \citenamefont {Katz},
  \citenamefont {Kenzie}, \citenamefont {Khoze}, \citenamefont {Klein},
  \citenamefont {Krauss}, \citenamefont {Kreps}, \citenamefont {Križan},
  \citenamefont {Krusche}, \citenamefont {Kwon}, \citenamefont {Lahav},
  \citenamefont {Laiho}, \citenamefont {Lellouch}, \citenamefont {Lesgourgues},
  \citenamefont {Liddle}, \citenamefont {Ligeti}, \citenamefont {Lin},
  \citenamefont {Lippmann}, \citenamefont {Liss}, \citenamefont {Littenberg},
  \citenamefont {Lourenço}, \citenamefont {Lugovsky}, \citenamefont
  {Lugovsky}, \citenamefont {Lusiani}, \citenamefont {Makida}, \citenamefont
  {Maltoni}, \citenamefont {Mannel}, \citenamefont {Manohar}, \citenamefont
  {Marciano}, \citenamefont {Masoni}, \citenamefont {Matthews}, \citenamefont
  {Meißner}, \citenamefont {Melzer-Pellmann}, \citenamefont {Mikhasenko},
  \citenamefont {Miller}, \citenamefont {Milstead}, \citenamefont {Mitchell},
  \citenamefont {Mönig}, \citenamefont {Molaro}, \citenamefont {Moortgat},
  \citenamefont {Moskovic}, \citenamefont {Nakamura}, \citenamefont {Narain},
  \citenamefont {Nason}, \citenamefont {Navas}, \citenamefont {Nelles},
  \citenamefont {Neubert}, \citenamefont {Nevski}, \citenamefont {Nir},
  \citenamefont {Olive}, \citenamefont {Patrignani}, \citenamefont {Peacock},
  \citenamefont {Petrov}, \citenamefont {Pianori}, \citenamefont {Pich},
  \citenamefont {Piepke}, \citenamefont {Pietropaolo}, \citenamefont {Pomarol},
  \citenamefont {Pordes}, \citenamefont {Profumo}, \citenamefont {Quadt},
  \citenamefont {Rabbertz}, \citenamefont {Rademacker}, \citenamefont
  {Raffelt}, \citenamefont {Ramsey-Musolf}, \citenamefont {Ratcliff},
  \citenamefont {Richardson}, \citenamefont {Ringwald}, \citenamefont
  {Robinson}, \citenamefont {Roesler}, \citenamefont {Rolli}, \citenamefont
  {Romaniouk}, \citenamefont {Rosenberg}, \citenamefont {Rosner}, \citenamefont
  {Rybka}, \citenamefont {Ryskin}, \citenamefont {Ryutin}, \citenamefont
  {Sakai}, \citenamefont {Sarkar}, \citenamefont {Sauli}, \citenamefont
  {Schneider}, \citenamefont {Schönert}, \citenamefont {Scholberg},
  \citenamefont {Schwartz}, \citenamefont {Schwiening}, \citenamefont {Scott},
  \citenamefont {Sefkow}, \citenamefont {Seljak}, \citenamefont {Sharma},
  \citenamefont {Sharpe}, \citenamefont {Shiltsev}, \citenamefont {Signorelli},
  \citenamefont {Silari}, \citenamefont {Simon}, \citenamefont {Sjöstrand},
  \citenamefont {Skands}, \citenamefont {Skwarnicki}, \citenamefont {Smoot},
  \citenamefont {Soffer}, \citenamefont {Sozzi}, \citenamefont {Spanier},
  \citenamefont {Spiering}, \citenamefont {Stahl}, \citenamefont {Stone},
  \citenamefont {Sumino}, \citenamefont {Syphers}, \citenamefont {Takahashi},
  \citenamefont {Tanabashi}, \citenamefont {Tanaka}, \citenamefont
  {Taševský}, \citenamefont {Terao}, \citenamefont {Terashi}, \citenamefont
  {Terning}, \citenamefont {Thorne}, \citenamefont {Titov}, \citenamefont
  {Tkachenko}, \citenamefont {Tovey}, \citenamefont {Trabelsi}, \citenamefont
  {Urquijo}, \citenamefont {Valencia}, \citenamefont {Van~de Water},
  \citenamefont {Varelas}, \citenamefont {Venanzoni}, \citenamefont {Verde},
  \citenamefont {Vivarelli}, \citenamefont {Vogel}, \citenamefont {Vogelsang},
  \citenamefont {Vorobyev}, \citenamefont {Wakely}, \citenamefont {Walkowiak},
  \citenamefont {Walter}, \citenamefont {Wands}, \citenamefont {Weinberg},
  \citenamefont {Weinberg}, \citenamefont {Wermes}, \citenamefont {White},
  \citenamefont {Wiencke}, \citenamefont {Willocq}, \citenamefont {Wohl},
  \citenamefont {Woody}, \citenamefont {Yao}, \citenamefont {Yokoyama},
  \citenamefont {Yoshida}, \citenamefont {Zanderighi}, \citenamefont {Zeller},
  \citenamefont {Zenin}, \citenamefont {Zhu}, \citenamefont {Zhu},
  \citenamefont {Zimmermann},\ and\ \citenamefont
  {Zyla}}]{10.1093/ptep/ptac097}%
  \BibitemOpen
  \bibfield  {author} {\bibinfo {author} {\bibfnamefont {P.~D.}\ \bibnamefont
  {Group}}, \bibinfo {author} {\bibfnamefont {R.~L.}\ \bibnamefont {Workman}},
  \bibinfo {author} {\bibfnamefont {V.~D.}\ \bibnamefont {Burkert}}, \bibinfo
  {author} {\bibfnamefont {V.}~\bibnamefont {Crede}}, \bibinfo {author}
  {\bibfnamefont {E.}~\bibnamefont {Klempt}}, \bibinfo {author} {\bibfnamefont
  {U.}~\bibnamefont {Thoma}}, \bibinfo {author} {\bibfnamefont
  {L.}~\bibnamefont {Tiator}}, \bibinfo {author} {\bibfnamefont
  {K.}~\bibnamefont {Agashe}}, \bibinfo {author} {\bibfnamefont
  {G.}~\bibnamefont {Aielli}}, \bibinfo {author} {\bibfnamefont {B.~C.}\
  \bibnamefont {Allanach}}, \bibinfo {author} {\bibfnamefont {C.}~\bibnamefont
  {Amsler}}, \bibinfo {author} {\bibfnamefont {M.}~\bibnamefont {Antonelli}},
  \bibinfo {author} {\bibfnamefont {E.~C.}\ \bibnamefont {Aschenauer}},
  \bibinfo {author} {\bibfnamefont {D.~M.}\ \bibnamefont {Asner}}, \bibinfo
  {author} {\bibfnamefont {H.}~\bibnamefont {Baer}}, \bibinfo {author}
  {\bibfnamefont {S.}~\bibnamefont {Banerjee}}, \bibinfo {author}
  {\bibfnamefont {R.~M.}\ \bibnamefont {Barnett}}, \bibinfo {author}
  {\bibfnamefont {L.}~\bibnamefont {Baudis}}, \bibinfo {author} {\bibfnamefont
  {C.~W.}\ \bibnamefont {Bauer}}, \bibinfo {author} {\bibfnamefont {J.~J.}\
  \bibnamefont {Beatty}}, \bibinfo {author} {\bibfnamefont {V.~I.}\
  \bibnamefont {Belousov}}, \bibinfo {author} {\bibfnamefont {J.}~\bibnamefont
  {Beringer}}, \bibinfo {author} {\bibfnamefont {A.}~\bibnamefont {Bettini}},
  \bibinfo {author} {\bibfnamefont {O.}~\bibnamefont {Biebel}}, \bibinfo
  {author} {\bibfnamefont {K.~M.}\ \bibnamefont {Black}}, \bibinfo {author}
  {\bibfnamefont {E.}~\bibnamefont {Blucher}}, \bibinfo {author} {\bibfnamefont
  {R.}~\bibnamefont {Bonventre}}, \bibinfo {author} {\bibfnamefont {V.~V.}\
  \bibnamefont {Bryzgalov}}, \bibinfo {author} {\bibfnamefont {O.}~\bibnamefont
  {Buchmuller}}, \bibinfo {author} {\bibfnamefont {M.~A.}\ \bibnamefont
  {Bychkov}}, \bibinfo {author} {\bibfnamefont {R.~N.}\ \bibnamefont {Cahn}},
  \bibinfo {author} {\bibfnamefont {M.}~\bibnamefont {Carena}}, \bibinfo
  {author} {\bibfnamefont {A.}~\bibnamefont {Ceccucci}}, \bibinfo {author}
  {\bibfnamefont {A.}~\bibnamefont {Cerri}}, \bibinfo {author} {\bibfnamefont
  {R.~S.}\ \bibnamefont {Chivukula}}, \bibinfo {author} {\bibfnamefont
  {G.}~\bibnamefont {Cowan}}, \bibinfo {author} {\bibfnamefont
  {K.}~\bibnamefont {Cranmer}}, \bibinfo {author} {\bibfnamefont
  {O.}~\bibnamefont {Cremonesi}}, \bibinfo {author} {\bibfnamefont
  {G.}~\bibnamefont {D'Ambrosio}}, \bibinfo {author} {\bibfnamefont
  {T.}~\bibnamefont {Damour}}, \bibinfo {author} {\bibfnamefont
  {D.}~\bibnamefont {de~Florian}}, \bibinfo {author} {\bibfnamefont
  {A.}~\bibnamefont {de~Gouvêa}}, \bibinfo {author} {\bibfnamefont
  {T.}~\bibnamefont {DeGrand}}, \bibinfo {author} {\bibfnamefont
  {P.}~\bibnamefont {de~Jong}}, \bibinfo {author} {\bibfnamefont
  {S.}~\bibnamefont {Demers}}, \bibinfo {author} {\bibfnamefont {B.~A.}\
  \bibnamefont {Dobrescu}}, \bibinfo {author} {\bibfnamefont {M.}~\bibnamefont
  {D'Onofrio}}, \bibinfo {author} {\bibfnamefont {M.}~\bibnamefont {Doser}},
  \bibinfo {author} {\bibfnamefont {H.~K.}\ \bibnamefont {Dreiner}}, \bibinfo
  {author} {\bibfnamefont {P.}~\bibnamefont {Eerola}}, \bibinfo {author}
  {\bibfnamefont {U.}~\bibnamefont {Egede}}, \bibinfo {author} {\bibfnamefont
  {S.}~\bibnamefont {Eidelman}}, \bibinfo {author} {\bibfnamefont {A.~X.}\
  \bibnamefont {El-Khadra}}, \bibinfo {author} {\bibfnamefont {J.}~\bibnamefont
  {Ellis}}, \bibinfo {author} {\bibfnamefont {S.~C.}\ \bibnamefont {Eno}},
  \bibinfo {author} {\bibfnamefont {J.}~\bibnamefont {Erler}}, \bibinfo
  {author} {\bibfnamefont {V.~V.}\ \bibnamefont {Ezhela}}, \bibinfo {author}
  {\bibfnamefont {W.}~\bibnamefont {Fetscher}}, \bibinfo {author}
  {\bibfnamefont {B.~D.}\ \bibnamefont {Fields}}, \bibinfo {author}
  {\bibfnamefont {A.}~\bibnamefont {Freitas}}, \bibinfo {author} {\bibfnamefont
  {H.}~\bibnamefont {Gallagher}}, \bibinfo {author} {\bibfnamefont
  {Y.}~\bibnamefont {Gershtein}}, \bibinfo {author} {\bibfnamefont
  {T.}~\bibnamefont {Gherghetta}}, \bibinfo {author} {\bibfnamefont {M.~C.}\
  \bibnamefont {Gonzalez-Garcia}}, \bibinfo {author} {\bibfnamefont
  {M.}~\bibnamefont {Goodman}}, \bibinfo {author} {\bibfnamefont
  {C.}~\bibnamefont {Grab}}, \bibinfo {author} {\bibfnamefont {A.~V.}\
  \bibnamefont {Gritsan}}, \bibinfo {author} {\bibfnamefont {C.}~\bibnamefont
  {Grojean}}, \bibinfo {author} {\bibfnamefont {D.~E.}\ \bibnamefont {Groom}},
  \bibinfo {author} {\bibfnamefont {M.}~\bibnamefont {Grünewald}}, \bibinfo
  {author} {\bibfnamefont {A.}~\bibnamefont {Gurtu}}, \bibinfo {author}
  {\bibfnamefont {T.}~\bibnamefont {Gutsche}}, \bibinfo {author} {\bibfnamefont
  {H.~E.}\ \bibnamefont {Haber}}, \bibinfo {author} {\bibfnamefont
  {M.}~\bibnamefont {Hamel}}, \bibinfo {author} {\bibfnamefont
  {C.}~\bibnamefont {Hanhart}}, \bibinfo {author} {\bibfnamefont
  {S.}~\bibnamefont {Hashimoto}}, \bibinfo {author} {\bibfnamefont
  {Y.}~\bibnamefont {Hayato}}, \bibinfo {author} {\bibfnamefont
  {A.}~\bibnamefont {Hebecker}}, \bibinfo {author} {\bibfnamefont
  {S.}~\bibnamefont {Heinemeyer}}, \bibinfo {author} {\bibfnamefont {J.~J.}\
  \bibnamefont {Hernández-Rey}}, \bibinfo {author} {\bibfnamefont
  {K.}~\bibnamefont {Hikasa}}, \bibinfo {author} {\bibfnamefont
  {J.}~\bibnamefont {Hisano}}, \bibinfo {author} {\bibfnamefont
  {A.}~\bibnamefont {Höcker}}, \bibinfo {author} {\bibfnamefont
  {J.}~\bibnamefont {Holder}}, \bibinfo {author} {\bibfnamefont
  {L.}~\bibnamefont {Hsu}}, \bibinfo {author} {\bibfnamefont {J.}~\bibnamefont
  {Huston}}, \bibinfo {author} {\bibfnamefont {T.}~\bibnamefont {Hyodo}},
  \bibinfo {author} {\bibfnamefont {A.}~\bibnamefont {Ianni}}, \bibinfo
  {author} {\bibfnamefont {M.}~\bibnamefont {Kado}}, \bibinfo {author}
  {\bibfnamefont {M.}~\bibnamefont {Karliner}}, \bibinfo {author}
  {\bibfnamefont {U.~F.}\ \bibnamefont {Katz}}, \bibinfo {author}
  {\bibfnamefont {M.}~\bibnamefont {Kenzie}}, \bibinfo {author} {\bibfnamefont
  {V.~A.}\ \bibnamefont {Khoze}}, \bibinfo {author} {\bibfnamefont {S.~R.}\
  \bibnamefont {Klein}}, \bibinfo {author} {\bibfnamefont {F.}~\bibnamefont
  {Krauss}}, \bibinfo {author} {\bibfnamefont {M.}~\bibnamefont {Kreps}},
  \bibinfo {author} {\bibfnamefont {P.}~\bibnamefont {Križan}}, \bibinfo
  {author} {\bibfnamefont {B.}~\bibnamefont {Krusche}}, \bibinfo {author}
  {\bibfnamefont {Y.}~\bibnamefont {Kwon}}, \bibinfo {author} {\bibfnamefont
  {O.}~\bibnamefont {Lahav}}, \bibinfo {author} {\bibfnamefont
  {J.}~\bibnamefont {Laiho}}, \bibinfo {author} {\bibfnamefont {L.~P.}\
  \bibnamefont {Lellouch}}, \bibinfo {author} {\bibfnamefont {J.}~\bibnamefont
  {Lesgourgues}}, \bibinfo {author} {\bibfnamefont {A.~R.}\ \bibnamefont
  {Liddle}}, \bibinfo {author} {\bibfnamefont {Z.}~\bibnamefont {Ligeti}},
  \bibinfo {author} {\bibfnamefont {C.-J.}\ \bibnamefont {Lin}}, \bibinfo
  {author} {\bibfnamefont {C.}~\bibnamefont {Lippmann}}, \bibinfo {author}
  {\bibfnamefont {T.~M.}\ \bibnamefont {Liss}}, \bibinfo {author}
  {\bibfnamefont {L.}~\bibnamefont {Littenberg}}, \bibinfo {author}
  {\bibfnamefont {C.}~\bibnamefont {Lourenço}}, \bibinfo {author}
  {\bibfnamefont {K.~S.}\ \bibnamefont {Lugovsky}}, \bibinfo {author}
  {\bibfnamefont {S.~B.}\ \bibnamefont {Lugovsky}}, \bibinfo {author}
  {\bibfnamefont {A.}~\bibnamefont {Lusiani}}, \bibinfo {author} {\bibfnamefont
  {Y.}~\bibnamefont {Makida}}, \bibinfo {author} {\bibfnamefont
  {F.}~\bibnamefont {Maltoni}}, \bibinfo {author} {\bibfnamefont
  {T.}~\bibnamefont {Mannel}}, \bibinfo {author} {\bibfnamefont {A.~V.}\
  \bibnamefont {Manohar}}, \bibinfo {author} {\bibfnamefont {W.~J.}\
  \bibnamefont {Marciano}}, \bibinfo {author} {\bibfnamefont {A.}~\bibnamefont
  {Masoni}}, \bibinfo {author} {\bibfnamefont {J.}~\bibnamefont {Matthews}},
  \bibinfo {author} {\bibfnamefont {U.-G.}\ \bibnamefont {Meißner}}, \bibinfo
  {author} {\bibfnamefont {I.-A.}\ \bibnamefont {Melzer-Pellmann}}, \bibinfo
  {author} {\bibfnamefont {M.}~\bibnamefont {Mikhasenko}}, \bibinfo {author}
  {\bibfnamefont {D.~J.}\ \bibnamefont {Miller}}, \bibinfo {author}
  {\bibfnamefont {D.}~\bibnamefont {Milstead}}, \bibinfo {author}
  {\bibfnamefont {R.~E.}\ \bibnamefont {Mitchell}}, \bibinfo {author}
  {\bibfnamefont {K.}~\bibnamefont {Mönig}}, \bibinfo {author} {\bibfnamefont
  {P.}~\bibnamefont {Molaro}}, \bibinfo {author} {\bibfnamefont
  {F.}~\bibnamefont {Moortgat}}, \bibinfo {author} {\bibfnamefont
  {M.}~\bibnamefont {Moskovic}}, \bibinfo {author} {\bibfnamefont
  {K.}~\bibnamefont {Nakamura}}, \bibinfo {author} {\bibfnamefont
  {M.}~\bibnamefont {Narain}}, \bibinfo {author} {\bibfnamefont
  {P.}~\bibnamefont {Nason}}, \bibinfo {author} {\bibfnamefont
  {S.}~\bibnamefont {Navas}}, \bibinfo {author} {\bibfnamefont
  {A.}~\bibnamefont {Nelles}}, \bibinfo {author} {\bibfnamefont
  {M.}~\bibnamefont {Neubert}}, \bibinfo {author} {\bibfnamefont
  {P.}~\bibnamefont {Nevski}}, \bibinfo {author} {\bibfnamefont
  {Y.}~\bibnamefont {Nir}}, \bibinfo {author} {\bibfnamefont {K.~A.}\
  \bibnamefont {Olive}}, \bibinfo {author} {\bibfnamefont {C.}~\bibnamefont
  {Patrignani}}, \bibinfo {author} {\bibfnamefont {J.~A.}\ \bibnamefont
  {Peacock}}, \bibinfo {author} {\bibfnamefont {V.~A.}\ \bibnamefont {Petrov}},
  \bibinfo {author} {\bibfnamefont {E.}~\bibnamefont {Pianori}}, \bibinfo
  {author} {\bibfnamefont {A.}~\bibnamefont {Pich}}, \bibinfo {author}
  {\bibfnamefont {A.}~\bibnamefont {Piepke}}, \bibinfo {author} {\bibfnamefont
  {F.}~\bibnamefont {Pietropaolo}}, \bibinfo {author} {\bibfnamefont
  {A.}~\bibnamefont {Pomarol}}, \bibinfo {author} {\bibfnamefont
  {S.}~\bibnamefont {Pordes}}, \bibinfo {author} {\bibfnamefont
  {S.}~\bibnamefont {Profumo}}, \bibinfo {author} {\bibfnamefont
  {A.}~\bibnamefont {Quadt}}, \bibinfo {author} {\bibfnamefont
  {K.}~\bibnamefont {Rabbertz}}, \bibinfo {author} {\bibfnamefont
  {J.}~\bibnamefont {Rademacker}}, \bibinfo {author} {\bibfnamefont
  {G.}~\bibnamefont {Raffelt}}, \bibinfo {author} {\bibfnamefont
  {M.}~\bibnamefont {Ramsey-Musolf}}, \bibinfo {author} {\bibfnamefont {B.~N.}\
  \bibnamefont {Ratcliff}}, \bibinfo {author} {\bibfnamefont {P.}~\bibnamefont
  {Richardson}}, \bibinfo {author} {\bibfnamefont {A.}~\bibnamefont
  {Ringwald}}, \bibinfo {author} {\bibfnamefont {D.~J.}\ \bibnamefont
  {Robinson}}, \bibinfo {author} {\bibfnamefont {S.}~\bibnamefont {Roesler}},
  \bibinfo {author} {\bibfnamefont {S.}~\bibnamefont {Rolli}}, \bibinfo
  {author} {\bibfnamefont {A.}~\bibnamefont {Romaniouk}}, \bibinfo {author}
  {\bibfnamefont {L.~J.}\ \bibnamefont {Rosenberg}}, \bibinfo {author}
  {\bibfnamefont {J.~L.}\ \bibnamefont {Rosner}}, \bibinfo {author}
  {\bibfnamefont {G.}~\bibnamefont {Rybka}}, \bibinfo {author} {\bibfnamefont
  {M.~G.}\ \bibnamefont {Ryskin}}, \bibinfo {author} {\bibfnamefont {R.~A.}\
  \bibnamefont {Ryutin}}, \bibinfo {author} {\bibfnamefont {Y.}~\bibnamefont
  {Sakai}}, \bibinfo {author} {\bibfnamefont {S.}~\bibnamefont {Sarkar}},
  \bibinfo {author} {\bibfnamefont {F.}~\bibnamefont {Sauli}}, \bibinfo
  {author} {\bibfnamefont {O.}~\bibnamefont {Schneider}}, \bibinfo {author}
  {\bibfnamefont {S.}~\bibnamefont {Schönert}}, \bibinfo {author}
  {\bibfnamefont {K.}~\bibnamefont {Scholberg}}, \bibinfo {author}
  {\bibfnamefont {A.~J.}\ \bibnamefont {Schwartz}}, \bibinfo {author}
  {\bibfnamefont {J.}~\bibnamefont {Schwiening}}, \bibinfo {author}
  {\bibfnamefont {D.}~\bibnamefont {Scott}}, \bibinfo {author} {\bibfnamefont
  {F.}~\bibnamefont {Sefkow}}, \bibinfo {author} {\bibfnamefont
  {U.}~\bibnamefont {Seljak}}, \bibinfo {author} {\bibfnamefont
  {V.}~\bibnamefont {Sharma}}, \bibinfo {author} {\bibfnamefont {S.~R.}\
  \bibnamefont {Sharpe}}, \bibinfo {author} {\bibfnamefont {V.}~\bibnamefont
  {Shiltsev}}, \bibinfo {author} {\bibfnamefont {G.}~\bibnamefont
  {Signorelli}}, \bibinfo {author} {\bibfnamefont {M.}~\bibnamefont {Silari}},
  \bibinfo {author} {\bibfnamefont {F.}~\bibnamefont {Simon}}, \bibinfo
  {author} {\bibfnamefont {T.}~\bibnamefont {Sjöstrand}}, \bibinfo {author}
  {\bibfnamefont {P.}~\bibnamefont {Skands}}, \bibinfo {author} {\bibfnamefont
  {T.}~\bibnamefont {Skwarnicki}}, \bibinfo {author} {\bibfnamefont {G.~F.}\
  \bibnamefont {Smoot}}, \bibinfo {author} {\bibfnamefont {A.}~\bibnamefont
  {Soffer}}, \bibinfo {author} {\bibfnamefont {M.~S.}\ \bibnamefont {Sozzi}},
  \bibinfo {author} {\bibfnamefont {S.}~\bibnamefont {Spanier}}, \bibinfo
  {author} {\bibfnamefont {C.}~\bibnamefont {Spiering}}, \bibinfo {author}
  {\bibfnamefont {A.}~\bibnamefont {Stahl}}, \bibinfo {author} {\bibfnamefont
  {S.~L.}\ \bibnamefont {Stone}}, \bibinfo {author} {\bibfnamefont
  {Y.}~\bibnamefont {Sumino}}, \bibinfo {author} {\bibfnamefont {M.~J.}\
  \bibnamefont {Syphers}}, \bibinfo {author} {\bibfnamefont {F.}~\bibnamefont
  {Takahashi}}, \bibinfo {author} {\bibfnamefont {M.}~\bibnamefont
  {Tanabashi}}, \bibinfo {author} {\bibfnamefont {J.}~\bibnamefont {Tanaka}},
  \bibinfo {author} {\bibfnamefont {M.}~\bibnamefont {Taševský}}, \bibinfo
  {author} {\bibfnamefont {K.}~\bibnamefont {Terao}}, \bibinfo {author}
  {\bibfnamefont {K.}~\bibnamefont {Terashi}}, \bibinfo {author} {\bibfnamefont
  {J.}~\bibnamefont {Terning}}, \bibinfo {author} {\bibfnamefont {R.~S.}\
  \bibnamefont {Thorne}}, \bibinfo {author} {\bibfnamefont {M.}~\bibnamefont
  {Titov}}, \bibinfo {author} {\bibfnamefont {N.~P.}\ \bibnamefont
  {Tkachenko}}, \bibinfo {author} {\bibfnamefont {D.~R.}\ \bibnamefont
  {Tovey}}, \bibinfo {author} {\bibfnamefont {K.}~\bibnamefont {Trabelsi}},
  \bibinfo {author} {\bibfnamefont {P.}~\bibnamefont {Urquijo}}, \bibinfo
  {author} {\bibfnamefont {G.}~\bibnamefont {Valencia}}, \bibinfo {author}
  {\bibfnamefont {R.}~\bibnamefont {Van~de Water}}, \bibinfo {author}
  {\bibfnamefont {N.}~\bibnamefont {Varelas}}, \bibinfo {author} {\bibfnamefont
  {G.}~\bibnamefont {Venanzoni}}, \bibinfo {author} {\bibfnamefont
  {L.}~\bibnamefont {Verde}}, \bibinfo {author} {\bibfnamefont
  {I.}~\bibnamefont {Vivarelli}}, \bibinfo {author} {\bibfnamefont
  {P.}~\bibnamefont {Vogel}}, \bibinfo {author} {\bibfnamefont
  {W.}~\bibnamefont {Vogelsang}}, \bibinfo {author} {\bibfnamefont
  {V.}~\bibnamefont {Vorobyev}}, \bibinfo {author} {\bibfnamefont {S.~P.}\
  \bibnamefont {Wakely}}, \bibinfo {author} {\bibfnamefont {W.}~\bibnamefont
  {Walkowiak}}, \bibinfo {author} {\bibfnamefont {C.~W.}\ \bibnamefont
  {Walter}}, \bibinfo {author} {\bibfnamefont {D.}~\bibnamefont {Wands}},
  \bibinfo {author} {\bibfnamefont {D.~H.}\ \bibnamefont {Weinberg}}, \bibinfo
  {author} {\bibfnamefont {E.~J.}\ \bibnamefont {Weinberg}}, \bibinfo {author}
  {\bibfnamefont {N.}~\bibnamefont {Wermes}}, \bibinfo {author} {\bibfnamefont
  {M.}~\bibnamefont {White}}, \bibinfo {author} {\bibfnamefont {L.~R.}\
  \bibnamefont {Wiencke}}, \bibinfo {author} {\bibfnamefont {S.}~\bibnamefont
  {Willocq}}, \bibinfo {author} {\bibfnamefont {C.~G.}\ \bibnamefont {Wohl}},
  \bibinfo {author} {\bibfnamefont {C.~L.}\ \bibnamefont {Woody}}, \bibinfo
  {author} {\bibfnamefont {W.-M.}\ \bibnamefont {Yao}}, \bibinfo {author}
  {\bibfnamefont {M.}~\bibnamefont {Yokoyama}}, \bibinfo {author}
  {\bibfnamefont {R.}~\bibnamefont {Yoshida}}, \bibinfo {author} {\bibfnamefont
  {G.}~\bibnamefont {Zanderighi}}, \bibinfo {author} {\bibfnamefont {G.~P.}\
  \bibnamefont {Zeller}}, \bibinfo {author} {\bibfnamefont {O.~V.}\
  \bibnamefont {Zenin}}, \bibinfo {author} {\bibfnamefont {R.-Y.}\ \bibnamefont
  {Zhu}}, \bibinfo {author} {\bibfnamefont {S.-L.}\ \bibnamefont {Zhu}},
  \bibinfo {author} {\bibfnamefont {F.}~\bibnamefont {Zimmermann}}, \ and\
  \bibinfo {author} {\bibfnamefont {P.~A.}\ \bibnamefont {Zyla}},\ }\href
  {\doibase 10.1093/ptep/ptac097} {\bibfield  {journal} {\bibinfo  {journal}
  {Progress of Theoretical and Experimental Physics}\ }\textbf {\bibinfo
  {volume} {2022}},\ \bibinfo {pages} {083C01} (\bibinfo {year} {2022})},\
  \Eprint
  {http://arxiv.org/abs/https://academic.oup.com/ptep/article-pdf/2022/8/083C01/49175539/ptac097.pdf}
  {https://academic.oup.com/ptep/article-pdf/2022/8/083C01/49175539/ptac097.pdf}
  \BibitemShut {NoStop}%
\bibitem [{\citenamefont {Lester}\ \emph {et~al.}(2007)\citenamefont {Lester},
  \citenamefont {Parker},\ and\ \citenamefont {White}}]{Lester:2006cf}%
  \BibitemOpen
  \bibfield  {author} {\bibinfo {author} {\bibfnamefont {C.~G.}\ \bibnamefont
  {Lester}}, \bibinfo {author} {\bibfnamefont {M.~A.}\ \bibnamefont {Parker}},
  \ and\ \bibinfo {author} {\bibfnamefont {M.~J.}\ \bibnamefont {White}},\
  }\href {\doibase 10.1088/1126-6708/2007/10/051} {\bibfield  {journal}
  {\bibinfo  {journal} {JHEP}\ }\textbf {\bibinfo {volume} {10}},\ \bibinfo
  {pages} {051} (\bibinfo {year} {2007})},\ \Eprint
  {http://arxiv.org/abs/hep-ph/0609298} {arXiv:hep-ph/0609298} \BibitemShut
  {NoStop}%
\bibitem [{\citenamefont {Franceschini}\ \emph {et~al.}(2023)\citenamefont
  {Franceschini}, \citenamefont {Kim}, \citenamefont {Kong}, \citenamefont
  {Matchev}, \citenamefont {Park},\ and\ \citenamefont
  {Shyamsundar}}]{Franceschini:2022vck}%
  \BibitemOpen
  \bibfield  {author} {\bibinfo {author} {\bibfnamefont {R.}~\bibnamefont
  {Franceschini}}, \bibinfo {author} {\bibfnamefont {D.}~\bibnamefont {Kim}},
  \bibinfo {author} {\bibfnamefont {K.}~\bibnamefont {Kong}}, \bibinfo {author}
  {\bibfnamefont {K.~T.}\ \bibnamefont {Matchev}}, \bibinfo {author}
  {\bibfnamefont {M.}~\bibnamefont {Park}}, \ and\ \bibinfo {author}
  {\bibfnamefont {P.}~\bibnamefont {Shyamsundar}},\ }\href {\doibase
  10.1103/RevModPhys.95.045004} {\bibfield  {journal} {\bibinfo  {journal}
  {Rev. Mod. Phys.}\ }\textbf {\bibinfo {volume} {95}},\ \bibinfo {pages}
  {045004} (\bibinfo {year} {2023})},\ \Eprint
  {http://arxiv.org/abs/2206.13431} {arXiv:2206.13431 [hep-ph]} \BibitemShut
  {NoStop}%
\bibitem [{\citenamefont {Caldwell}\ \emph {et~al.}(2022)\citenamefont
  {Caldwell} \emph {et~al.}}]{Caldwell:2022qsj}%
  \BibitemOpen
  \bibfield  {author} {\bibinfo {author} {\bibfnamefont {R.}~\bibnamefont
  {Caldwell}} \emph {et~al.},\ }\href {\doibase 10.1007/s10714-022-03027-x}
  {\bibfield  {journal} {\bibinfo  {journal} {Gen. Rel. Grav.}\ }\textbf
  {\bibinfo {volume} {54}},\ \bibinfo {pages} {156} (\bibinfo {year} {2022})},\
  \Eprint {http://arxiv.org/abs/2203.07972} {arXiv:2203.07972 [gr-qc]}
  \BibitemShut {NoStop}%
\bibitem [{\citenamefont {Cepeda}\ \emph {et~al.}(2019)\citenamefont {Cepeda}
  \emph {et~al.}}]{Cepeda:2019klc}%
  \BibitemOpen
  \bibfield  {author} {\bibinfo {author} {\bibfnamefont {M.}~\bibnamefont
  {Cepeda}} \emph {et~al.},\ }\href {\doibase 10.23731/CYRM-2019-007.221}
  {\bibfield  {journal} {\bibinfo  {journal} {CERN Yellow Rep. Monogr.}\
  }\textbf {\bibinfo {volume} {7}},\ \bibinfo {pages} {221} (\bibinfo {year}
  {2019})},\ \Eprint {http://arxiv.org/abs/1902.00134} {arXiv:1902.00134
  [hep-ph]} \BibitemShut {NoStop}%
\bibitem [{\citenamefont {Aad}\ \emph {et~al.}(2023)\citenamefont {Aad} \emph
  {et~al.}}]{ATLAS:2022jtk}%
  \BibitemOpen
  \bibfield  {author} {\bibinfo {author} {\bibfnamefont {G.}~\bibnamefont
  {Aad}} \emph {et~al.} (\bibinfo {collaboration} {ATLAS}),\ }\href {\doibase
  10.1016/j.physletb.2023.137745} {\bibfield  {journal} {\bibinfo  {journal}
  {Phys. Lett. B}\ }\textbf {\bibinfo {volume} {843}},\ \bibinfo {pages}
  {137745} (\bibinfo {year} {2023})},\ \Eprint
  {http://arxiv.org/abs/2211.01216} {arXiv:2211.01216 [hep-ex]} \BibitemShut
  {NoStop}%
\bibitem [{\citenamefont {Tumasyan}\ \emph {et~al.}(2022)\citenamefont
  {Tumasyan} \emph {et~al.}}]{CMS:2022dwd}%
  \BibitemOpen
  \bibfield  {author} {\bibinfo {author} {\bibfnamefont {A.}~\bibnamefont
  {Tumasyan}} \emph {et~al.} (\bibinfo {collaboration} {CMS}),\ }\href
  {\doibase 10.1038/s41586-022-04892-x} {\bibfield  {journal} {\bibinfo
  {journal} {Nature}\ }\textbf {\bibinfo {volume} {607}},\ \bibinfo {pages}
  {60} (\bibinfo {year} {2022})},\ \Eprint {http://arxiv.org/abs/2207.00043}
  {arXiv:2207.00043 [hep-ex]} \BibitemShut {NoStop}%
\bibitem [{\citenamefont
  {Collaboration}(2023)}]{atlascollaboration2023studies}%
  \BibitemOpen
  \bibfield  {author} {\bibinfo {author} {\bibfnamefont {A.}~\bibnamefont
  {Collaboration}},\ }\href@noop {} {\enquote {\bibinfo {title} {Studies of new
  higgs boson interactions through nonresonant $hh$ production in the
  $b\bar{b}\gamma\gamma$ final state in $pp$ collisions at $\sqrt{s}=13$ tev
  with the atlas detector},}\ } (\bibinfo {year} {2023}),\ \Eprint
  {http://arxiv.org/abs/2310.12301} {arXiv:2310.12301 [hep-ex]} \BibitemShut
  {NoStop}%
\bibitem [{\citenamefont {Kim}\ \emph {et~al.}(2019)\citenamefont {Kim},
  \citenamefont {Kong}, \citenamefont {Matchev},\ and\ \citenamefont
  {Park}}]{Kim:2018cxf}%
  \BibitemOpen
  \bibfield  {author} {\bibinfo {author} {\bibfnamefont {J.~H.}\ \bibnamefont
  {Kim}}, \bibinfo {author} {\bibfnamefont {K.}~\bibnamefont {Kong}}, \bibinfo
  {author} {\bibfnamefont {K.~T.}\ \bibnamefont {Matchev}}, \ and\ \bibinfo
  {author} {\bibfnamefont {M.}~\bibnamefont {Park}},\ }\href {\doibase
  10.1103/PhysRevLett.122.091801} {\bibfield  {journal} {\bibinfo  {journal}
  {Phys. Rev. Lett.}\ }\textbf {\bibinfo {volume} {122}},\ \bibinfo {pages}
  {091801} (\bibinfo {year} {2019})},\ \Eprint
  {http://arxiv.org/abs/1807.11498} {arXiv:1807.11498 [hep-ph]} \BibitemShut
  {NoStop}%
\bibitem [{\citenamefont {Roloff}\ \emph {et~al.}(2020)\citenamefont {Roloff},
  \citenamefont {Schnoor}, \citenamefont {Simoniello},\ and\ \citenamefont
  {Xu}}]{Roloff:2019crr}%
  \BibitemOpen
  \bibfield  {author} {\bibinfo {author} {\bibfnamefont {P.}~\bibnamefont
  {Roloff}}, \bibinfo {author} {\bibfnamefont {U.}~\bibnamefont {Schnoor}},
  \bibinfo {author} {\bibfnamefont {R.}~\bibnamefont {Simoniello}}, \ and\
  \bibinfo {author} {\bibfnamefont {B.}~\bibnamefont {Xu}} (\bibinfo
  {collaboration} {CLICdp}),\ }\href {\doibase 10.1140/epjc/s10052-020-08567-7}
  {\bibfield  {journal} {\bibinfo  {journal} {Eur. Phys. J. C}\ }\textbf
  {\bibinfo {volume} {80}},\ \bibinfo {pages} {1010} (\bibinfo {year}
  {2020})},\ \Eprint {http://arxiv.org/abs/1901.05897} {arXiv:1901.05897
  [hep-ex]} \BibitemShut {NoStop}%
\bibitem [{\citenamefont {Contino}\ \emph {et~al.}(2014)\citenamefont
  {Contino}, \citenamefont {Grojean}, \citenamefont {Pappadopulo},
  \citenamefont {Rattazzi},\ and\ \citenamefont {Thamm}}]{Contino:2013gna}%
  \BibitemOpen
  \bibfield  {author} {\bibinfo {author} {\bibfnamefont {R.}~\bibnamefont
  {Contino}}, \bibinfo {author} {\bibfnamefont {C.}~\bibnamefont {Grojean}},
  \bibinfo {author} {\bibfnamefont {D.}~\bibnamefont {Pappadopulo}}, \bibinfo
  {author} {\bibfnamefont {R.}~\bibnamefont {Rattazzi}}, \ and\ \bibinfo
  {author} {\bibfnamefont {A.}~\bibnamefont {Thamm}},\ }\href {\doibase
  10.1007/JHEP02(2014)006} {\bibfield  {journal} {\bibinfo  {journal} {JHEP}\
  }\textbf {\bibinfo {volume} {02}},\ \bibinfo {pages} {006} (\bibinfo {year}
  {2014})},\ \Eprint {http://arxiv.org/abs/1309.7038} {arXiv:1309.7038
  [hep-ph]} \BibitemShut {NoStop}%
\bibitem [{\citenamefont {Zhao}\ \emph {et~al.}(2019)\citenamefont {Zhao},
  \citenamefont {Adeli}, \citenamefont {Honnorat}, \citenamefont {Leng},\ and\
  \citenamefont {Pohl}}]{DBLP:journals/corr/abs-1904-05948}%
  \BibitemOpen
  \bibfield  {author} {\bibinfo {author} {\bibfnamefont {Q.}~\bibnamefont
  {Zhao}}, \bibinfo {author} {\bibfnamefont {E.}~\bibnamefont {Adeli}},
  \bibinfo {author} {\bibfnamefont {N.}~\bibnamefont {Honnorat}}, \bibinfo
  {author} {\bibfnamefont {T.}~\bibnamefont {Leng}}, \ and\ \bibinfo {author}
  {\bibfnamefont {K.~M.}\ \bibnamefont {Pohl}},\ }\href
  {http://arxiv.org/abs/1904.05948} {\bibfield  {journal} {\bibinfo  {journal}
  {CoRR}\ }\textbf {\bibinfo {volume} {abs/1904.05948}} (\bibinfo {year}
  {2019})},\ \Eprint {http://arxiv.org/abs/1904.05948} {1904.05948}
  \BibitemShut {NoStop}%
\bibitem [{\citenamefont {Kingma}\ and\ \citenamefont
  {Welling}(2019)}]{Kingma_2019}%
  \BibitemOpen
  \bibfield  {author} {\bibinfo {author} {\bibfnamefont {D.~P.}\ \bibnamefont
  {Kingma}}\ and\ \bibinfo {author} {\bibfnamefont {M.}~\bibnamefont
  {Welling}},\ }\href {\doibase 10.1561/2200000056} {\bibfield  {journal}
  {\bibinfo  {journal} {Foundations and Trends® in Machine Learning}\ }\textbf
  {\bibinfo {volume} {12}},\ \bibinfo {pages} {307–392} (\bibinfo {year}
  {2019})}\BibitemShut {NoStop}%
\bibitem [{\citenamefont {Csiszar}(1975)}]{KL}%
  \BibitemOpen
  \bibfield  {author} {\bibinfo {author} {\bibfnamefont {I.}~\bibnamefont
  {Csiszar}},\ }\href {\doibase 10.1214/aop/1176996454} {\bibfield  {journal}
  {\bibinfo  {journal} {The Annals of Probability}\ }\textbf {\bibinfo {volume}
  {3}},\ \bibinfo {pages} {146 } (\bibinfo {year} {1975})}\BibitemShut
  {NoStop}%
\bibitem [{\citenamefont {Alwall}\ \emph {et~al.}(2014)\citenamefont {Alwall},
  \citenamefont {Frederix}, \citenamefont {Frixione}, \citenamefont {Hirschi},
  \citenamefont {Maltoni}, \citenamefont {Mattelaer}, \citenamefont {Shao},
  \citenamefont {Stelzer}, \citenamefont {Torrielli},\ and\ \citenamefont
  {Zaro}}]{Alwall:2014hca}%
  \BibitemOpen
  \bibfield  {author} {\bibinfo {author} {\bibfnamefont {J.}~\bibnamefont
  {Alwall}}, \bibinfo {author} {\bibfnamefont {R.}~\bibnamefont {Frederix}},
  \bibinfo {author} {\bibfnamefont {S.}~\bibnamefont {Frixione}}, \bibinfo
  {author} {\bibfnamefont {V.}~\bibnamefont {Hirschi}}, \bibinfo {author}
  {\bibfnamefont {F.}~\bibnamefont {Maltoni}}, \bibinfo {author} {\bibfnamefont
  {O.}~\bibnamefont {Mattelaer}}, \bibinfo {author} {\bibfnamefont {H.~S.}\
  \bibnamefont {Shao}}, \bibinfo {author} {\bibfnamefont {T.}~\bibnamefont
  {Stelzer}}, \bibinfo {author} {\bibfnamefont {P.}~\bibnamefont {Torrielli}},
  \ and\ \bibinfo {author} {\bibfnamefont {M.}~\bibnamefont {Zaro}},\ }\href
  {\doibase 10.1007/JHEP07(2014)079} {\bibfield  {journal} {\bibinfo  {journal}
  {JHEP}\ }\textbf {\bibinfo {volume} {07}},\ \bibinfo {pages} {079} (\bibinfo
  {year} {2014})},\ \Eprint {http://arxiv.org/abs/1405.0301} {arXiv:1405.0301
  [hep-ph]} \BibitemShut {NoStop}%
\bibitem [{\citenamefont {Sjostrand}\ \emph {et~al.}(2008)\citenamefont
  {Sjostrand}, \citenamefont {Mrenna},\ and\ \citenamefont
  {Skands}}]{Sjostrand:2007gs}%
  \BibitemOpen
  \bibfield  {author} {\bibinfo {author} {\bibfnamefont {T.}~\bibnamefont
  {Sjostrand}}, \bibinfo {author} {\bibfnamefont {S.}~\bibnamefont {Mrenna}}, \
  and\ \bibinfo {author} {\bibfnamefont {P.~Z.}\ \bibnamefont {Skands}},\
  }\href {\doibase 10.1016/j.cpc.2008.01.036} {\bibfield  {journal} {\bibinfo
  {journal} {Comput. Phys. Commun.}\ }\textbf {\bibinfo {volume} {178}},\
  \bibinfo {pages} {852} (\bibinfo {year} {2008})},\ \Eprint
  {http://arxiv.org/abs/0710.3820} {arXiv:0710.3820 [hep-ph]} \BibitemShut
  {NoStop}%
\bibitem [{\citenamefont {Cacciari}\ \emph {et~al.}(2012)\citenamefont
  {Cacciari}, \citenamefont {Salam},\ and\ \citenamefont
  {Soyez}}]{Cacciari:2011ma}%
  \BibitemOpen
  \bibfield  {author} {\bibinfo {author} {\bibfnamefont {M.}~\bibnamefont
  {Cacciari}}, \bibinfo {author} {\bibfnamefont {G.~P.}\ \bibnamefont {Salam}},
  \ and\ \bibinfo {author} {\bibfnamefont {G.}~\bibnamefont {Soyez}},\ }\href
  {\doibase 10.1140/epjc/s10052-012-1896-2} {\bibfield  {journal} {\bibinfo
  {journal} {Eur. Phys. J. C}\ }\textbf {\bibinfo {volume} {72}},\ \bibinfo
  {pages} {1896} (\bibinfo {year} {2012})},\ \Eprint
  {http://arxiv.org/abs/1111.6097} {arXiv:1111.6097 [hep-ph]} \BibitemShut
  {NoStop}%
\bibitem [{\citenamefont {Mangano}\ \emph {et~al.}(2007)\citenamefont
  {Mangano}, \citenamefont {Moretti}, \citenamefont {Piccinini},\ and\
  \citenamefont {Treccani}}]{Mangano:2006rw}%
  \BibitemOpen
  \bibfield  {author} {\bibinfo {author} {\bibfnamefont {M.~L.}\ \bibnamefont
  {Mangano}}, \bibinfo {author} {\bibfnamefont {M.}~\bibnamefont {Moretti}},
  \bibinfo {author} {\bibfnamefont {F.}~\bibnamefont {Piccinini}}, \ and\
  \bibinfo {author} {\bibfnamefont {M.}~\bibnamefont {Treccani}},\ }\href
  {\doibase 10.1088/1126-6708/2007/01/013} {\bibfield  {journal} {\bibinfo
  {journal} {JHEP}\ }\textbf {\bibinfo {volume} {01}},\ \bibinfo {pages} {013}
  (\bibinfo {year} {2007})},\ \Eprint {http://arxiv.org/abs/hep-ph/0611129}
  {arXiv:hep-ph/0611129} \BibitemShut {NoStop}%
\bibitem [{\citenamefont {Barr}(2006)}]{Barr:2005dz}%
  \BibitemOpen
  \bibfield  {author} {\bibinfo {author} {\bibfnamefont {A.~J.}\ \bibnamefont
  {Barr}},\ }\href {\doibase 10.1088/1126-6708/2006/02/042} {\bibfield
  {journal} {\bibinfo  {journal} {JHEP}\ }\textbf {\bibinfo {volume} {02}},\
  \bibinfo {pages} {042} (\bibinfo {year} {2006})},\ \Eprint
  {http://arxiv.org/abs/hep-ph/0511115} {arXiv:hep-ph/0511115} \BibitemShut
  {NoStop}%
\bibitem [{\citenamefont {Virtanen}\ \emph {et~al.}(2020)\citenamefont
  {Virtanen}, \citenamefont {Gommers}, \citenamefont {Oliphant}, \citenamefont
  {Haberland}, \citenamefont {Reddy}, \citenamefont {Cournapeau}, \citenamefont
  {Burovski}, \citenamefont {Peterson}, \citenamefont {Weckesser},
  \citenamefont {Bright}, \citenamefont {{van der Walt}}, \citenamefont
  {Brett}, \citenamefont {Wilson}, \citenamefont {Millman}, \citenamefont
  {Mayorov}, \citenamefont {Nelson}, \citenamefont {Jones}, \citenamefont
  {Kern}, \citenamefont {Larson}, \citenamefont {Carey}, \citenamefont {Polat},
  \citenamefont {Feng}, \citenamefont {Moore}, \citenamefont {{VanderPlas}},
  \citenamefont {Laxalde}, \citenamefont {Perktold}, \citenamefont {Cimrman},
  \citenamefont {Henriksen}, \citenamefont {Quintero}, \citenamefont {Harris},
  \citenamefont {Archibald}, \citenamefont {Ribeiro}, \citenamefont
  {Pedregosa}, \citenamefont {{van Mulbregt}},\ and\ \citenamefont {{SciPy 1.0
  Contributors}}}]{2020SciPy-NMeth}%
  \BibitemOpen
  \bibfield  {author} {\bibinfo {author} {\bibfnamefont {P.}~\bibnamefont
  {Virtanen}}, \bibinfo {author} {\bibfnamefont {R.}~\bibnamefont {Gommers}},
  \bibinfo {author} {\bibfnamefont {T.~E.}\ \bibnamefont {Oliphant}}, \bibinfo
  {author} {\bibfnamefont {M.}~\bibnamefont {Haberland}}, \bibinfo {author}
  {\bibfnamefont {T.}~\bibnamefont {Reddy}}, \bibinfo {author} {\bibfnamefont
  {D.}~\bibnamefont {Cournapeau}}, \bibinfo {author} {\bibfnamefont
  {E.}~\bibnamefont {Burovski}}, \bibinfo {author} {\bibfnamefont
  {P.}~\bibnamefont {Peterson}}, \bibinfo {author} {\bibfnamefont
  {W.}~\bibnamefont {Weckesser}}, \bibinfo {author} {\bibfnamefont
  {J.}~\bibnamefont {Bright}}, \bibinfo {author} {\bibfnamefont {S.~J.}\
  \bibnamefont {{van der Walt}}}, \bibinfo {author} {\bibfnamefont
  {M.}~\bibnamefont {Brett}}, \bibinfo {author} {\bibfnamefont
  {J.}~\bibnamefont {Wilson}}, \bibinfo {author} {\bibfnamefont {K.~J.}\
  \bibnamefont {Millman}}, \bibinfo {author} {\bibfnamefont {N.}~\bibnamefont
  {Mayorov}}, \bibinfo {author} {\bibfnamefont {A.~R.~J.}\ \bibnamefont
  {Nelson}}, \bibinfo {author} {\bibfnamefont {E.}~\bibnamefont {Jones}},
  \bibinfo {author} {\bibfnamefont {R.}~\bibnamefont {Kern}}, \bibinfo {author}
  {\bibfnamefont {E.}~\bibnamefont {Larson}}, \bibinfo {author} {\bibfnamefont
  {C.~J.}\ \bibnamefont {Carey}}, \bibinfo {author} {\bibfnamefont
  {{\.I}.}~\bibnamefont {Polat}}, \bibinfo {author} {\bibfnamefont
  {Y.}~\bibnamefont {Feng}}, \bibinfo {author} {\bibfnamefont {E.~W.}\
  \bibnamefont {Moore}}, \bibinfo {author} {\bibfnamefont {J.}~\bibnamefont
  {{VanderPlas}}}, \bibinfo {author} {\bibfnamefont {D.}~\bibnamefont
  {Laxalde}}, \bibinfo {author} {\bibfnamefont {J.}~\bibnamefont {Perktold}},
  \bibinfo {author} {\bibfnamefont {R.}~\bibnamefont {Cimrman}}, \bibinfo
  {author} {\bibfnamefont {I.}~\bibnamefont {Henriksen}}, \bibinfo {author}
  {\bibfnamefont {E.~A.}\ \bibnamefont {Quintero}}, \bibinfo {author}
  {\bibfnamefont {C.~R.}\ \bibnamefont {Harris}}, \bibinfo {author}
  {\bibfnamefont {A.~M.}\ \bibnamefont {Archibald}}, \bibinfo {author}
  {\bibfnamefont {A.~H.}\ \bibnamefont {Ribeiro}}, \bibinfo {author}
  {\bibfnamefont {F.}~\bibnamefont {Pedregosa}}, \bibinfo {author}
  {\bibfnamefont {P.}~\bibnamefont {{van Mulbregt}}}, \ and\ \bibinfo {author}
  {\bibnamefont {{SciPy 1.0 Contributors}}},\ }\href {\doibase
  10.1038/s41592-019-0686-2} {\bibfield  {journal} {\bibinfo  {journal} {Nature
  Methods}\ }\textbf {\bibinfo {volume} {17}},\ \bibinfo {pages} {261}
  (\bibinfo {year} {2020})}\BibitemShut {NoStop}%
\bibitem [{\citenamefont {Profumo}\ \emph {et~al.}(2007)\citenamefont
  {Profumo}, \citenamefont {Ramsey-Musolf},\ and\ \citenamefont
  {Shaughnessy}}]{Profumo:2007wc}%
  \BibitemOpen
  \bibfield  {author} {\bibinfo {author} {\bibfnamefont {S.}~\bibnamefont
  {Profumo}}, \bibinfo {author} {\bibfnamefont {M.~J.}\ \bibnamefont
  {Ramsey-Musolf}}, \ and\ \bibinfo {author} {\bibfnamefont {G.}~\bibnamefont
  {Shaughnessy}},\ }\href {\doibase 10.1088/1126-6708/2007/08/010} {\bibfield
  {journal} {\bibinfo  {journal} {JHEP}\ }\textbf {\bibinfo {volume} {08}},\
  \bibinfo {pages} {010} (\bibinfo {year} {2007})},\ \Eprint
  {http://arxiv.org/abs/0705.2425} {arXiv:0705.2425 [hep-ph]} \BibitemShut
  {NoStop}%
\bibitem [{\citenamefont {Profumo}\ \emph {et~al.}(2015)\citenamefont
  {Profumo}, \citenamefont {Ramsey-Musolf}, \citenamefont {Wainwright},\ and\
  \citenamefont {Winslow}}]{Profumo:2014opa}%
  \BibitemOpen
  \bibfield  {author} {\bibinfo {author} {\bibfnamefont {S.}~\bibnamefont
  {Profumo}}, \bibinfo {author} {\bibfnamefont {M.~J.}\ \bibnamefont
  {Ramsey-Musolf}}, \bibinfo {author} {\bibfnamefont {C.~L.}\ \bibnamefont
  {Wainwright}}, \ and\ \bibinfo {author} {\bibfnamefont {P.}~\bibnamefont
  {Winslow}},\ }\href {\doibase 10.1103/PhysRevD.91.035018} {\bibfield
  {journal} {\bibinfo  {journal} {Phys. Rev. D}\ }\textbf {\bibinfo {volume}
  {91}},\ \bibinfo {pages} {035018} (\bibinfo {year} {2015})},\ \Eprint
  {http://arxiv.org/abs/1407.5342} {arXiv:1407.5342 [hep-ph]} \BibitemShut
  {NoStop}%
\bibitem [{\citenamefont {Baldi}\ \emph {et~al.}(2016)\citenamefont {Baldi},
  \citenamefont {Cranmer}, \citenamefont {Faucett}, \citenamefont {Sadowski},\
  and\ \citenamefont {Whiteson}}]{Baldi:2016fzo}%
  \BibitemOpen
  \bibfield  {author} {\bibinfo {author} {\bibfnamefont {P.}~\bibnamefont
  {Baldi}}, \bibinfo {author} {\bibfnamefont {K.}~\bibnamefont {Cranmer}},
  \bibinfo {author} {\bibfnamefont {T.}~\bibnamefont {Faucett}}, \bibinfo
  {author} {\bibfnamefont {P.}~\bibnamefont {Sadowski}}, \ and\ \bibinfo
  {author} {\bibfnamefont {D.}~\bibnamefont {Whiteson}},\ }\href {\doibase
  10.1140/epjc/s10052-016-4099-4} {\bibfield  {journal} {\bibinfo  {journal}
  {Eur. Phys. J. C}\ }\textbf {\bibinfo {volume} {76}},\ \bibinfo {pages} {235}
  (\bibinfo {year} {2016})},\ \Eprint {http://arxiv.org/abs/1601.07913}
  {arXiv:1601.07913 [hep-ex]} \BibitemShut {NoStop}%
\bibitem [{\citenamefont {Pedregosa}\ \emph {et~al.}(2011)\citenamefont
  {Pedregosa}, \citenamefont {Varoquaux}, \citenamefont {Gramfort},
  \citenamefont {Michel}, \citenamefont {Thirion}, \citenamefont {Grisel},
  \citenamefont {Blondel}, \citenamefont {Prettenhofer}, \citenamefont {Weiss},
  \citenamefont {Dubourg}, \citenamefont {Vanderplas}, \citenamefont {Passos},
  \citenamefont {Cournapeau}, \citenamefont {Brucher}, \citenamefont {Perrot},\
  and\ \citenamefont {Duchesnay}}]{scikit-learn}%
  \BibitemOpen
  \bibfield  {author} {\bibinfo {author} {\bibfnamefont {F.}~\bibnamefont
  {Pedregosa}}, \bibinfo {author} {\bibfnamefont {G.}~\bibnamefont
  {Varoquaux}}, \bibinfo {author} {\bibfnamefont {A.}~\bibnamefont {Gramfort}},
  \bibinfo {author} {\bibfnamefont {V.}~\bibnamefont {Michel}}, \bibinfo
  {author} {\bibfnamefont {B.}~\bibnamefont {Thirion}}, \bibinfo {author}
  {\bibfnamefont {O.}~\bibnamefont {Grisel}}, \bibinfo {author} {\bibfnamefont
  {M.}~\bibnamefont {Blondel}}, \bibinfo {author} {\bibfnamefont
  {P.}~\bibnamefont {Prettenhofer}}, \bibinfo {author} {\bibfnamefont
  {R.}~\bibnamefont {Weiss}}, \bibinfo {author} {\bibfnamefont
  {V.}~\bibnamefont {Dubourg}}, \bibinfo {author} {\bibfnamefont
  {J.}~\bibnamefont {Vanderplas}}, \bibinfo {author} {\bibfnamefont
  {A.}~\bibnamefont {Passos}}, \bibinfo {author} {\bibfnamefont
  {D.}~\bibnamefont {Cournapeau}}, \bibinfo {author} {\bibfnamefont
  {M.}~\bibnamefont {Brucher}}, \bibinfo {author} {\bibfnamefont
  {M.}~\bibnamefont {Perrot}}, \ and\ \bibinfo {author} {\bibfnamefont
  {E.}~\bibnamefont {Duchesnay}},\ }\href@noop {} {\bibfield  {journal}
  {\bibinfo  {journal} {Journal of Machine Learning Research}\ }\textbf
  {\bibinfo {volume} {12}},\ \bibinfo {pages} {2825} (\bibinfo {year}
  {2011})}\BibitemShut {NoStop}%
\bibitem [{\citenamefont {Chollet}\ \emph {et~al.}(2015)\citenamefont {Chollet}
  \emph {et~al.}}]{chollet2015keras}%
  \BibitemOpen
  \bibfield  {author} {\bibinfo {author} {\bibfnamefont {F.}~\bibnamefont
  {Chollet}} \emph {et~al.},\ }\href {https://github.com/fchollet/keras}
  {\enquote {\bibinfo {title} {Keras},}\ } (\bibinfo {year} {2015})\BibitemShut
  {NoStop}%
\bibitem [{\citenamefont {Abadi}\ \emph {et~al.}(2015)\citenamefont {Abadi},
  \citenamefont {Agarwal}, \citenamefont {Barham}, \citenamefont {Brevdo},
  \citenamefont {Chen}, \citenamefont {Citro}, \citenamefont {Corrado},
  \citenamefont {Davis}, \citenamefont {Dean}, \citenamefont {Devin},
  \citenamefont {Ghemawat}, \citenamefont {Goodfellow}, \citenamefont {Harp},
  \citenamefont {Irving}, \citenamefont {Isard}, \citenamefont {Jia},
  \citenamefont {Jozefowicz}, \citenamefont {Kaiser}, \citenamefont {Kudlur},
  \citenamefont {Levenberg}, \citenamefont {Man\'{e}}, \citenamefont {Monga},
  \citenamefont {Moore}, \citenamefont {Murray}, \citenamefont {Olah},
  \citenamefont {Schuster}, \citenamefont {Shlens}, \citenamefont {Steiner},
  \citenamefont {Sutskever}, \citenamefont {Talwar}, \citenamefont {Tucker},
  \citenamefont {Vanhoucke}, \citenamefont {Vasudevan}, \citenamefont
  {Vi\'{e}gas}, \citenamefont {Vinyals}, \citenamefont {Warden}, \citenamefont
  {Wattenberg}, \citenamefont {Wicke}, \citenamefont {Yu},\ and\ \citenamefont
  {Zheng}}]{tensorflow2015-whitepaper}%
  \BibitemOpen
  \bibfield  {author} {\bibinfo {author} {\bibfnamefont {M.}~\bibnamefont
  {Abadi}}, \bibinfo {author} {\bibfnamefont {A.}~\bibnamefont {Agarwal}},
  \bibinfo {author} {\bibfnamefont {P.}~\bibnamefont {Barham}}, \bibinfo
  {author} {\bibfnamefont {E.}~\bibnamefont {Brevdo}}, \bibinfo {author}
  {\bibfnamefont {Z.}~\bibnamefont {Chen}}, \bibinfo {author} {\bibfnamefont
  {C.}~\bibnamefont {Citro}}, \bibinfo {author} {\bibfnamefont {G.~S.}\
  \bibnamefont {Corrado}}, \bibinfo {author} {\bibfnamefont {A.}~\bibnamefont
  {Davis}}, \bibinfo {author} {\bibfnamefont {J.}~\bibnamefont {Dean}},
  \bibinfo {author} {\bibfnamefont {M.}~\bibnamefont {Devin}}, \bibinfo
  {author} {\bibfnamefont {S.}~\bibnamefont {Ghemawat}}, \bibinfo {author}
  {\bibfnamefont {I.}~\bibnamefont {Goodfellow}}, \bibinfo {author}
  {\bibfnamefont {A.}~\bibnamefont {Harp}}, \bibinfo {author} {\bibfnamefont
  {G.}~\bibnamefont {Irving}}, \bibinfo {author} {\bibfnamefont
  {M.}~\bibnamefont {Isard}}, \bibinfo {author} {\bibfnamefont
  {Y.}~\bibnamefont {Jia}}, \bibinfo {author} {\bibfnamefont {R.}~\bibnamefont
  {Jozefowicz}}, \bibinfo {author} {\bibfnamefont {L.}~\bibnamefont {Kaiser}},
  \bibinfo {author} {\bibfnamefont {M.}~\bibnamefont {Kudlur}}, \bibinfo
  {author} {\bibfnamefont {J.}~\bibnamefont {Levenberg}}, \bibinfo {author}
  {\bibfnamefont {D.}~\bibnamefont {Man\'{e}}}, \bibinfo {author}
  {\bibfnamefont {R.}~\bibnamefont {Monga}}, \bibinfo {author} {\bibfnamefont
  {S.}~\bibnamefont {Moore}}, \bibinfo {author} {\bibfnamefont
  {D.}~\bibnamefont {Murray}}, \bibinfo {author} {\bibfnamefont
  {C.}~\bibnamefont {Olah}}, \bibinfo {author} {\bibfnamefont {M.}~\bibnamefont
  {Schuster}}, \bibinfo {author} {\bibfnamefont {J.}~\bibnamefont {Shlens}},
  \bibinfo {author} {\bibfnamefont {B.}~\bibnamefont {Steiner}}, \bibinfo
  {author} {\bibfnamefont {I.}~\bibnamefont {Sutskever}}, \bibinfo {author}
  {\bibfnamefont {K.}~\bibnamefont {Talwar}}, \bibinfo {author} {\bibfnamefont
  {P.}~\bibnamefont {Tucker}}, \bibinfo {author} {\bibfnamefont
  {V.}~\bibnamefont {Vanhoucke}}, \bibinfo {author} {\bibfnamefont
  {V.}~\bibnamefont {Vasudevan}}, \bibinfo {author} {\bibfnamefont
  {F.}~\bibnamefont {Vi\'{e}gas}}, \bibinfo {author} {\bibfnamefont
  {O.}~\bibnamefont {Vinyals}}, \bibinfo {author} {\bibfnamefont
  {P.}~\bibnamefont {Warden}}, \bibinfo {author} {\bibfnamefont
  {M.}~\bibnamefont {Wattenberg}}, \bibinfo {author} {\bibfnamefont
  {M.}~\bibnamefont {Wicke}}, \bibinfo {author} {\bibfnamefont
  {Y.}~\bibnamefont {Yu}}, \ and\ \bibinfo {author} {\bibfnamefont
  {X.}~\bibnamefont {Zheng}},\ }\href {https://www.tensorflow.org/} {\enquote
  {\bibinfo {title} {{TensorFlow}: Large-scale machine learning on
  heterogeneous systems},}\ } (\bibinfo {year} {2015}),\ \bibinfo {note}
  {software available from tensorflow.org}\BibitemShut {NoStop}%
\bibitem [{\citenamefont {{Loshchilov}}\ and\ \citenamefont
  {{Hutter}}(2017)}]{2017arXiv171105101L}%
  \BibitemOpen
  \bibfield  {author} {\bibinfo {author} {\bibfnamefont {I.}~\bibnamefont
  {{Loshchilov}}}\ and\ \bibinfo {author} {\bibfnamefont {F.}~\bibnamefont
  {{Hutter}}},\ }\href {\doibase 10.48550/arXiv.1711.05101} {\bibfield
  {journal} {\bibinfo  {journal} {arXiv e-prints}\ ,\ \bibinfo {eid}
  {arXiv:1711.05101}} (\bibinfo {year} {2017})},\ \Eprint
  {http://arxiv.org/abs/1711.05101} {arXiv:1711.05101 [cs.LG]} \BibitemShut
  {NoStop}%
\bibitem [{\citenamefont {Dawson}\ \emph {et~al.}(2013)\citenamefont {Dawson},
  \citenamefont {Furlan},\ and\ \citenamefont {Lewis}}]{PhysRevD.87.014007}%
  \BibitemOpen
  \bibfield  {author} {\bibinfo {author} {\bibfnamefont {S.}~\bibnamefont
  {Dawson}}, \bibinfo {author} {\bibfnamefont {E.}~\bibnamefont {Furlan}}, \
  and\ \bibinfo {author} {\bibfnamefont {I.}~\bibnamefont {Lewis}},\ }\href
  {\doibase 10.1103/PhysRevD.87.014007} {\bibfield  {journal} {\bibinfo
  {journal} {Phys. Rev. D}\ }\textbf {\bibinfo {volume} {87}},\ \bibinfo
  {pages} {014007} (\bibinfo {year} {2013})}\BibitemShut {NoStop}%
\bibitem [{\citenamefont {Alves}\ and\ \citenamefont
  {Yamaguchi}(2022)}]{Alves:2022gnw}%
  \BibitemOpen
  \bibfield  {author} {\bibinfo {author} {\bibfnamefont {A.}~\bibnamefont
  {Alves}}\ and\ \bibinfo {author} {\bibfnamefont {C.~H.}\ \bibnamefont
  {Yamaguchi}},\ }\href {\doibase 10.1140/epjc/s10052-022-10714-1} {\bibfield
  {journal} {\bibinfo  {journal} {Eur. Phys. J. C}\ }\textbf {\bibinfo {volume}
  {82}},\ \bibinfo {pages} {746} (\bibinfo {year} {2022})},\ \Eprint
  {http://arxiv.org/abs/2203.03662} {arXiv:2203.03662 [hep-ph]} \BibitemShut
  {NoStop}%
\bibitem [{\citenamefont {Erdmann}\ \emph {et~al.}(2019)\citenamefont
  {Erdmann}, \citenamefont {Kallage}, \citenamefont {Kröninger},\ and\
  \citenamefont {Nackenhorst}}]{Erdmann_2019}%
  \BibitemOpen
  \bibfield  {author} {\bibinfo {author} {\bibfnamefont {J.}~\bibnamefont
  {Erdmann}}, \bibinfo {author} {\bibfnamefont {T.}~\bibnamefont {Kallage}},
  \bibinfo {author} {\bibfnamefont {K.}~\bibnamefont {Kröninger}}, \ and\
  \bibinfo {author} {\bibfnamefont {O.}~\bibnamefont {Nackenhorst}},\ }\href
  {\doibase 10.1088/1748-0221/14/11/p11015} {\bibfield  {journal} {\bibinfo
  {journal} {Journal of Instrumentation}\ }\textbf {\bibinfo {volume} {14}},\
  \bibinfo {pages} {P11015–P11015} (\bibinfo {year} {2019})}\BibitemShut
  {NoStop}%
\bibitem [{\citenamefont {Erdmann}\ \emph {et~al.}(2017)\citenamefont
  {Erdmann}, \citenamefont {Fischer},\ and\ \citenamefont
  {Rieger}}]{Erdmann_2017}%
  \BibitemOpen
  \bibfield  {author} {\bibinfo {author} {\bibfnamefont {M.}~\bibnamefont
  {Erdmann}}, \bibinfo {author} {\bibfnamefont {B.}~\bibnamefont {Fischer}}, \
  and\ \bibinfo {author} {\bibfnamefont {M.}~\bibnamefont {Rieger}},\ }\href
  {\doibase 10.1088/1748-0221/12/08/p08020} {\bibfield  {journal} {\bibinfo
  {journal} {Journal of Instrumentation}\ }\textbf {\bibinfo {volume} {12}},\
  \bibinfo {pages} {P08020–P08020} (\bibinfo {year} {2017})}\BibitemShut
  {NoStop}%
\bibitem [{\citenamefont {de~Favereau}\ \emph {et~al.}(2014)\citenamefont
  {de~Favereau}, \citenamefont {Delaere}, \citenamefont {Demin}, \citenamefont
  {Giammanco}, \citenamefont {Lema\^\i{}tre}, \citenamefont {Mertens},\ and\
  \citenamefont {Selvaggi}}]{deFavereau:2013fsa}%
  \BibitemOpen
  \bibfield  {author} {\bibinfo {author} {\bibfnamefont {J.}~\bibnamefont
  {de~Favereau}}, \bibinfo {author} {\bibfnamefont {C.}~\bibnamefont
  {Delaere}}, \bibinfo {author} {\bibfnamefont {P.}~\bibnamefont {Demin}},
  \bibinfo {author} {\bibfnamefont {A.}~\bibnamefont {Giammanco}}, \bibinfo
  {author} {\bibfnamefont {V.}~\bibnamefont {Lema\^\i{}tre}}, \bibinfo {author}
  {\bibfnamefont {A.}~\bibnamefont {Mertens}}, \ and\ \bibinfo {author}
  {\bibfnamefont {M.}~\bibnamefont {Selvaggi}} (\bibinfo {collaboration}
  {DELPHES 3}),\ }\href {\doibase 10.1007/JHEP02(2014)057} {\bibfield
  {journal} {\bibinfo  {journal} {JHEP}\ }\textbf {\bibinfo {volume} {02}},\
  \bibinfo {pages} {057} (\bibinfo {year} {2014})},\ \Eprint
  {http://arxiv.org/abs/1307.6346} {arXiv:1307.6346 [hep-ex]} \BibitemShut
  {NoStop}%
\bibitem [{\citenamefont {Huang}\ \emph {et~al.}(2017)\citenamefont {Huang},
  \citenamefont {No}, \citenamefont {Perni\'e}, \citenamefont {Ramsey-Musolf},
  \citenamefont {Safonov}, \citenamefont {Spannowsky},\ and\ \citenamefont
  {Winslow}}]{Huang:2017jws}%
  \BibitemOpen
  \bibfield  {author} {\bibinfo {author} {\bibfnamefont {T.}~\bibnamefont
  {Huang}}, \bibinfo {author} {\bibfnamefont {J.~M.}\ \bibnamefont {No}},
  \bibinfo {author} {\bibfnamefont {L.}~\bibnamefont {Perni\'e}}, \bibinfo
  {author} {\bibfnamefont {M.}~\bibnamefont {Ramsey-Musolf}}, \bibinfo {author}
  {\bibfnamefont {A.}~\bibnamefont {Safonov}}, \bibinfo {author} {\bibfnamefont
  {M.}~\bibnamefont {Spannowsky}}, \ and\ \bibinfo {author} {\bibfnamefont
  {P.}~\bibnamefont {Winslow}},\ }\href {\doibase 10.1103/PhysRevD.96.035007}
  {\bibfield  {journal} {\bibinfo  {journal} {Phys. Rev. D}\ }\textbf {\bibinfo
  {volume} {96}},\ \bibinfo {pages} {035007} (\bibinfo {year} {2017})},\
  \Eprint {http://arxiv.org/abs/1701.04442} {arXiv:1701.04442 [hep-ph]}
  \BibitemShut {NoStop}%
\bibitem [{\citenamefont {Elagin}\ \emph {et~al.}(2011)\citenamefont {Elagin},
  \citenamefont {Murat}, \citenamefont {Pranko},\ and\ \citenamefont
  {Safonov}}]{Elagin:2010aw}%
  \BibitemOpen
  \bibfield  {author} {\bibinfo {author} {\bibfnamefont {A.}~\bibnamefont
  {Elagin}}, \bibinfo {author} {\bibfnamefont {P.}~\bibnamefont {Murat}},
  \bibinfo {author} {\bibfnamefont {A.}~\bibnamefont {Pranko}}, \ and\ \bibinfo
  {author} {\bibfnamefont {A.}~\bibnamefont {Safonov}},\ }\href {\doibase
  10.1016/j.nima.2011.07.009} {\bibfield  {journal} {\bibinfo  {journal} {Nucl.
  Instrum. Meth. A}\ }\textbf {\bibinfo {volume} {654}},\ \bibinfo {pages}
  {481} (\bibinfo {year} {2011})},\ \Eprint {http://arxiv.org/abs/1012.4686}
  {arXiv:1012.4686 [hep-ex]} \BibitemShut {NoStop}%
\end{thebibliography}%
\end{document}